\documentclass[10pt,journal,compsoc]{IEEEtran}
\usepackage{graphicx}
\usepackage{wrapfig}
\usepackage{amsmath}
\usepackage{subfig}
\usepackage{esvect}
\usepackage[utf8]{inputenc}
\usepackage{color}
\usepackage[square,sort,comma,numbers]{natbib}
\usepackage{amssymb}

\usepackage[numbers]{natbib}
\usepackage{empheq} 
\usepackage{amsmath}
\usepackage{commath}
\usepackage{indentfirst}
\usepackage{amsfonts}
\usepackage{xcolor}
\DeclareMathOperator*{\argmax}{argmax}
\DeclareMathOperator*{\const}{const}
\begin{document}

\title{InversionNet: A Real-Time and Accurate Full Waveform Inversion with CNNs and continuous CRFs}

\author{Yue Wu$^1$ and Youzuo Lin$^{1,*}$
\IEEEcompsocitemizethanks{\IEEEcompsocthanksitem Y. Wu and Y. Lin are with the Geophysics Group, Earth and Environmental Sciences Division, Los Alamos National Laboratory, Los Alamos,
NM, 87544 USA. \protect \\

\IEEEcompsocthanksitem \textbf{*}  Correspondence to: Y. Lin, ylin@lanl.gov.}
}

\markboth{IEEE Transactions on Computational Imaging}%
{Shell \MakeLowercase{\textit{et al.}}: Bare Demo of IEEEtran.cls for IEEE Journals}


\IEEEtitleabstractindextext{
\begin{abstract}
Full-waveform inversion problems are usually formulated as optimization problems, where the forward-wave propagation operator $f$ maps the subsurface velocity structures to seismic signals. The existing computational methods for solving full-waveform inversion are not only computationally expensive, but also yields low-resolution results because of the ill-posedness and cycle skipping issues of full-waveform inversion. To resolve those issues, we employ machine-learning techniques to solve the full-waveform inversion. Specifically, we focus on applying the convolutional neural network~(CNN) to directly derive the inversion operator $f^{-1}$ so that the velocity structure can be obtained without knowing the forward operator $f$. We build a convolutional neural network with an encoder-decoder structure to model the correspondence from seismic data to subsurface velocity structures. Furthermore, we employ the conditional random field~(CRF) on top of the CNN to generate structural predictions by modeling the interactions between different locations on the velocity model. Our numerical examples using synthetic seismic reflection data show that the propose CNN-CRF model significantly improve the accuracy of the velocity inversion while the computational time is reduced.
\end{abstract}

}

\maketitle

\IEEEdisplaynontitleabstractindextext

\IEEEpeerreviewmaketitle

\ifCLASSOPTIONcompsoc
\IEEEraisesectionheading{\section{Introduction}\label{sec:introduction}}
\else
\section{Introduction}
\label{sec:introduction}
\fi

\IEEEPARstart{F}{ull}-waveform inversion~(FWI) plays an important role in various applications such as subsurface characterization in geoscience~\citep{introduction-2014-Virieux, Virieux-2009-Overview}, breast cancer detection in medicine~\cite{Ultrasound-2016-Lin, Ultrasound-2012-Lin}, etc. The numerical implementations of FWI can be in either the time domain or the frequency domain~\citep{Lin2015, Quantifying-2015-Lin, Guitton-2012-Blocky, Hu-2009-Simultaneous,  Vigh-2008-Comparisons}. FWI is a non-linear and ill-posed inverse problem and computationally expensive to solve~\citep{Virieux-2009-Overview}. There may exist many local minima when solving the minimization problem of inversion, making the technique less robust. To mitigate the ill-posedness of the problem, many approaches have been proposed and developed in recent years. The popular methods include: regularization-based techniques \citep{Ultrasound-2016-Lin, Ultrasound-2014-Lin, Ultrasound-2013-Lin, Ultrasound-2012-Lin, Hu-2009-Simultaneous,Burstedde-2009-Algorithmic,Ramirez-2010-Regularization,Guitton-2012-Blocky}, dynamic warping techniques \citep{Ma-2013-Wave, Qin-2013-Velocity}, prior information-based methods \citep{Ma-2012-Full,Ma-2012-Projected,Zhang-2013-double}, multiscale inversion approaches \citep{Bunks-1995-Multiscale,Tran-2013-Sinkhole}, and preconditioning methods \citep{Tang-2010-Preconditioning, Guitton-2012-Preconditioning}.

In recent years, with the largely increased computational power and the revitalization of deep neural networks~\citep{AlexNet,VGG,ResNet}, there is a surging trend of using data-driven methods for solving inverse problems in many scientific domains~\citep{Deep-2018-Lucas, Deep-2017-Jin}. Meanwhile, machine learning and deep learning methods have also drawn much attention in inverse problems applications~\citep{Deep-2018-Lucas, Randomization-2018-Lin,  Dr2-2017-Yao, Towards-2017-Lin, Schnetzler-2017-Use,  Accurate-2015-Kim}. In general, those different deep-learning based methods for solving inverse problems can be categorized into four types: 1)~to learn an end-to-end regression with vanilla convolutional neural network~(CNN), 2)~to learn higher-level representation, 3)~to gradual refinement of inversion procedure, and 4)~to incorporate with analytical methods and to learn a denoiser. The idea behind the first category is that a fully-connected neural network with a large number of neurons in its hidden layer has the ability to represent any functions, which is also known as the universal approximation theorem~\citep{Universal-1990-Hornik}. Examples of works that use the vanilla CNN include the work from \citet{Natural-2008-Jain}, where they use a five-layer CNN to denoise an image subjected to Gaussian noise. More recently,  \citet{Restoring-2013-Eigen} trained a CNN with three layers for denoising photographs that showed windows covered with dirt and rain. A common use of CNNs is to learn a compressed representation prior to constructing an output image. Several existing works use the effectiveness of autoencoders to learn relevant features to solve inverse problems in imaging. As an example, \citet{Coupled-2017-Zeng} employ the autoencoder's representation-learning capability to learn useful representations of low-resolution and high-resolution images. A shallow neural network is then trained to learn a correspondence between the learned low-resolution representation and the high-resolution representation. In the third category, CNNs are used to learn a residual between two or more layers by the skip connection from the input of the residual block to its output. This network structure is particularly well suited to inverse problems such as image restorations when the input and the output images share similar content. The work of \citet{Dr2-2017-Yao} and \citet{Accurate-2015-Kim} both belong to this category. Another type of research effort to solve inverse problems using neural networks is to incorporate analytical solutions. An example of this idea is LISTA~\cite{Learning-2010-Gregor}. Its basic idea is to start with an analytical approach and an associated inference algorithm and unfold the inference iterations as layers in a deep network. 

Provided with all the above relevant work,  there are some similarities between our inverse problems and the aforementioned inverse problems. All these work including ours are to infer the unknown from the known data. However, there are some unique characteristics associated with our inverse problems. In our inverse problems, the governing equation relating the recorded data and the velocity model is a wave equation, which describes the wave phenomenon and its propagation in the medium. To our knowledge, there are limited research works employing neural networks to solve FWI for a reconstruction. The only research works demonstrating the potential of deep learning in solving FWI problems include the work of~\citet{deep_learning_prior_seismic_image_fwi} and \citet{deep_inverse}. Specifically, \citet{deep_learning_prior_seismic_image_fwi} utilizes neural networks to generate some prior knowledge, which is used to inject into the conventional FWI iteration.  \citet{deep_inverse} uses recurrent neural network~(RNN) to solve the forward wave propagation modeling. Different from both of those research, in this work we developed a novel deep convolutional neural networks architecture (called ``InversionNet'') for the direct reconstruction of full-waveform inversion provided with seismic measurements.  

Our InversionNet is a data-driven model that learns a mapping from seismic waves to the subsurface velocity models. The architecture of our InversionNet is built upon CNNs due to the fact that CNNs have made substantial breakthroughs in processing image data. Considering the discrepancy of dimension size between seismic datasets and subsurface velocity models, we design an encoder-decoder CNN such that the encoder learns an abstract representation of the seismic data, which is then used by the decoder to produce a subsurface velocity model. Similar ideas can be found in biomedical image segmentation~\citep{UNet-2015-Ronneberger}.

One major challenge of FWI is to capture the subsurface structure, that is, the location of boundaries of layers and faults. Such structures can be reflected by the velocity model where values within each layer and the fault are nearly constant. However, these physics characteristics are difficult to capture by CNNs trained with per-pixel losses~(e.g., L1 or L2 losses). To address this issue, we couple the CNN with a conditional random field~(CRF) to generate velocity models with enhanced structural details. The potential of CRFs has been demonstrated in several computer vision domains including semantic segmentation~\citep{crf_nips2011, Deeplabv1, crfasrnn_iccv2015}, depth estimation~\citep{crf_multiscale_depth, crf_depth_liu} and remote sensing applications~\citep{CRF_RemoteSensing_2010, crf_aaai13}. CRFs are composed of a unary potential on individual nodes~(pixels or superpixels) and a pairwise potential on nodes that are connected. The nodes in the graph are usually enriched with low-level features such as color vectors and color histogram vectors. In our problem, low-level features of the input seismic data cannot translate to the velocity model so we instead use deep features from the decoder to represent nodes. Meanwhile, different strategies can be applied to build edges in the graph. \citet{CRF_RemoteSensing_2010, crf_multiscale_depth} and \citet{crf_depth_liu} model pairwise potential on neighboring nodes to enforce smoothness. \citet{crf_nips2011, crfasrnn_iccv2015, crf_aaai13}, and \citet{Deeplabv1} construct fully connected graphs where each node is connected to all other nodes in the graph so that long-range dependencies can also be captured. We find that the long-range dependencies on velocity models are not as significant as it is on image data. For effectiveness and efficiency considerations, we propose a \textit{locally connected} setting where each node is connected with all other nodes within a $d\times d$ window. 

We apply our methods to synthetic velocity models and seismic reflection data to numerically validate the performance of our InversionNet. As baseline methods, we compare our methods to two different physics-driven FWI methods: one with with advanced regularization techniques, which are recently developed in~\citet{Building-2017-Lin, Quantifying-2015-Lin, Lin2015}, and the other using energy-weighted preconditioning technique~\cite{Zhang_seg_2012}. Through comparison, we observe that our novel data-driven inversion method not only yields accurate inversion results but also significantly improves the computational efficiency. 

In the following sections, we first briefly describe the fundamentals of physics-driven versus data-driven methods, and deep neural networks~(Section~\ref{sec:background}). We then develop and discuss our novel inversion method - inversionNet~(Section~\ref{sec:inversionnet}). Section~\ref{sec:Experiments} describes the data we tested on, experimental setup, and experimental results we obtained. Finally, concluding remarks are presented in the Conclusions Section.

\section{Background}
\label{sec:background}

\subsection{Physics-Driven Techniques}

The physics-driven methods are those to infer subsurface model provided with governing physics and equations. Take the seismic exploration as an example. Seismic waves are mechanical perturbations that travel in the medium at a speed governed by the acoustic/elastic impedance of the medium in which they are traveling. In the time-domain, the acoustic-wave equation is given by 
\begin{equation}
\left [ \frac{1}{K(\mathbf{r})} \frac{\partial ^2}{\partial t ^2} 
- \nabla  \cdot \left ( \frac{1}{\rho (\mathbf{r})}\,\, \nabla \right 
) \right ]
p(\mathbf{r}, t) = s(\mathbf{r},\, t),
\label{eq:Forward}
\end{equation}
where $\rho (\mathbf{r})$ is the density at spatial location 
$\mathbf{r}$, $K(\mathbf{r})$ is the bulk modulus, $s(\mathbf{r},\, 
t)$ is the source term, $p(\mathbf{r}, t)$ is the pressure wavefield, 
and $t$ represents time.

The forward modeling problems in Eq.~\eqref{eq:Forward} can be written as
\begin{equation}
P = f(\mathbf{m}),
\label{eq:ForwardLinearM}
\end{equation}
where $P$ is the pressure wavefield for the acoustic case or the displacement wavefields for the elastic case, $f$ is the forward acoustic or elastic-wave modeling operator, and $\mathbf{m}$ is the velocity model parameter vector, including the density and compressional- and shear-wave velocities.  We use a time-domain stagger-grid finite-difference scheme to solve the acoustic- or elastic-wave equation. Throughout this paper, we consider only constant density acoustic or elastic media. 

The inverse problem of Eq.~(\ref{eq:ForwardLinearM}) is usually posed 
as a minimization problem~\cite{introduction-2014-Virieux, Virieux-2009-Overview}
\begin{equation}
E(\mathbf{m}) = \underset{\mathbf{m}}{\operatorname{min}} \left 
\{\left \| \mathbf{x} - f(\mathbf{m})\right \| _2 ^2 + \lambda\, R(\mathbf{m}) 
\right \},
\label{eq:MisFit}
\end{equation}
where $\mathbf{x}$ represents a recorded/field waveform dataset, 
$f(\mathbf{m})$ is the corresponding forward modeling result, $ \left \| \mathbf{x} - f(\mathbf{m})\right \| _2 ^2$ is the data misfit,  $||\cdot ||_2$ stands for the $\text{L}_2$ norm, $\lambda$ is a regularization parameter and $R(\mathbf{m})$ is the regularization term. The Tikhonov regularization and total-variation~(TV) regularization are the most commonly used.  The Tikhonov 
regularization is formulated as
\begin{equation}
E(\mathbf{m}) =  \underset{\mathbf{m}}{\operatorname{min}} \left 
 \{\left \| \mathbf{x} - f(\mathbf{m})\right \|  _2 ^2 + \lambda\, \| H \mathbf{m} \| ^2_2 \right \},
\label{eq:tikhonov}
\end{equation}
where the matrix $H$ is usually defined as a high-pass filtering 
operator, or an identity matrix. The Tikhonov regularization is an 
$\text{L}_2$-norm-based regularization and is best suited
for a smooth model $\mathbf{m}$. Waveform inversion with the Tikhonov 
regularization produces blurred interfaces for piecewise-constant 
velocity models.  To help preserve sharp interfaces in subsurface 
structures, total-variation (TV) regularization~\citep{Osher-1992-Nonlinear} has been incorprated into FWI, leading to 
\begin{equation}
E(\mathbf{m}) =  \underset{\mathbf{m}}{\operatorname{min}} \left 
 \{||\mathbf{x}  - f(\mathbf{m})|| _2 ^2 + \lambda \left\| \mathbf{m}
    \right\|_{\text{TV}} \right \},
\label{eq:tv}
\end{equation}
where the TV-norm for a 2D model is defined as
\begin{equation}
\| \mathbf{m} \|_{\text{TV}} = \sum_{1 \leq i, j \leq n}   
\sqrt{|(\nabla_x \mathbf{m})_{i, j}|^2 + |(\nabla_z \mathbf{m})_{i, 
j}|^2},
\label{eq:tvDef} \end{equation}
where $(\nabla _x \mathbf{m})_{i, j} = m_{i+1, j} - m_{i, j}$ and 
$(\nabla _z \mathbf{m})_{i, j}  = m_{i, j+1} - m_{i, j}$ are the 
spatial derivatives at a  spatial grid point $(i,j)$ in a Cartesian 
coordinate~$(x,z)$. The regularization
parameter $\lambda$ in eq.~(\ref{eq:tikhonov}) and eq.~(\ref{eq:tv}) plays 
an important of role of balancing the trade-off between the regularization
term and the data-misfit term.  Too much
regularization may be imposed on inversion if $\lambda$ is too large.    
Conversely, too small $\lambda$ may produce under-regularized 
inversion results. \citet{Lin2015} further developed a FWI with a modified total-variation~(MTV) regularization, which yields better results comparing to the FWI with conventional TV regularization term. The formulation of FWI with MTV regularization can be posed as
\begin{equation}
E(\mathbf{m}, \mathbf{u}) = \underset{\mathbf{m}, \mathbf{u}}{\operatorname{min}} \left 
\{\left \| \mathbf{x} - f(\mathbf{m})\right \| _2 ^2 + \lambda_1\, \| \mathbf{m} - \mathbf{u} \|_2 ^2  + \lambda _2 \left\| \mathbf{u}\right\|_{\text{TV}}
\right \}.
\label{eq:MTV}
\end{equation}

The current physics-driven computational techniques to infer the velocity model is based on gradient-based optimization methods, which are computationally expensive and often yield unsatisfactory resolution in identifying small structures~\citep{Building-2017-Lin, Quantifying-2015-Lin}.  In recent years, with the significantly improved computational power, machine learning and data mining have been successfully employed to various domains from science to engineering. In the next section, we provide a different perspective (data-driven approach) of obtaining velocity models from seismic measurements.

\subsection{Data-Driven Techniques}

In this paper, we adopt a data-driven approach, which means that we employ machine learning techniques directly to infer the velocity model and that no underlying  
physics is utilized. Specifically, suppose one has historical \emph{seismic measurement}. Overall, the idea of data-driven approach independent of applications can be illustrated as
\begin{equation*}
\displaystyle    \mathrm{\textbf{Seismic Measurements}}  \xrightarrow[]{f^{-1}} \mathrm{\textbf{Velocity Models}} .
\end{equation*}

For FWI problems, we feed a large amount of seismic data into the machine and train them to predict the corresponding velocity models. When the size of the training dataset is sufficiently large, the mapping from the seismic data to the velocity model can be correctly learned. Once the training phase is completed, the machine can predict the velocity model from new seismic data. 

\begin{figure*}[t]
\begin{center}
\includegraphics[width=\textwidth]{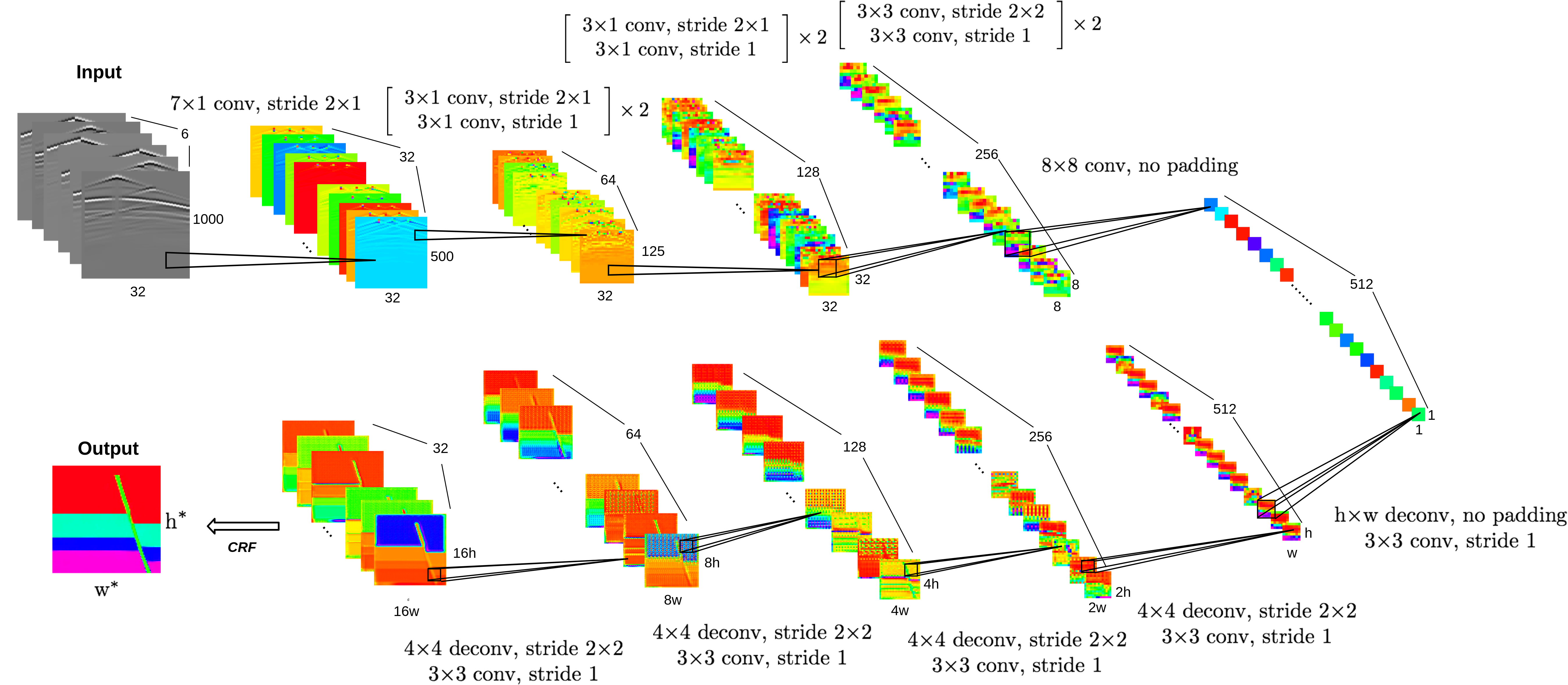}
\caption{An illustration of the proposed framework of the data-driven model. The CNN has an encoder-decoder architecture. The encoder~(the top pipeline) is primarily built with convolution layers, which extract high-level features from the input seismic data and compress them into a single high-dimensional vector. The decoder~(the bottom pipeline) then translates those features into velocity models through a set of deconvolution layers. The specification of each layer is provided in the figure. We build a locally connected CRF on top of the final feature map to generate the final predictions. }
\label{fig:overall_architecture}
\end{center}
\end{figure*}

With two different categories of methods introduced (``Data-Driven Methods'' V.S. ``Physics-Driven Methods''), it is worthwhile to mention the distinct differences between these two approaches. The problem of recovering the inherent parameters of a system (i.e. inverse problem) can be posed as the problem of regressing those parameters (even thousands) from the input measurements. However, unlike conventional optimization solutions, machine learning solutions have a strong data dependency, which is more severe when the regressing parameters are statistically independent. Though in practice the parameters exhibit strong correlations, the data requirement even for that case is quite high. In contrast, physics-driven methods are usually formulated as inverse problems where a solution vector can be calculated, without an explicit need for training data. 


\section{Methodology}
\label{sec:inversionnet}
The forward modeling of full-waveform inversion can be posed as 
\begin{equation}
f(\mathbf{m}) = \mathbf{x},
\label{eq:fwi_forward}
\end{equation}
where $f$ is the forward wave propagation operator, $\mathbf{m}$ is the subsurface  model, and $\mathbf{x}$ is the seismic data. In this work, our InversionNet is to directly obtain an approximation of $f^{-1}$ mapping from $\mathbf{x}$ to $\mathbf{m}$. We design our network to have an encoder-decoder architecture since our goal is to translate the data from one domain to other. Generally speaking, the encoder can be applied to extract high-level features from the input data and significantly reduce the data dimension. Then, the decoder is capable of translating those features into other domains according to our needs. The proposed InversionNet architecture with detail information of each layer is illustrated in Fig.~\ref{fig:overall_architecture}. All dimensions indicated in Fig.~\ref{fig:overall_architecture} are based on the dataset we use to evaluate our model. These dimensions may change when using other datasets, but the same methodology can be applied. A CRF is built on top of the decoder to produce structural predictions.

\subsection{Encoder} The encoder~(the top pipeline) includes a set of convolution blocks denoted by ``conv'' in Fig.~\ref{fig:overall_architecture}. Each convolution block consists of a convolution operation, batch normalization(BN)~\citep{BN} and ReLU~\citep{ReLU, LeakyReLU}, which are formulated as 
\begin{align}
	\mathbf{x}^{(l+1)} & = \text{ReLU}(\text{BN}(\text{Conv}(\mathbf{x}^{(l)}))), \label{eq:conv_block} \\
	\text{Conv}(\mathbf{x})_{(i, j)} & = \sum_{m}\sum_{n}\sum_{c}K_{m, n, c} \cdot \mathbf{x}_{(s-1)\times i+m, (s-1)\times j+n, c}, \label{eq:conv}\\
	\text{ReLU}(\mathbf{x}) & = 
	\begin{cases}
	\mathbf{x} & \text{if} \quad \mathbf{x} \geq 0 \\
	\alpha \mathbf{x} & \text{if} \quad \mathbf{x} < 0,
	\end{cases} \label{eq:relu} \\
	\text{BN}_{\gamma, \beta}(\mathbf{x}_{i, j, c}) & = \gamma \left(\frac{\mathbf{x}_{i, j, c} - \mu_{\mathcal{B}}}{\sqrt{\sigma_{\mathcal{B}}^2 + \epsilon}} \right) + \beta, 
	\label{eq:bn}
\end{align} 
where both the input image $\mathbf{x}$ and kernel $K$ are 3D tensors with the first two dimensions indicating the spatial location, $s$ denotes the stride between each sliding location of the kernel, $\gamma$ and $\beta$ are two trainable parameters, $\mu_{\mathcal{B}}$ and $\sigma_{\mathcal{B}}^2$ are the mean and variance calculated with all values on the same feature map over the mini-batch, and $\epsilon$ is a small constant added for numerical stability. 

The spatial dimensions of the convolution kernels and strides are given in Fig.~\ref{fig:overall_architecture}. Layers in brackets are repeated twice (weights are not shared). Initial convolutions are 1D, which is because the time dimension is greatly larger so we start with incorporating temporal features of the seismic wave. We do not pad zeros in the last convolution layer so that the feature map can be compressed into a single vector. This is reasonable since it is unnecessary to preserve the temporal and spatial correlations in the seismic data. 

\subsection{Decoder} The decoder~(the bottom pipeline) consists of mixed convolution and deconvolution blocks. Deconvolution~(a.k.a. transposed convolution) produces outputs with a larger size than the input, which can be achieved by padding zeros on the input feature map. ``deconv'' in Fig.~\ref{fig:overall_architecture} denotes a deconvolution block that replaces the convolution in Eq.~\eqref{eq:conv_block} with deconvolution. In each deconvolution block, we apply $4 \times 4$ kernels with stride 2 on the input feature map to double the resolution, followed by a regular convolution layer with $3 \times 3$ kernels to refine the upsampled feature maps. 


\subsection{Conditional Random Fields}

We build locally connected CRF on the final feature map from the decoder to model the interaction between output values on the velocity model. 

A CRF is defined by a Gibbs distribution 
\begin{align}
P(\mathbf{y}|\mathbf{x}) & = \frac{1}{Z(\mathbf{x})} \exp (-E(\mathbf{y} | \mathbf{x})), \label{eq:gibbs_distribution}\\
E(\mathbf{y} | \mathbf{x}) & = \sum_{c \in C(\mathcal{G})} \phi_c(\mathbf{y_{c}} | \mathbf{x}),\\
Z(\mathbf{x}) & = \int_{\mathbf{y}} \exp(-E(\mathbf{y} | \mathbf{x})) \text{d}\mathbf{y},
\end{align}
where $\mathbf{y} = \{y_1,...,y_n\}$ and $\mathbf{x} = \{\mathbf{x}_{1},...,\mathbf{x}_{n}\}$ are two sets of variables, $\mathcal{G} = (\mathcal{V}, \mathcal{E})$ is a graph defined on $\mathbf{x}$ with a set of cliques $C(\mathcal{G})$, each clique $c$ has a potential $\phi_{c}$ and $E(\mathbf{y} | \mathbf{x})$ is an energy function summing up all potentials, and $Z(x)$ is a normalizing constant. An inference is made by the maximum a posteriori~(MAP) $\mathbf{y^{*}} = \argmax_{\mathbf{y}} P(\mathbf{y}|\mathbf{x})$. The parameters in the CRF can be optimized by maximizing $\log P(\mathbf{y}|\mathbf{x})$.

In our problem, $\mathbf{x}$ ranges over all velocity models of size $n$ and $\mathbf{y}$ ranges over all possible velocity values. The velocity values are implicitly conditioned on each velocity model. The energy function of a CRF consists of a unary potential $\phi_{u}$ and a pairwise potential $\phi_{p}$
\begin{equation}
	E(\mathbf{y} | \mathbf{x}) = \sum_{i \in \mathcal{V}} \phi_{u}(y_{i} | \mathbf{x}) + \sum_{i \in \mathcal{V}} \sum_{j \in \mathcal{N}_{i}} \phi_{p}(y_{i}, y_{j} | \mathbf{x}) \label{eq:energy_function_fccrf},
\end{equation}
where $\mathcal{N}_{i}$ denotes a set of nodes connected to $y_{i}$.

The unary potential models a mapping between the input and each individual output $y_{i}$. The pairwise potential models the interaction between outputs $y_{i}$ and $y_{j}$. We define $\phi_{u}$, $\phi_{p}$ as
\begin{align}
\phi_{u}(y_{i} | \mathbf{x}) & = (y_{i} - z_{i})^{2}, \label{eq:unary_potential}\\
\phi_{p}(y_{i}, y_{j} | \mathbf{x}) & = w \cdot k(\mathbf{f}_{i}, \mathbf{f}_{j}) (y_{i} - y_{j})^{2}, \label{eq:pairwise_potential}
\end{align}
where $z_{i},...,z_{n}$ are velocity values predicted by the CNN, and $w$ is a weight to be learned. $k$ is similarity function defined as 
\begin{equation}
	k(\mathbf{f}_{i}, \mathbf{f}_{j}) = \exp(-\lambda_{1}||\mathbf{I}_{i} - \mathbf{I}_{j}|| - \lambda_{2}||\mathbf{p}_{i} - \mathbf{p}_{j}||),
\end{equation}
where $\mathbf{I}$ is the feature vector from the final feature map generated by the decoder, $\mathbf{p}$ is the position vector, $\lambda^{(1)}$ and $\lambda^{(2)}$ are hyperparameters. 

\subsubsection{Approximate Inference}

The exact inference on the proposed CRF representation $P(\mathbf{y}|\mathbf{x})$ requires $O(n^{3})$ complexity as it needs to compute the inverse of a large matrix~\citep{CRF_RemoteSensing_2010, crf_depth_liu}. We instead apply mean field theory to compute a distribution $Q(\mathbf{y}|\mathbf{x})$ that can be factorized as $Q(\mathbf{y}|\mathbf{x}) =\prod_{i} Q_{i}(y_{i}|\mathbf{x})$ to approximate $P(\mathbf{y}|\mathbf{x})$ by minimizing Kullback-Leibler~(KL) divergence between $P$ and $Q$~\citep{Bishop}. The optimal $Q(\mathbf{y}|\mathbf{x})$ has the form
\begin{equation}
	\log Q_{i}(y_{i}|\mathbf{x}) = \mathbb{E}_{j\in \mathcal{N}_{i}}[\log P(\mathbf{y}|\mathbf{x})] + \const,
	\label{eq:optimal_q_i_general_form} 
\end{equation}
where $\mathbb{E}_{j \in \mathcal{N}_{i}}$ denotes the expectation of $\log P(\mathbf{y}|\mathbf{x})$ under distributions $Q_{j}(y_{j}|\mathbf{x})$ for $j \in \mathcal{N}_{i}$.
Combine Eq.~\eqref{eq:gibbs_distribution}, \eqref{eq:energy_function_fccrf}, \eqref{eq:unary_potential}, \eqref{eq:pairwise_potential} and \eqref{eq:optimal_q_i_general_form}, we have
\begin{equation}
	\begin{split}
		\log Q_{i}(y_{i}|\mathbf{x}) & = (y_{i} - z_{i})^{2} + w \sum_{j \in \mathcal{N}_{i}} k(\mathbf{f}_{i}, \mathbf{f}_{j}) (y_{i} - y_{j})^{2} \\
		& = (1 + w \sum_{j \in \mathcal{N}_{i}} k(\mathbf{f}_{i}, \mathbf{f}_{j})) y_{i}^{2} \\
		& - 2(z_{i} + w \sum_{j \in \mathcal{N}_{i}} k(\mathbf{f}_{i}, \mathbf{f}_{j}) \mathbb{E}[y_{j}]) y_{i} + \const.
	\end{split}
\end{equation}
Since $Q_{i}(y_{i}|\mathbf{x})$ is a quadratic function w.r.t $y_{j}$, it can be represented by a Gaussian distribution with
\begin{align}
	\mu_{i} & = \frac{z_{i} + w \sum_{j \in \mathcal{N}_{i}} k(\mathbf{f}_{i}, \mathbf{f}_{j}) \mu_{j}}{1 + w \sum_{j \in \mathcal{N}_{i}} k(\mathbf{f}_{i}, \mathbf{f}_{j})}, \label{eq:mean_update}\\
	\sigma_{i}^{2} & = \frac{1}{2(1 + w \sum_{j \in \mathcal{N}_{i}} k(\mathbf{f}_{i}, \mathbf{f}_{j}))} \label{eq:variance_update}.
\end{align}
We enforce $w \ge 0$ to make each $Q_{i}(y_{i} | \mathbf{x})$ a valid distribution, since $k(\mathbf{f}_{i}, \mathbf{f}_{j}) > 0$. To obtain the optimal solution for each $\mu$, we iteratively calculate $Q_{1}(y_{1}|\mathbf{x})$,...,$Q_{n}(y_{n}|\mathbf{x})$ using Eq.\eqref{eq:mean_update} and Eq.\eqref{eq:variance_update} until the convergence criterion is satisfied. We use the unary prediction $z$ as the initial guess for $\mu$.

In inference phase, we perform MAP on each factorized distributions $Q_{i}$ to obtain $y_{i}$
\begin{equation}
	\begin{split}
		y_{i} & = \argmax_{y_{i}} Q_{i}(y_{i} | \mathbf{x}) \\ 
			  & = \mu_{i}.
	\end{split}
\end{equation}

\subsubsection{Learning}
\label{sec:crf_learning}
We aim to find an optimal parameter $w$ to maximize the log-likelihood $ \log P(\mathbf{y}|\mathbf{x})$. By utilizing $Q(\mathbf{y}|\mathbf{x})$, we can instead efficiently optimize the approximate log-likelihood:
\begin{equation}
	 \mathcal{L} (Q \text{;}\ w) = \sum_{i \in \mathcal{V}} \log Q_{i}(y_{i} | \mathbf{x}) \label{eq:log_likelihood}.
\end{equation}
The optimal $w^{\star}$ can be learned by the gradient ascent algorithm. Taking the derivative w.r.t $w$ in Eq.~\eqref{eq:log_likelihood}, we have
\begin{equation}
	\frac{\partial \mathcal{L} (Q \text{;}\ w)}{\partial w} = \sum_{i \in \mathcal{V}} ( - \frac{\partial E_{i}(y_{i} | \mathbf{x})}{\partial w} - \frac{\partial \log Z_{i}}{\partial w}).
\end{equation}
The derivative w.r.t $E_{i}(y_{i} | \mathbf{x})$ can be calculated from Eq.~\eqref{eq:energy_function_fccrf}
\begin{equation}
	\frac{\partial E_{i}(y_{i} | \mathbf{x})}{\partial w} = \sum_{j \in \mathcal{N}_{i}} k(\mathbf{f}_{i}, \mathbf{f}_{j}) (y_{i} - y_{j}) ^ {2}.
	\label{eq:derivative_logE}
\end{equation}
The derivative w.r.t $\log Z_{i}$ is
\begin{equation}
\begin{split}
	\frac{\partial \log Z_{i}}{\partial w} & = - \int_{y_{i}} \frac{1}{Z_{i}} \exp(-E_{i}(y_{i} | \mathbf{x})) \frac{\partial E_{i}(y_{i} | \mathbf{x})}{\partial w} \text{d}y_{i},\\
	& = - \int_{y_{i}} Q_{i}(y_{i} | \mathbf{x}) \frac{\partial E_{i}(y_{i} | \mathbf{x})}{\partial w} \text{d}y_{i}, \\
	& = - \sum_{j \in \mathcal{N}_{i}} k(\mathbf{f}_{i}, \mathbf{f}_{j}) \mathbb{E}_{y_{i} \thicksim Q_{i}} [(y_{i} - y_{j}) ^ {2}].
\end{split}
\label{eq:derivative_logZ}
\end{equation}
Combine Eq.~\eqref{eq:derivative_logE} and~\eqref{eq:derivative_logZ}, we have
\begin{equation}
\begin{split}
	\frac{\partial \mathcal{L} (Q \text{;}\ w)}{\partial w} & = \sum_{i \in \mathcal{V}} \sum_{j \in \mathcal{N}_{i}} k(\mathbf{f}_{i}, \mathbf{f}_{j}) (\mathbb{E}_{y_{i} \thicksim Q_{i}} [(y_{i} - y_{j}) ^ {2}] - (y_{i} - y_{j})^{2}), \\
	& = \sum_{i \in \mathcal{V}} \sum_{j \in \mathcal{N}_{i}} k(\mathbf{f}_{i}, \mathbf{f}_{j}) (\mu_{i}^{2} + \sigma_{i}^{2} - 2\mu_{i}y_{j} -y_{i}^{2} + 2y_{i}y_{j}).
\end{split}
\end{equation}
We apply the projected gradient ascent to project $w$ to $0$ whenever the constraint $w \ge 0$ is violated.

We initialize $w^{(0)}$ to 0. Since we make $z$ as the initial guess for $\mu$, we can directly calculate $w^{(1)}$ with $\mu_{i}=z_{i}$ and $\sigma^{2}=0.5$. The values of hyperparameters $\lambda^{(1)}$ and $\lambda^{(2)}$ can be found with the grid search on a validation set. 

\subsubsection{Computational Cost Analysis}
For both inference and learning phases, it requires to iterate over all nodes and their connecting nodes. The complexity is $\mathcal{O}(cd^{2}n)$, where $c$ is the number of mean field iterations, $d$ is the window size of the locally connected CRF, and $n$ is the number of nodes. Since $c << n$ and $d^{2} << n$, the overall complexity of the CRF is $\mathcal{O}(n)$.

\section{Experimental Settings}
\label{sec:Experiments}

\subsection{Data}

\begin{figure*}
\centering
\centerline{
\subfloat{\includegraphics[width=.1\linewidth]{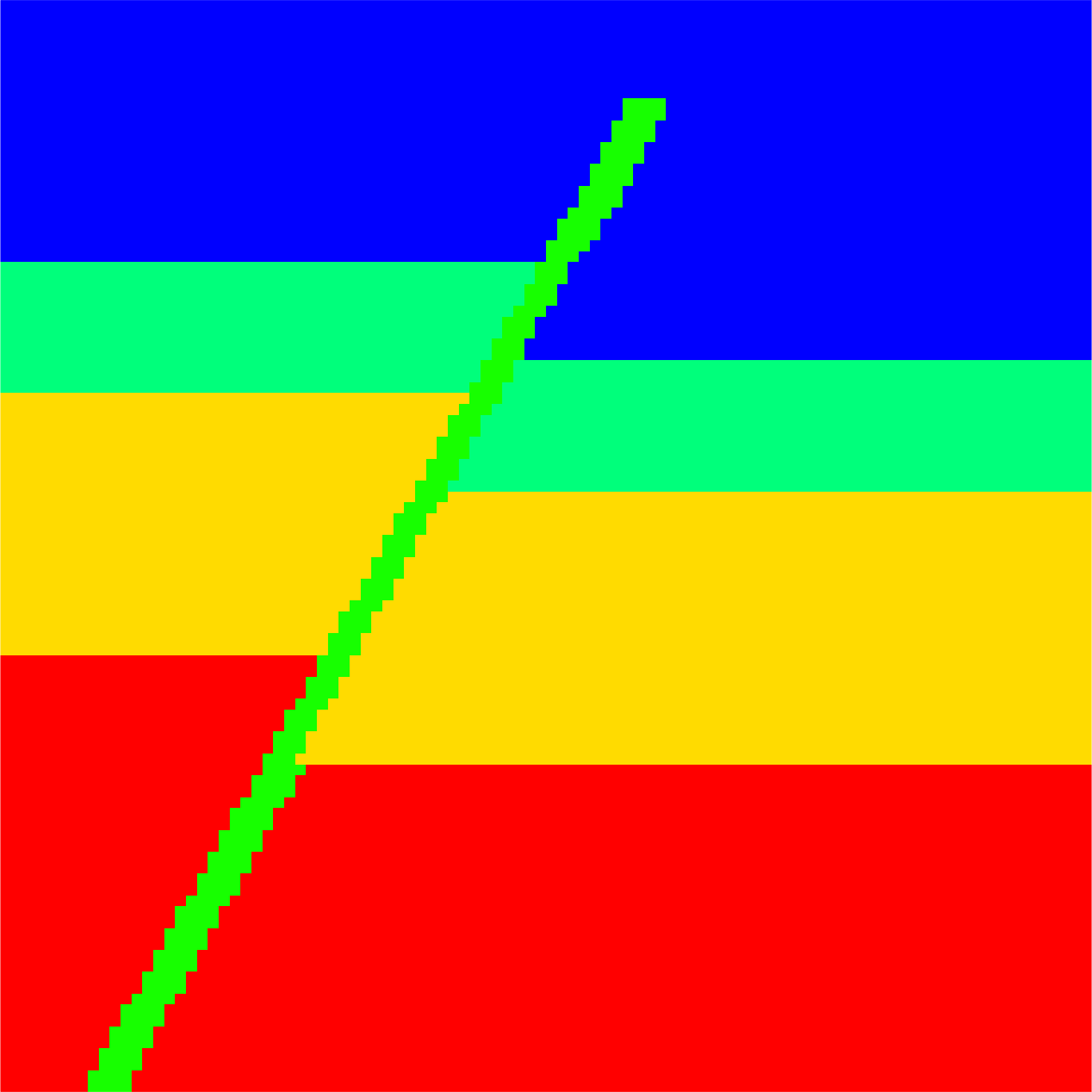}}
\hspace{0.05cm}
\subfloat{\includegraphics[width=.1\linewidth]{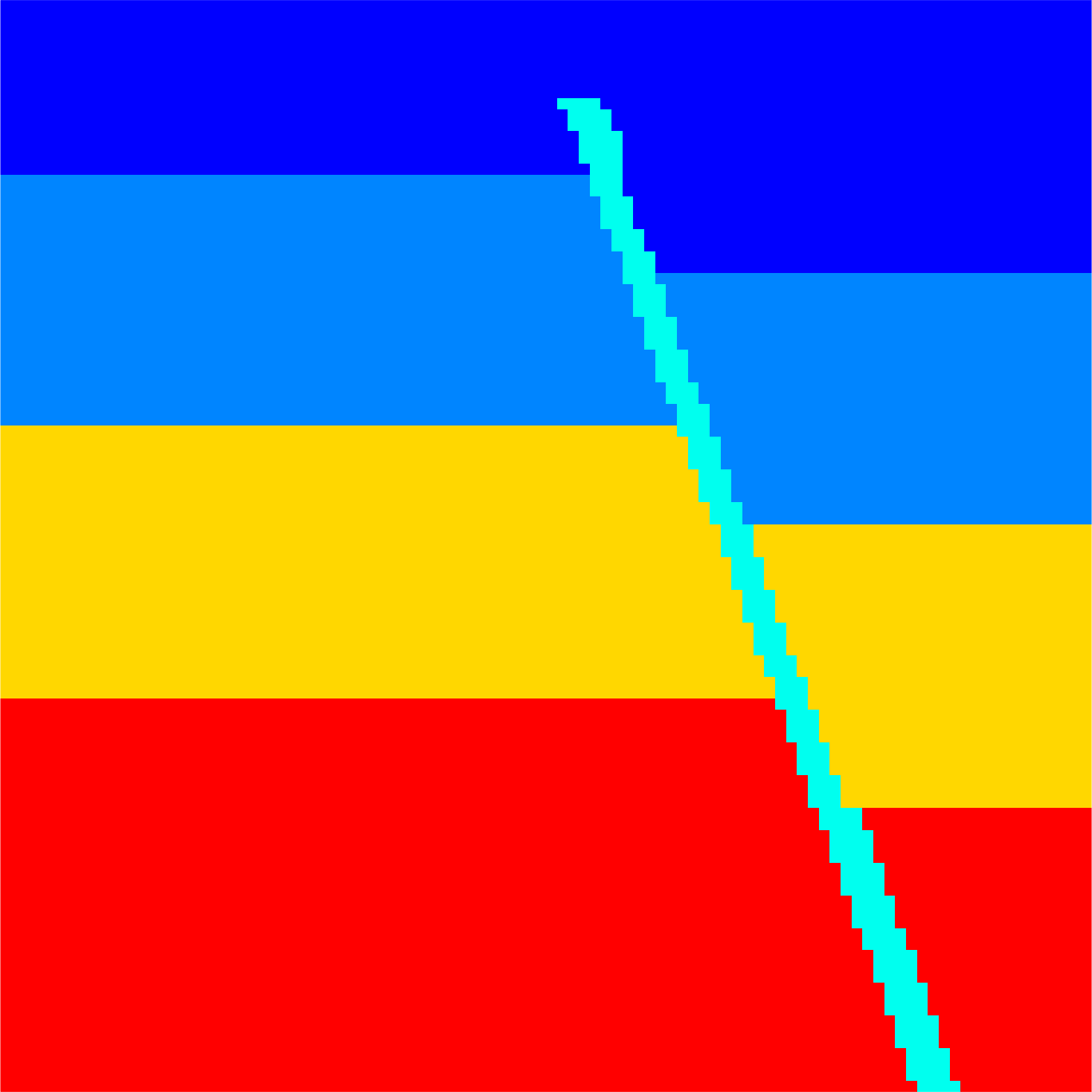}}
\hspace{0.05cm}
\subfloat{\includegraphics[width=.1\linewidth]{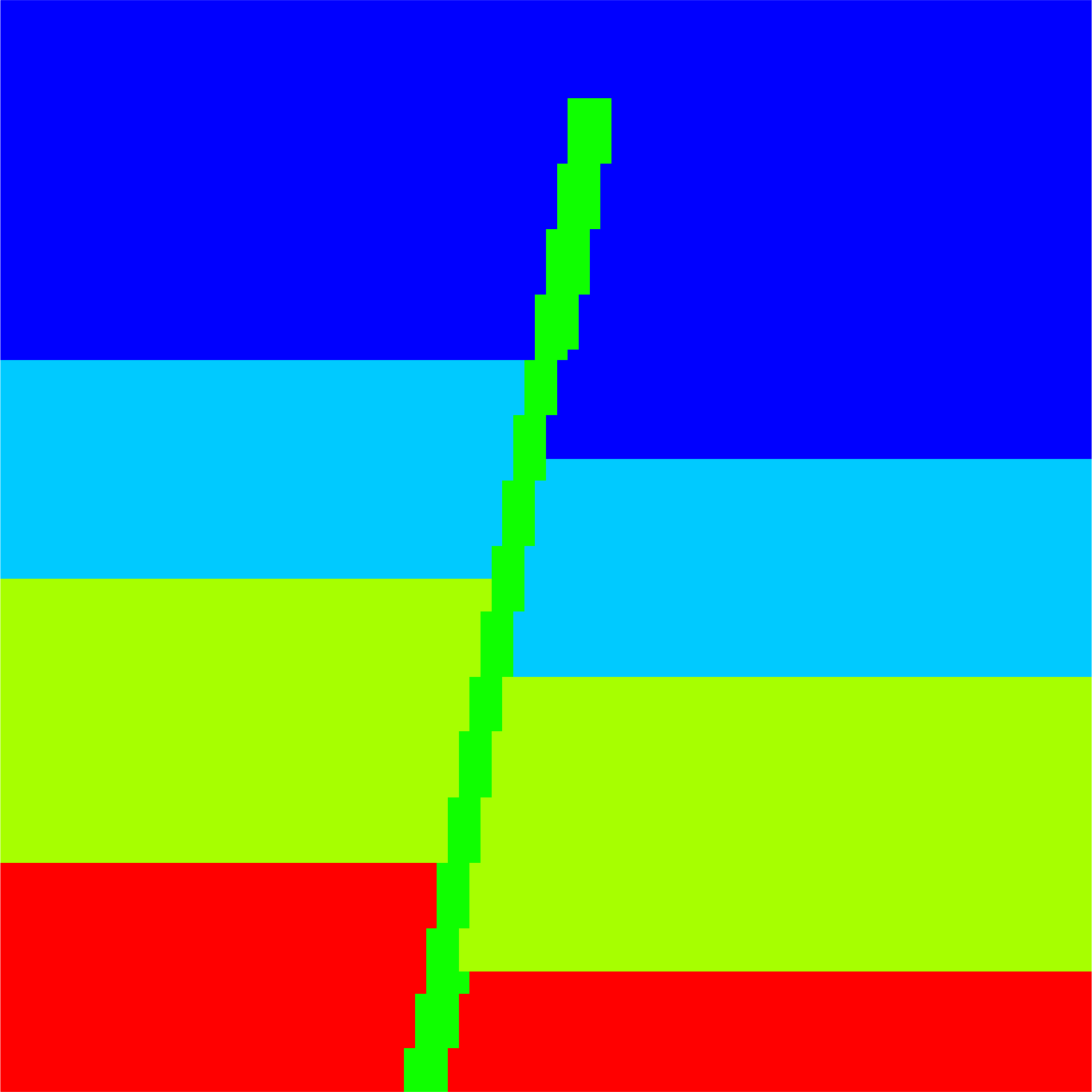}}
\hspace{0.05cm}
\subfloat{\includegraphics[width=.1\linewidth]{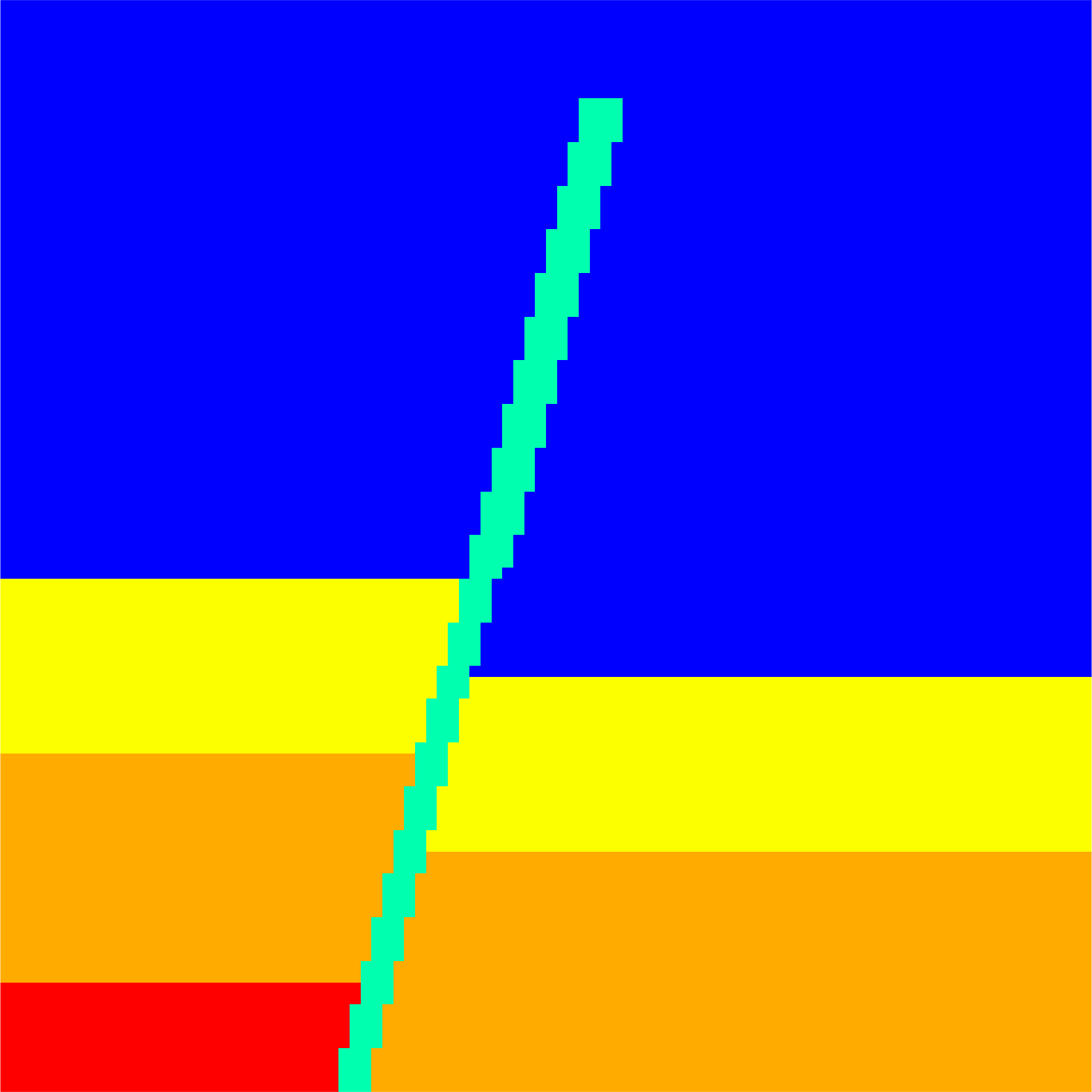}}
\hspace{0.05cm}
\subfloat{\includegraphics[width=.1\linewidth]{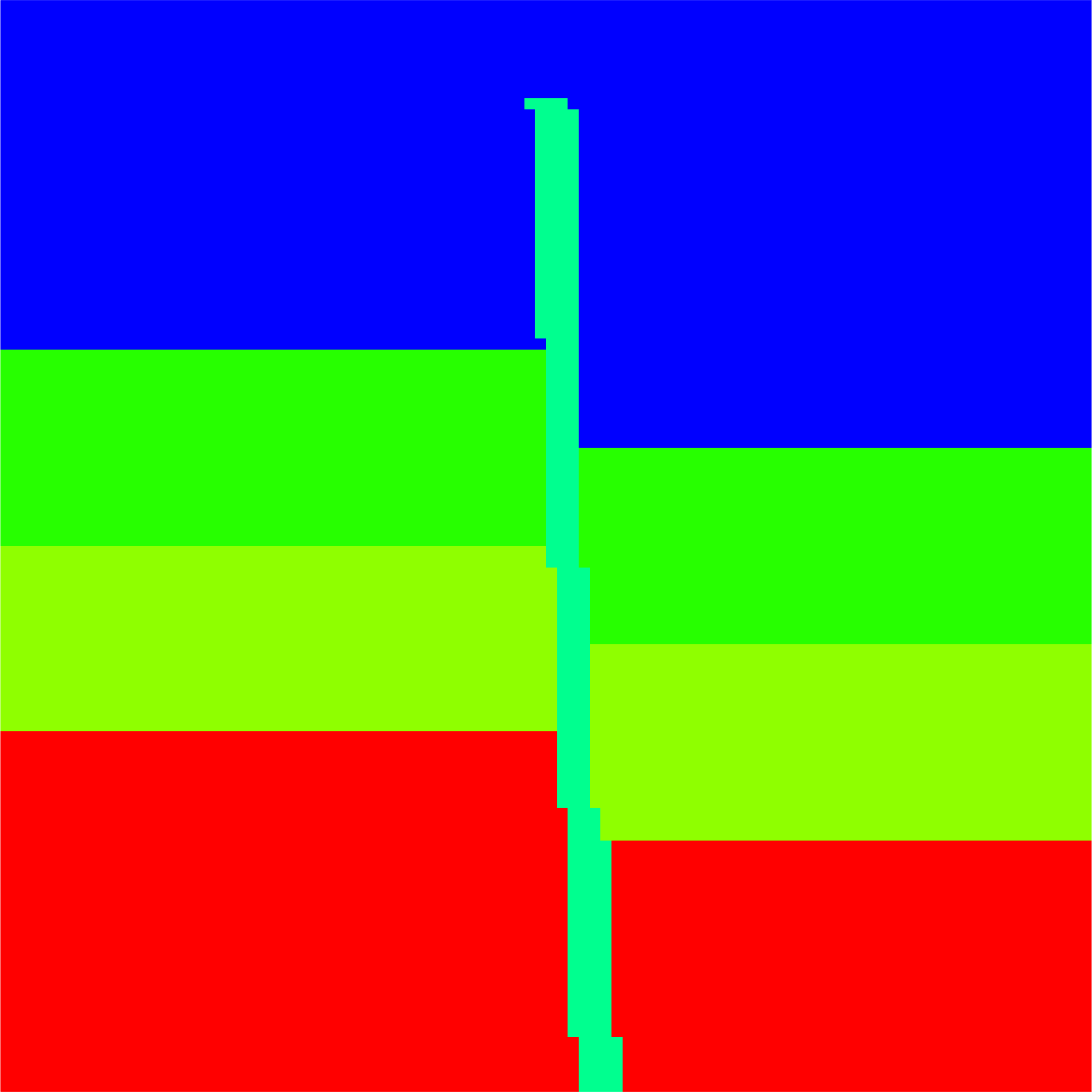}}
\hspace{0.3cm}
\subfloat{\includegraphics[width=.1\linewidth]{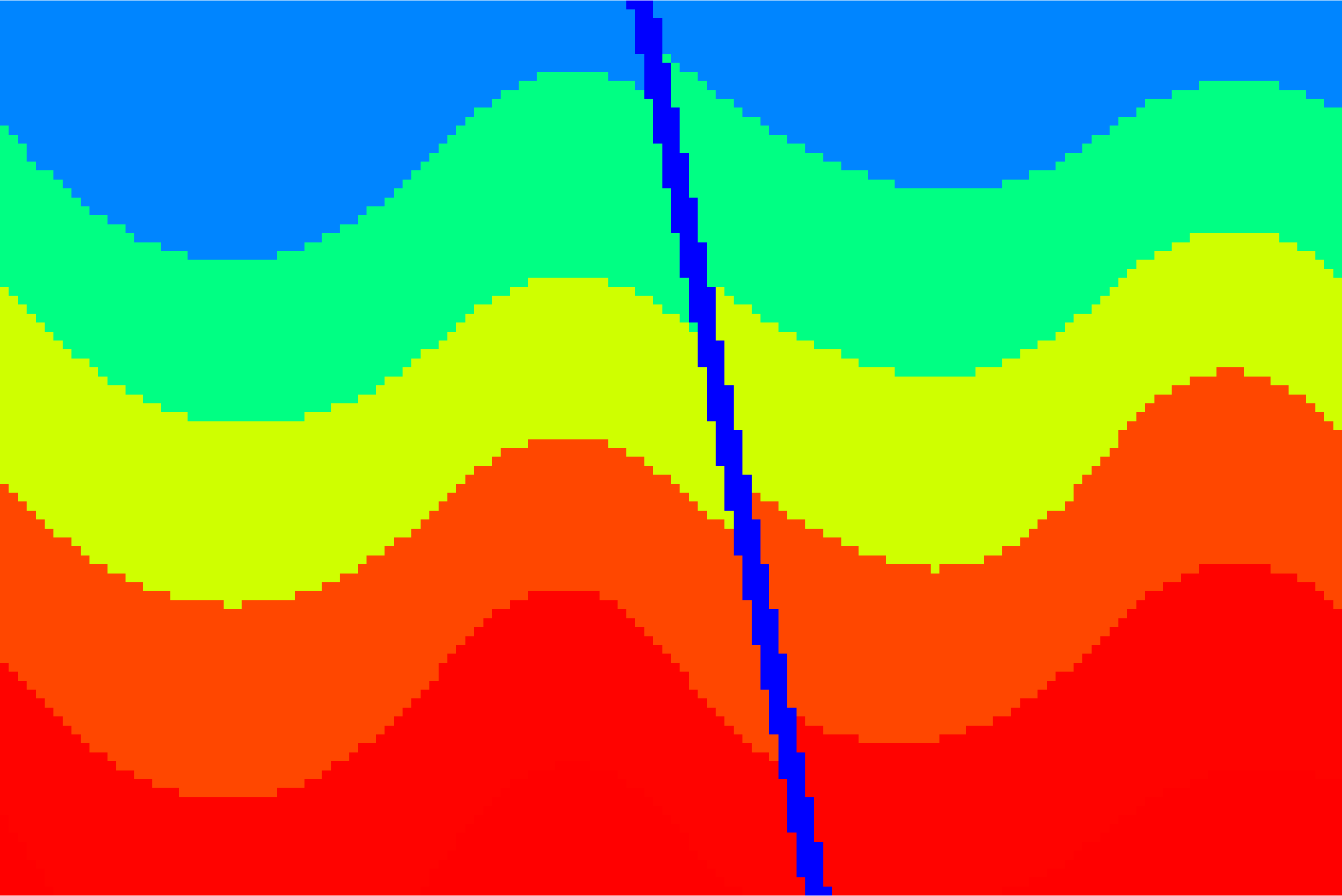}}
\hspace{0.05cm}
\subfloat{\includegraphics[width=.1\linewidth]{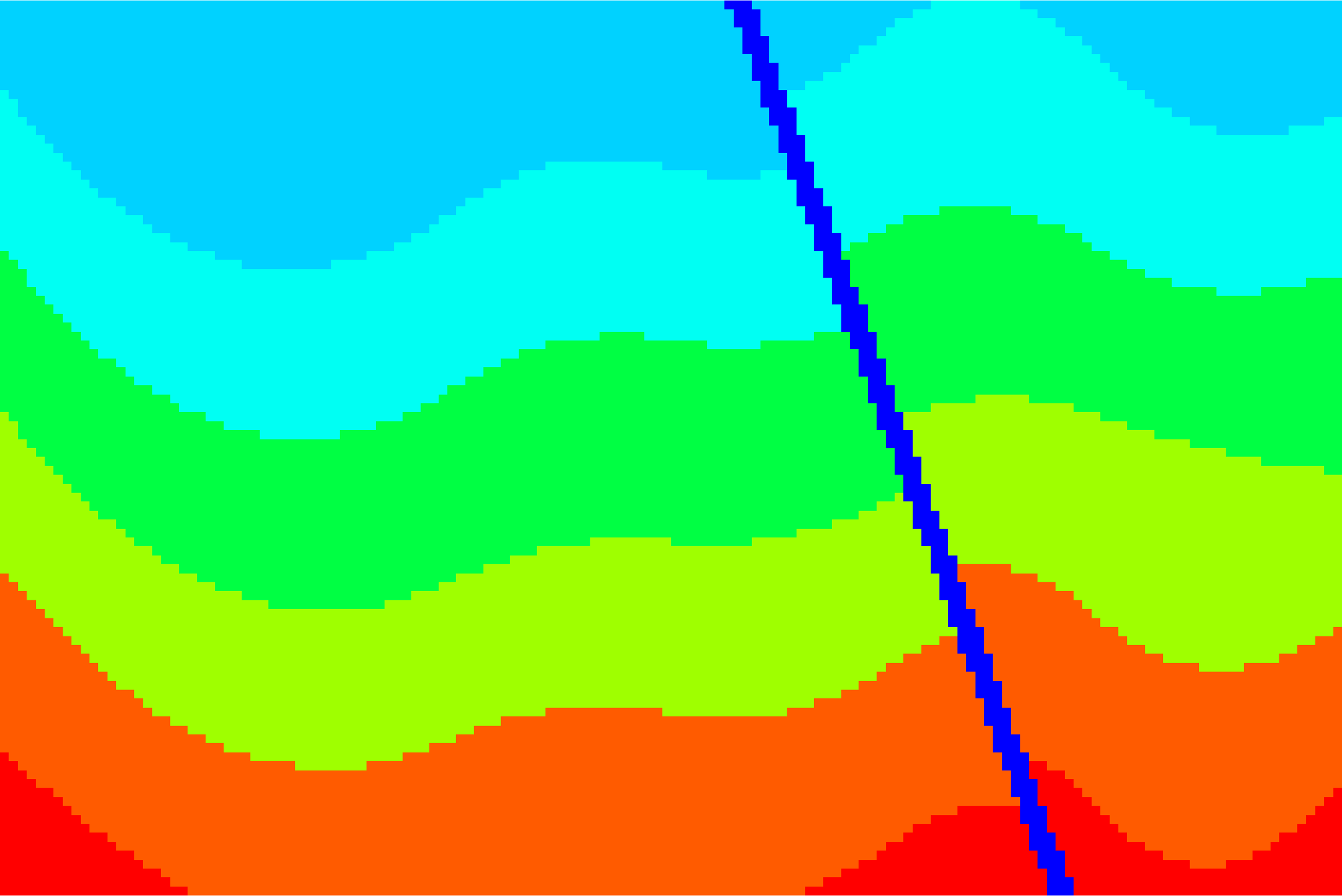}}
\hspace{0.05cm}
\subfloat{\includegraphics[width=.1\linewidth]{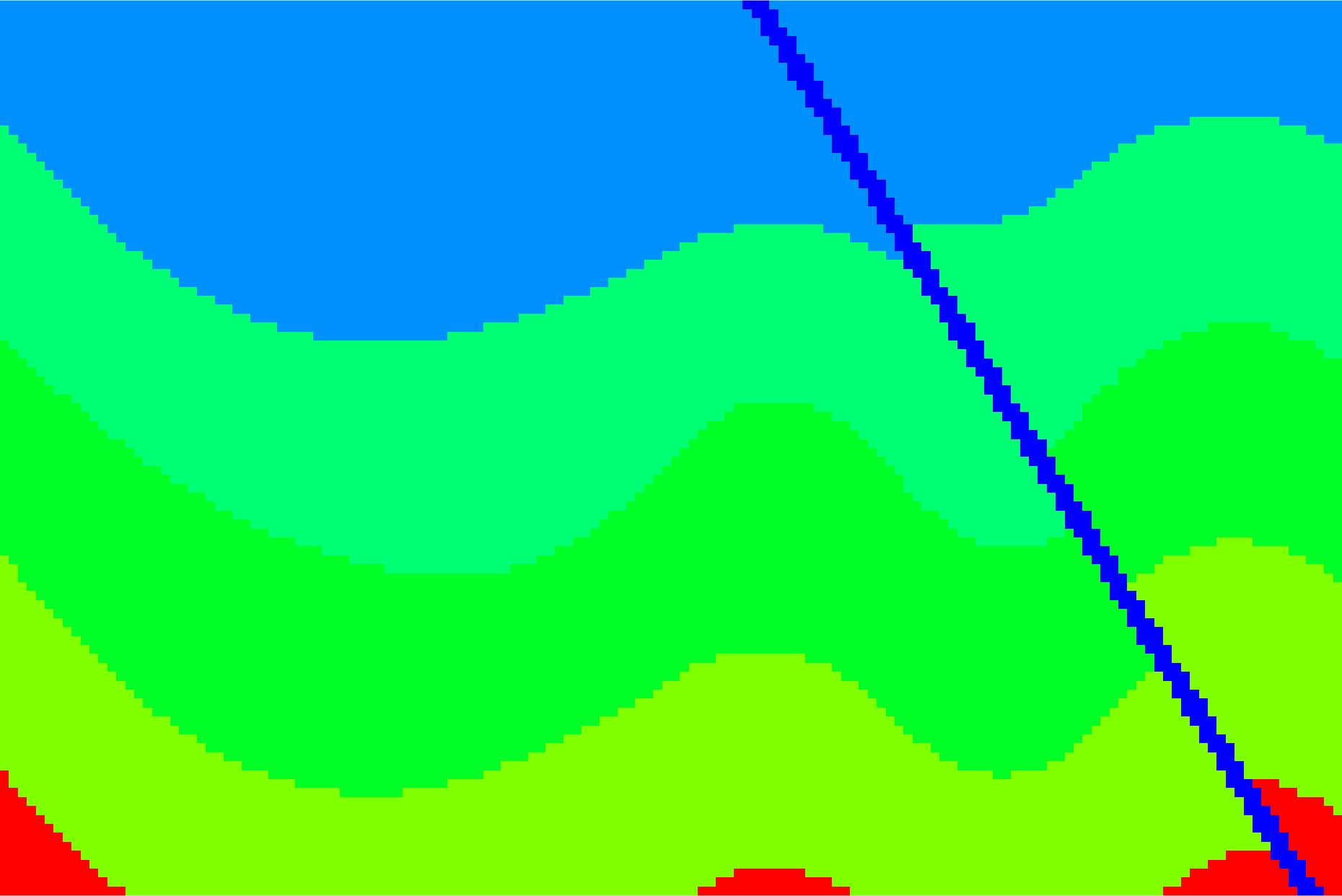}}
\hspace{0.05cm}
\subfloat{\includegraphics[width=.1\linewidth]{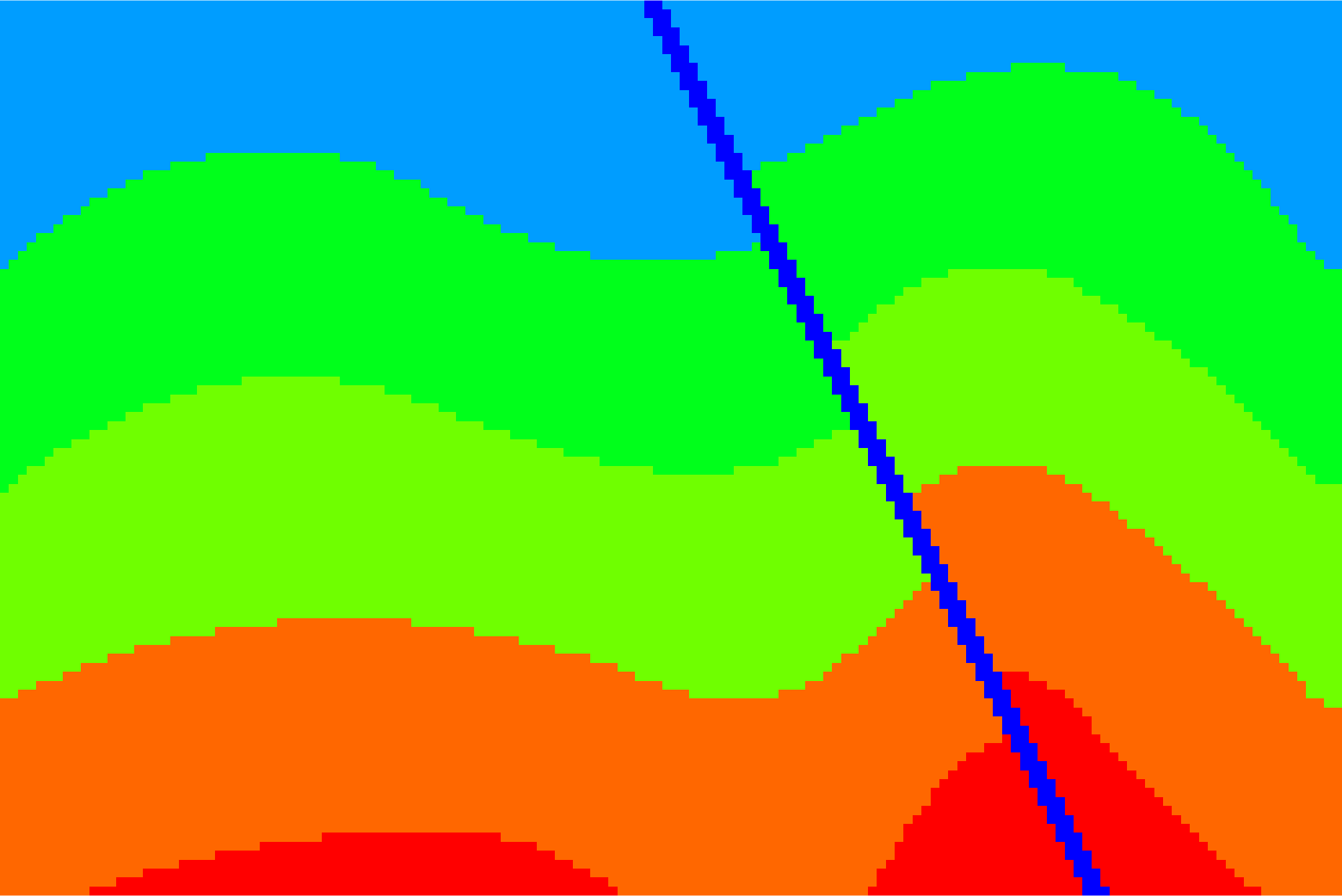}}
\hspace{0.05cm}
\subfloat{\includegraphics[width=.1\linewidth]{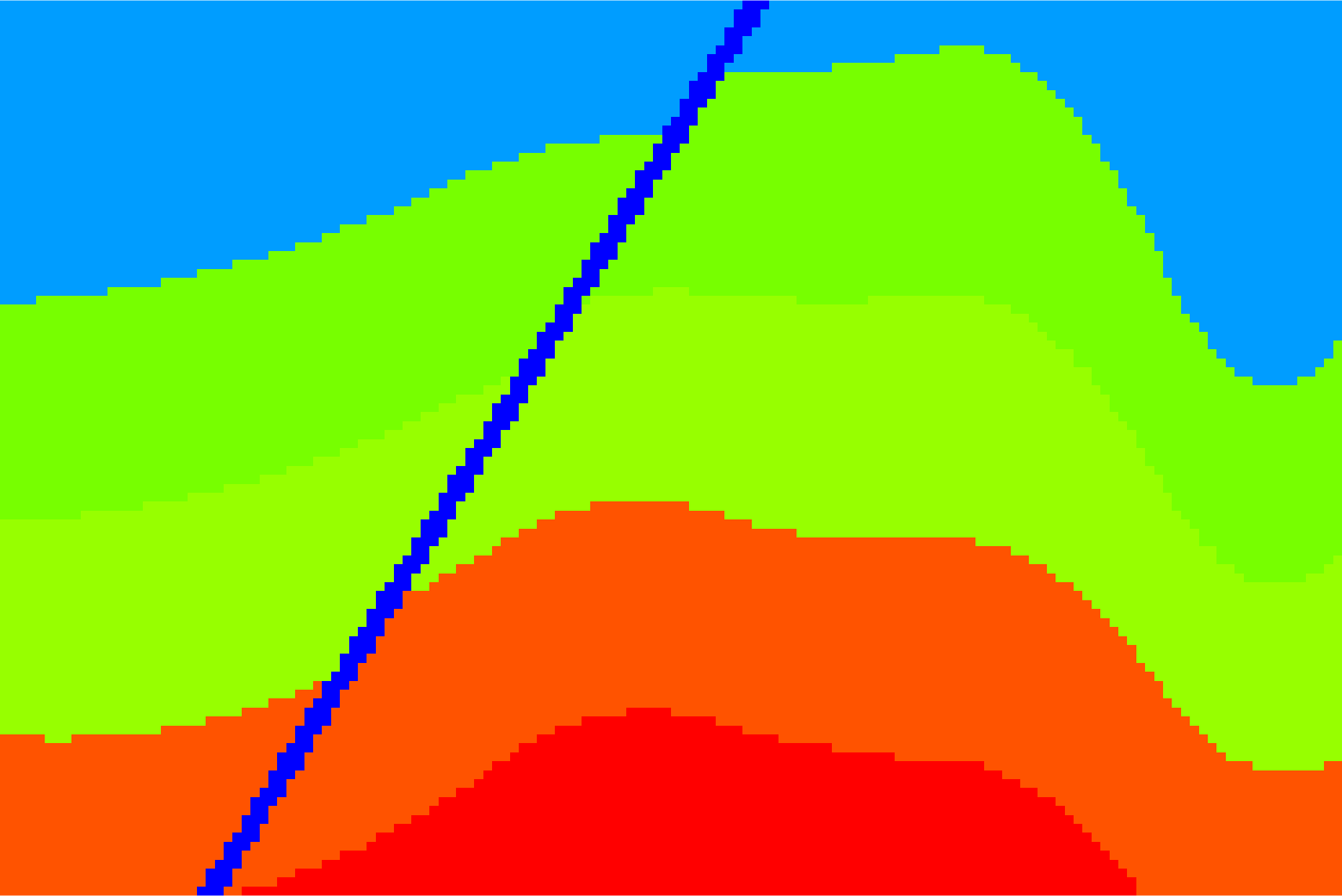}}}
\vspace{0.1cm}
\centerline{
\subfloat{\includegraphics[width=.1\linewidth]{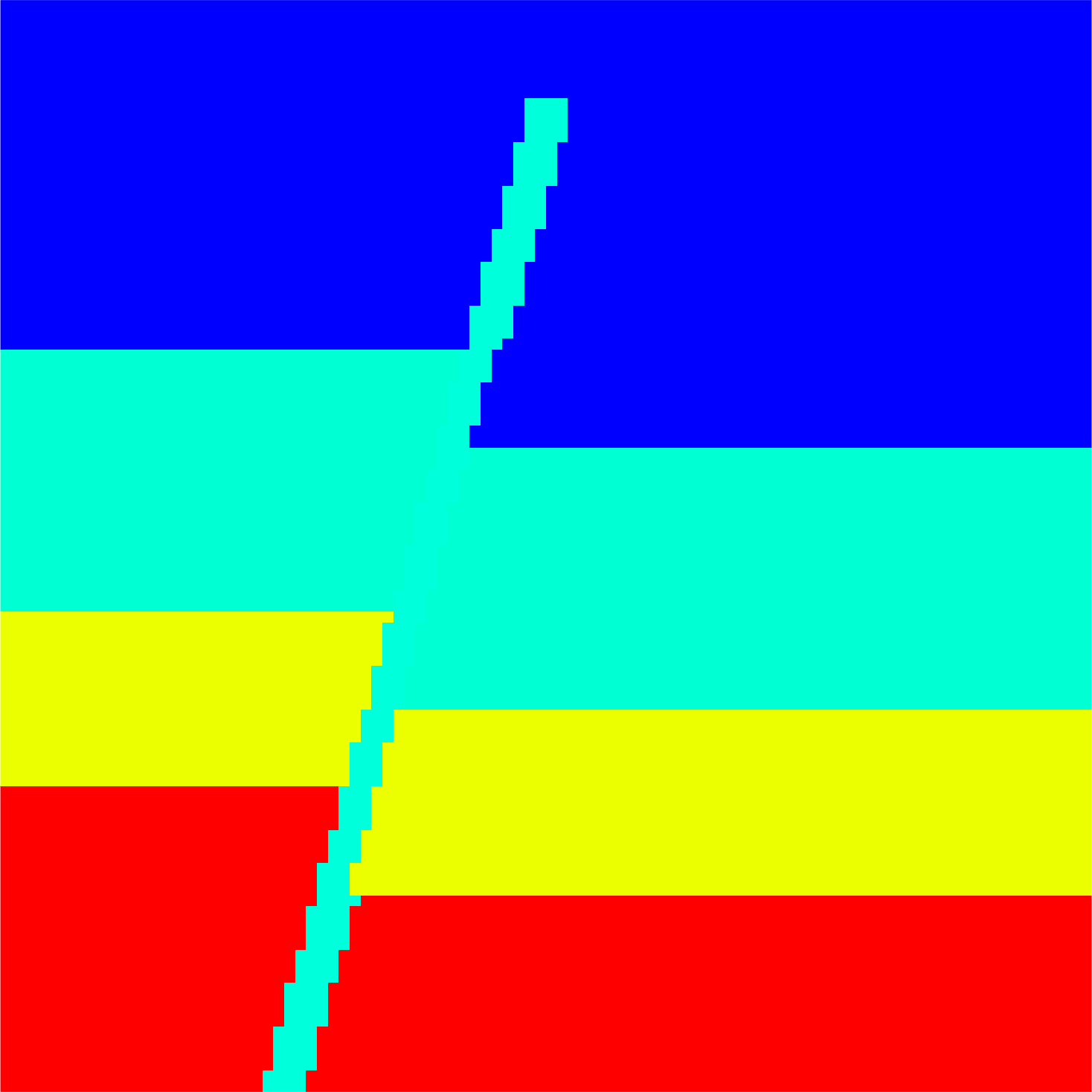}}
\hspace{0.05cm}
\subfloat{\includegraphics[width=.1\linewidth]{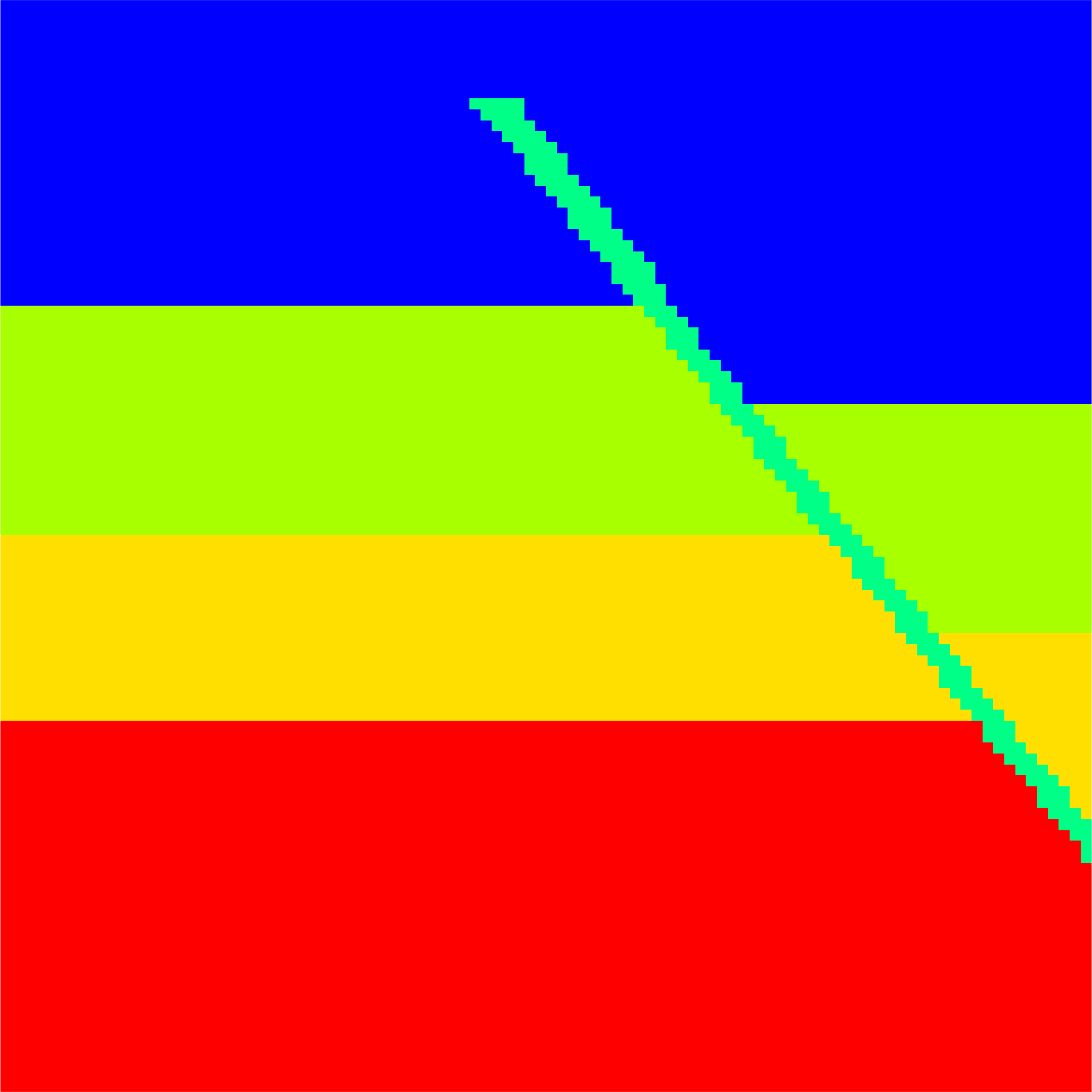}}
\hspace{0.05cm}
\subfloat{\includegraphics[width=.1\linewidth]{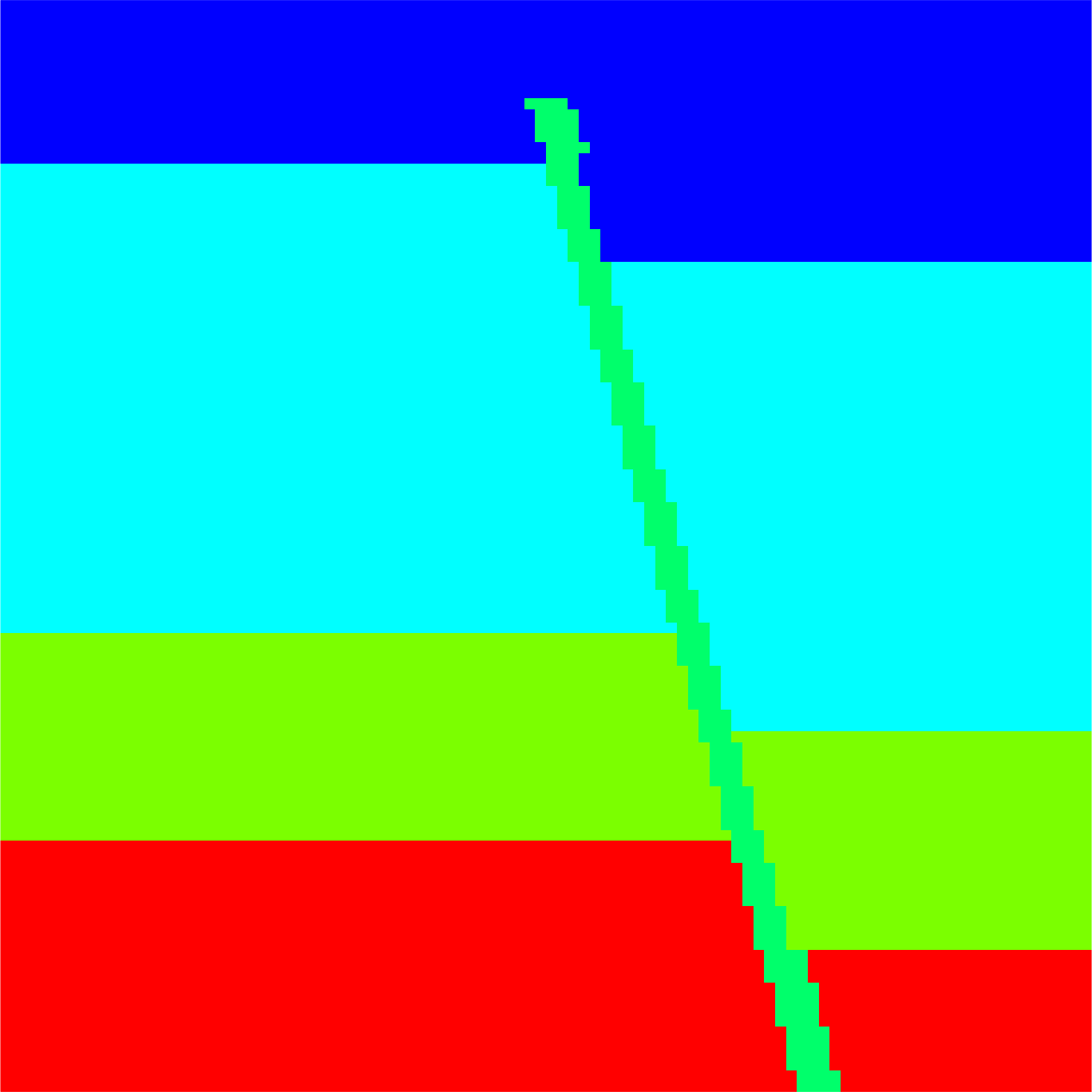}}
\hspace{0.05cm}
\subfloat{\includegraphics[width=.1\linewidth]{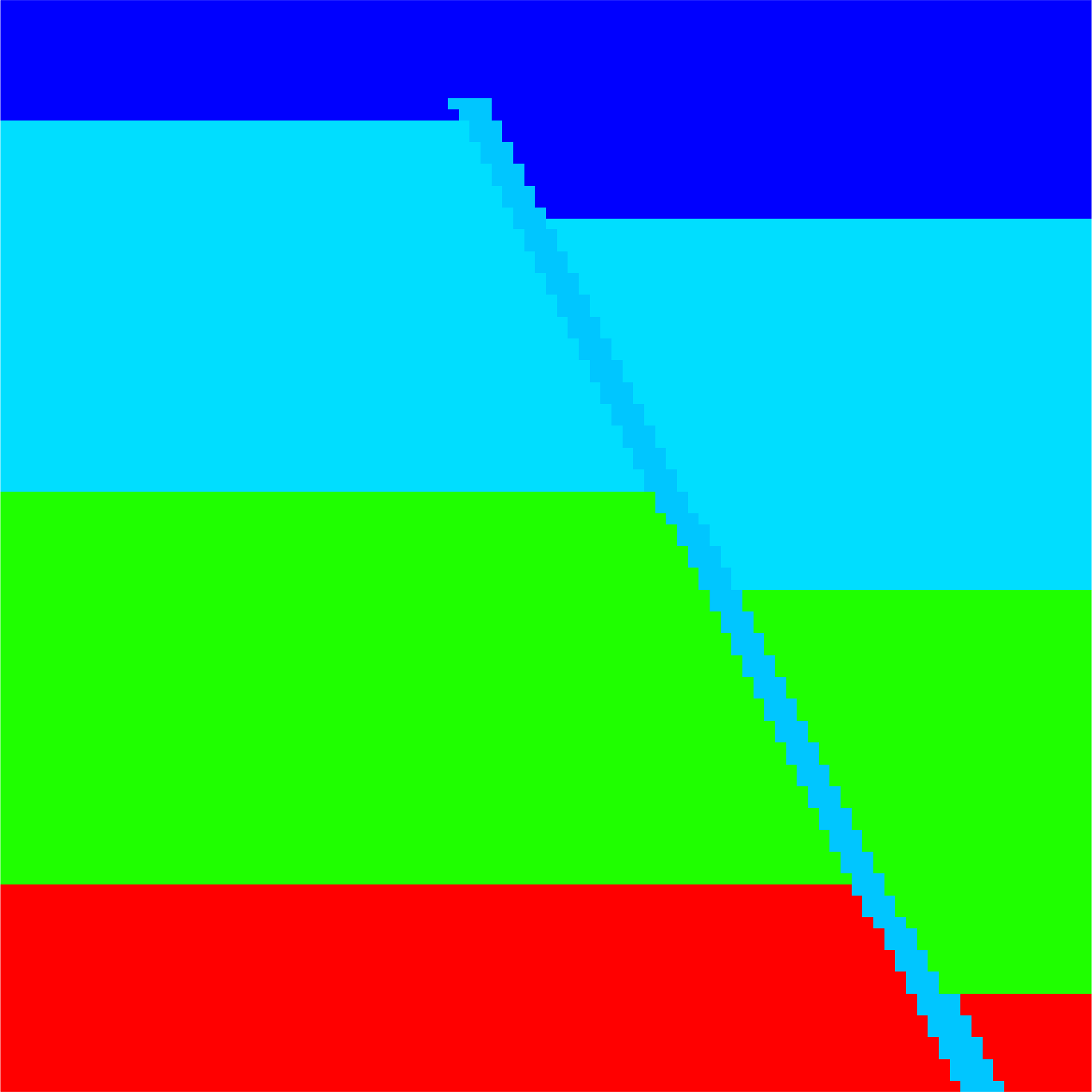}}
\hspace{0.05cm}
\subfloat{\includegraphics[width=.1\linewidth]{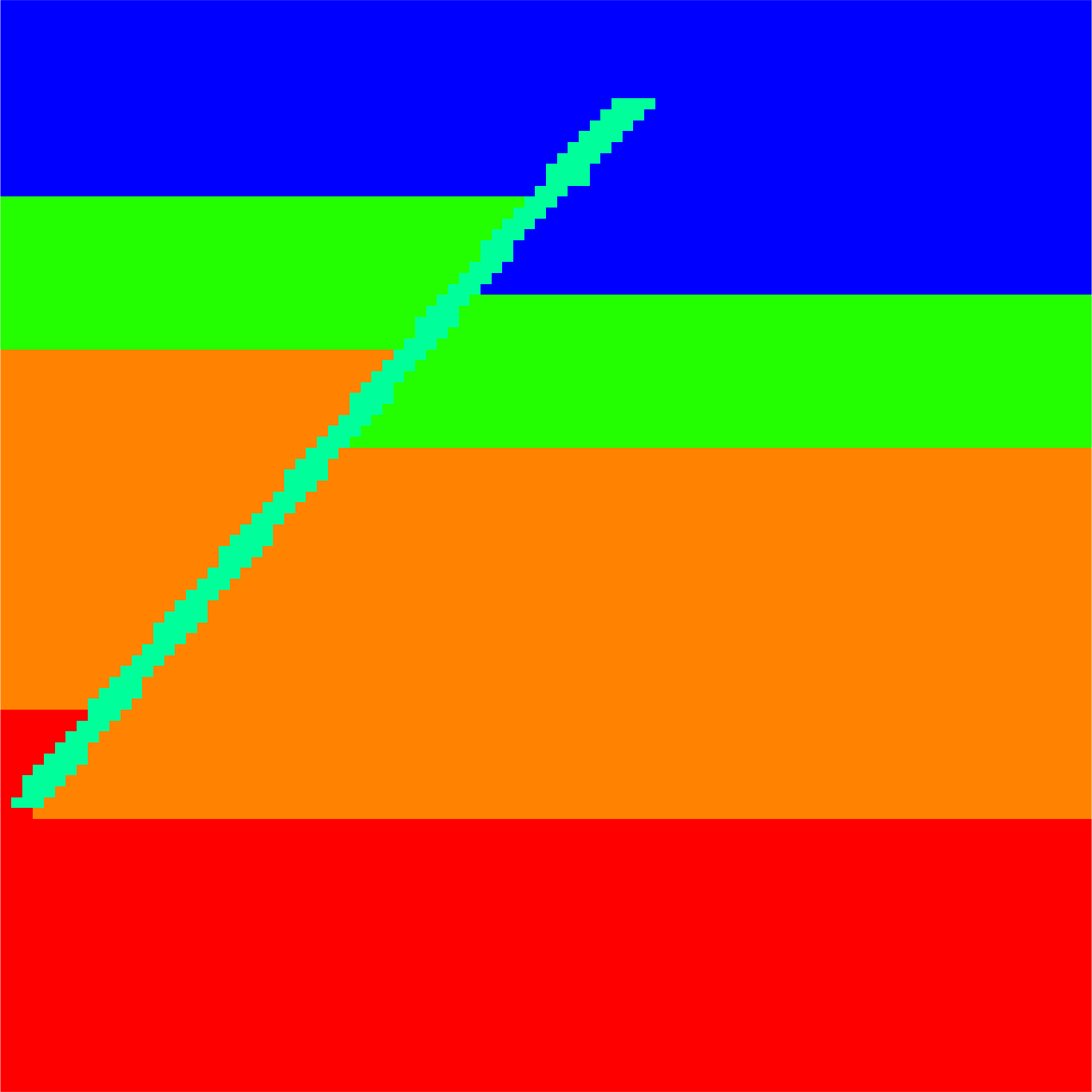}}
\hspace{0.3cm}
\subfloat{\includegraphics[width=.1\linewidth]{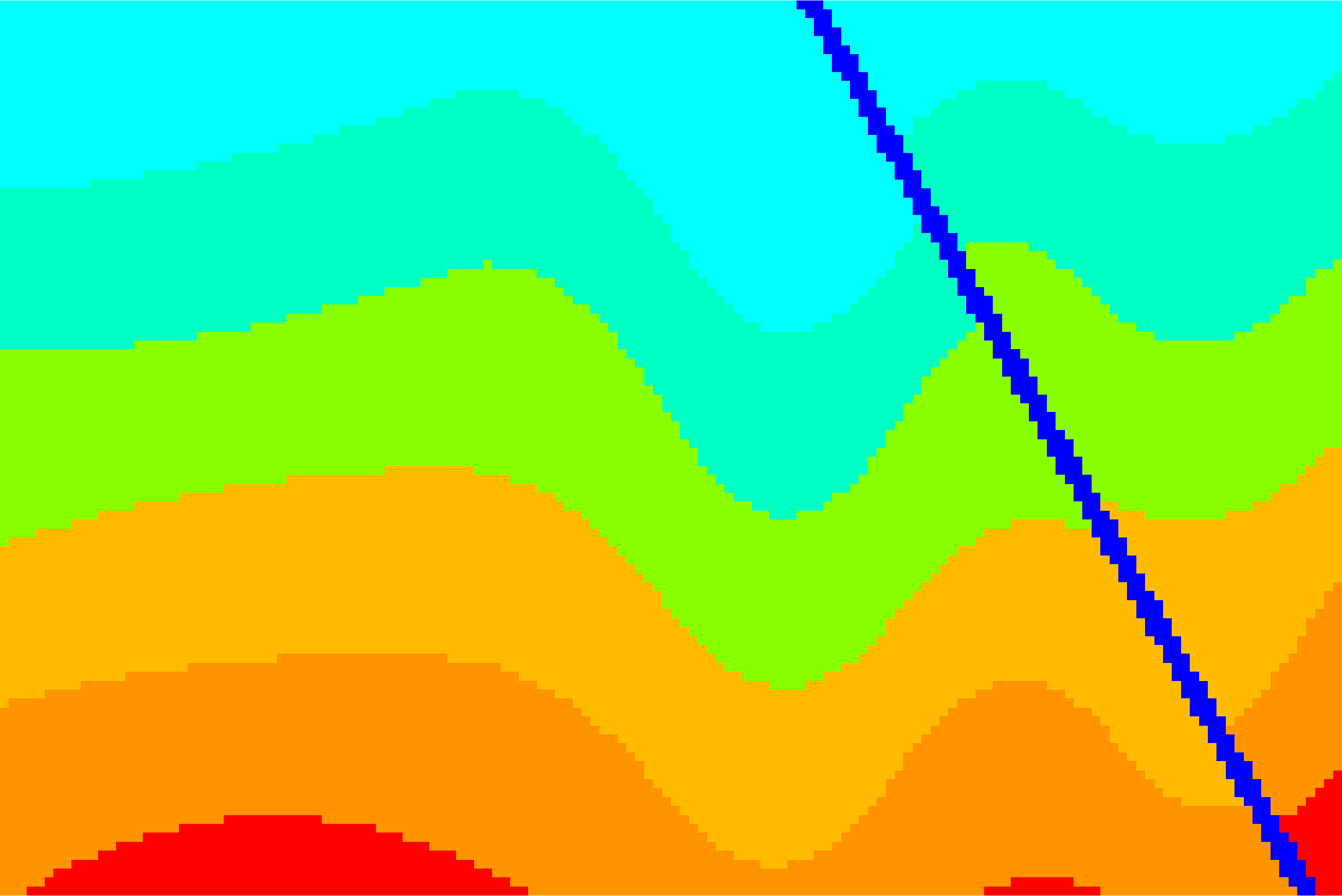}}
\hspace{0.05cm}
\subfloat{\includegraphics[width=.1\linewidth]{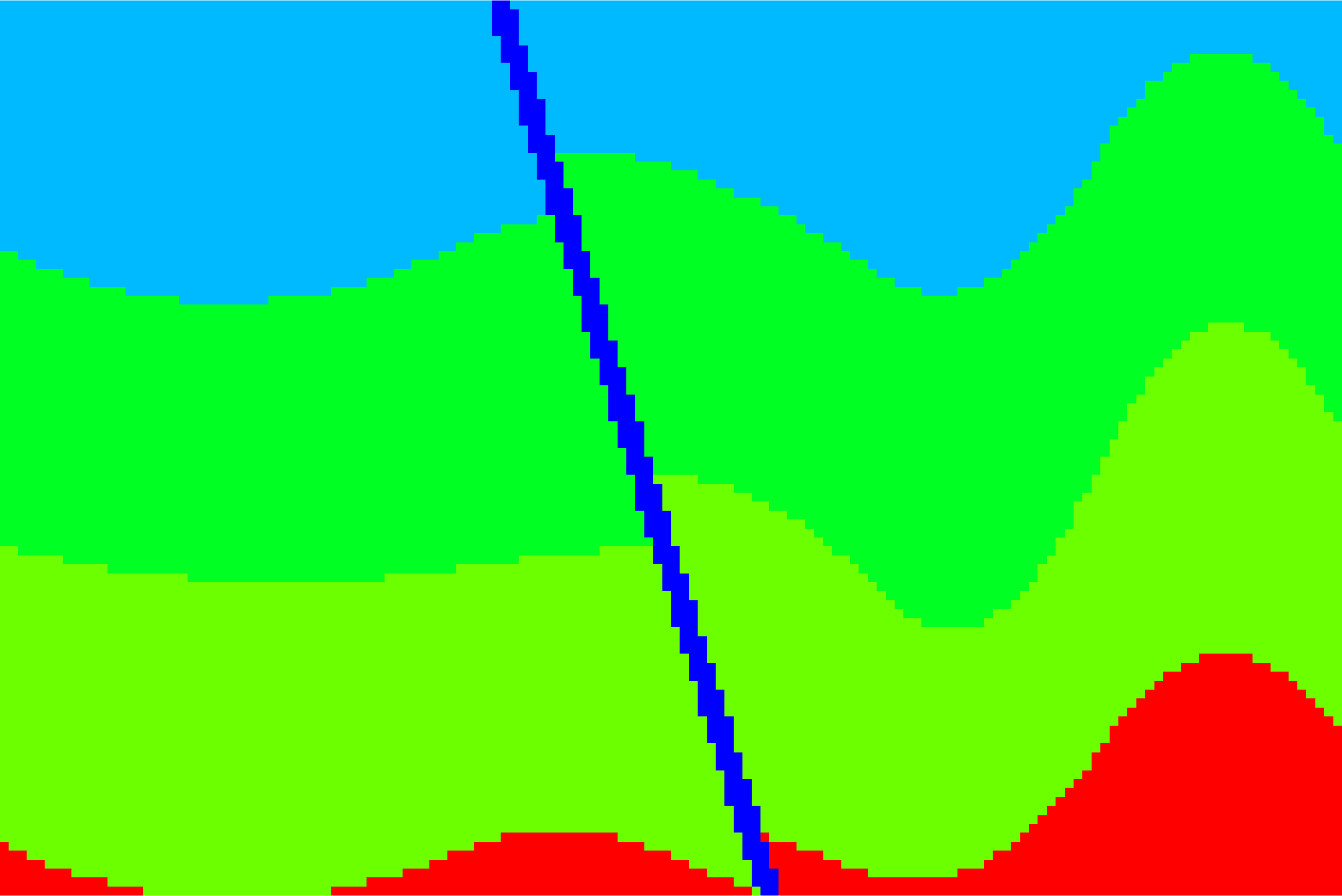}}
\hspace{0.05cm}
\subfloat{\includegraphics[width=.1\linewidth]{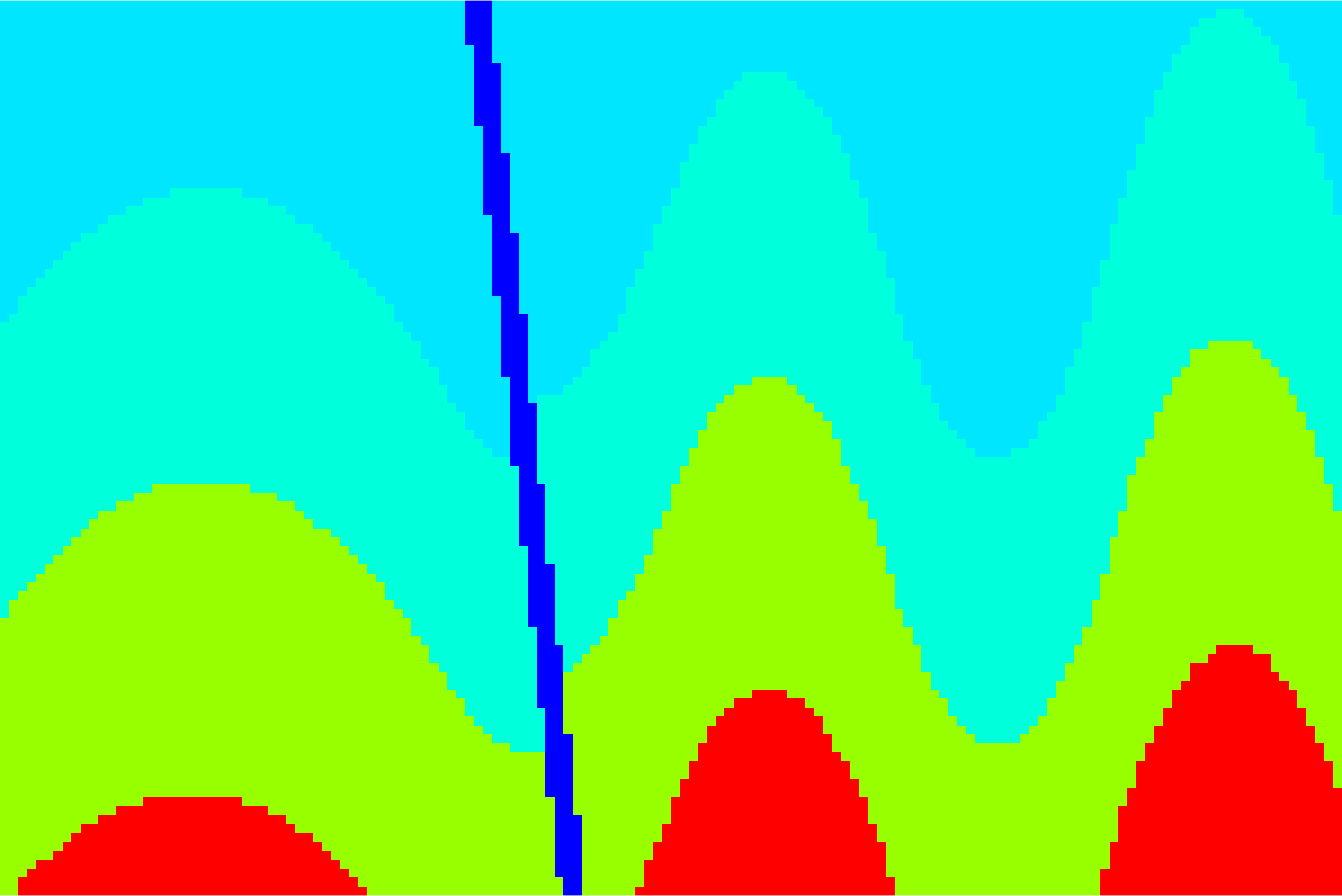}}
\hspace{0.05cm}
\subfloat{\includegraphics[width=.1\linewidth]{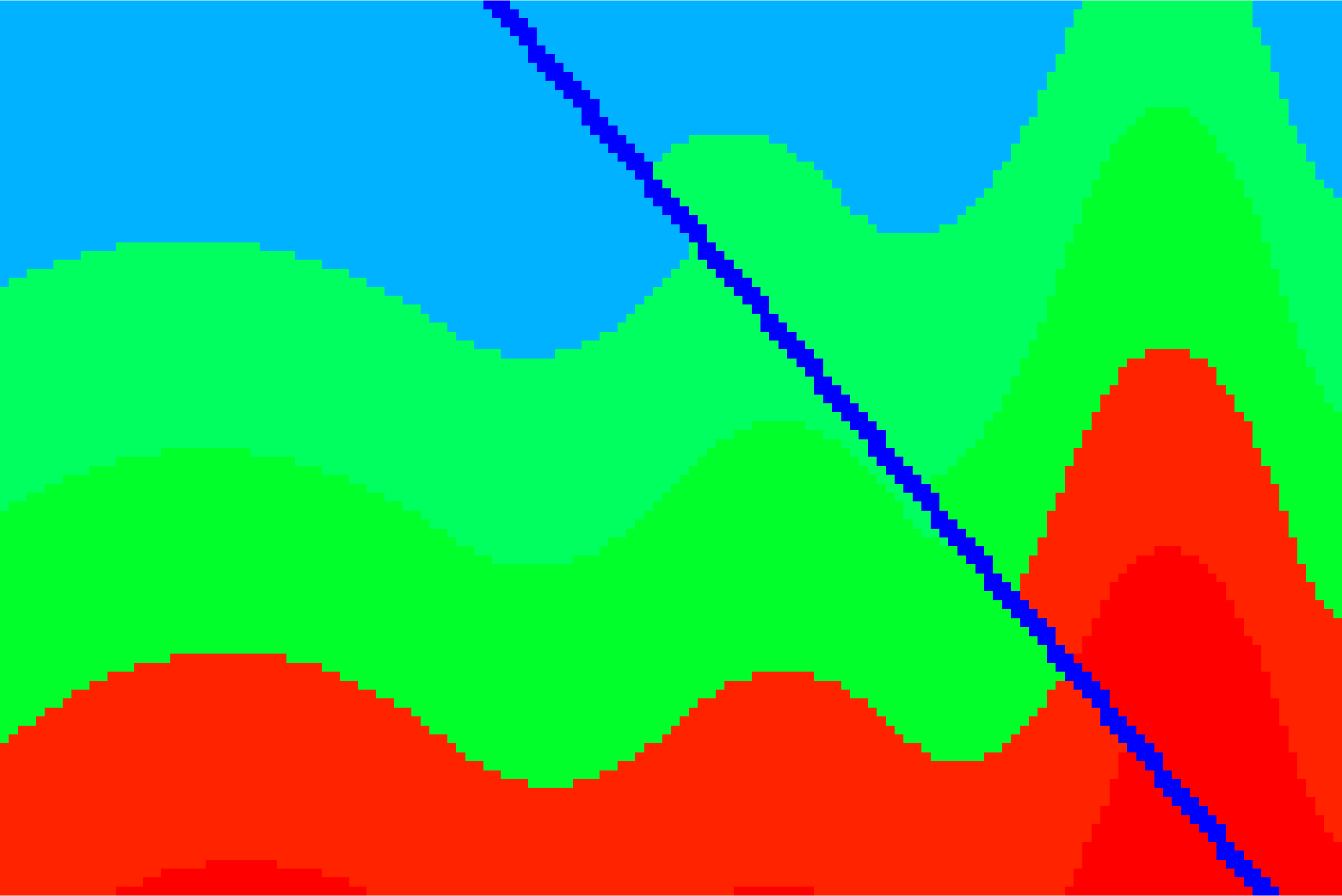}}
\hspace{0.05cm}
\subfloat{\includegraphics[width=.1\linewidth]{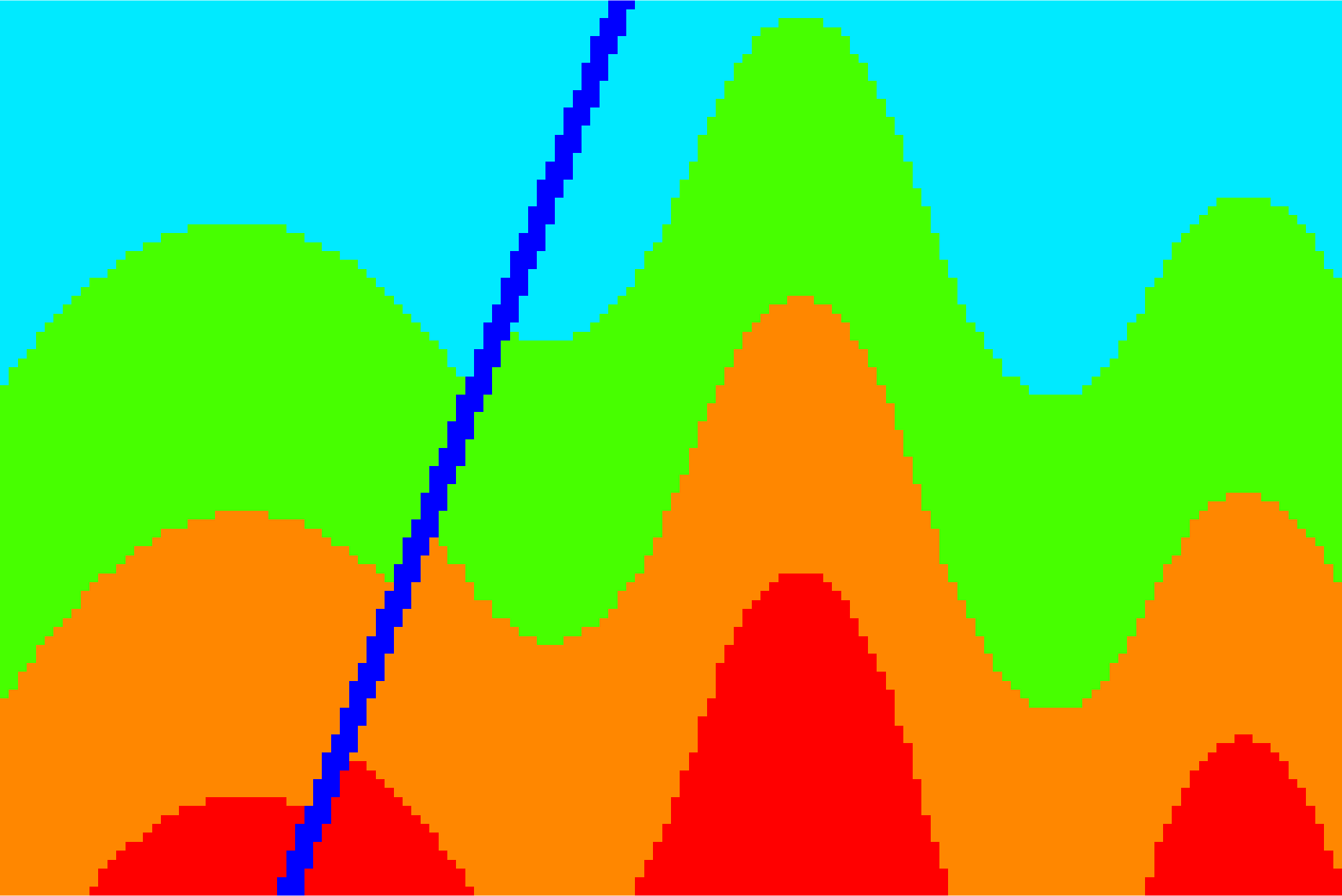}}}
\vspace{0.1cm}
\caption{Two database of velocity models. The FlatVel~(\textbf{Left} panel) consists of $36,000$ models of size $100 \times 100$ grid points. The CurvedVel~(\textbf{Right} panel) consists of $50,000$ models of size $100 \times 150$ grid points.}
\label{fig:TrainingSeismicModel}
\end{figure*}

\begin{figure*}
\centering
\centerline{
\subfloat{\includegraphics[width=.15\linewidth]{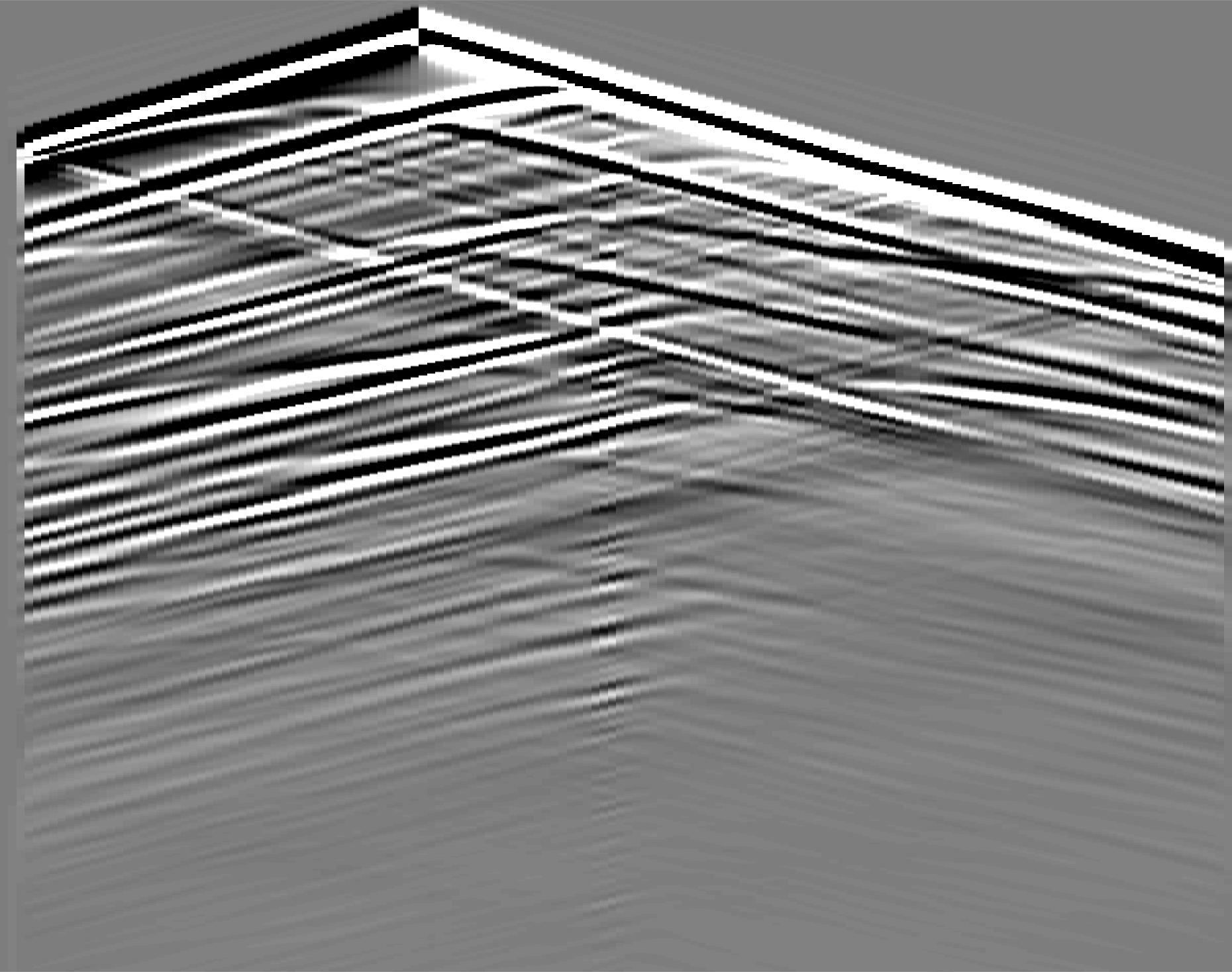}}
\hspace{0.05cm}
\subfloat{\includegraphics[width=.15\linewidth]{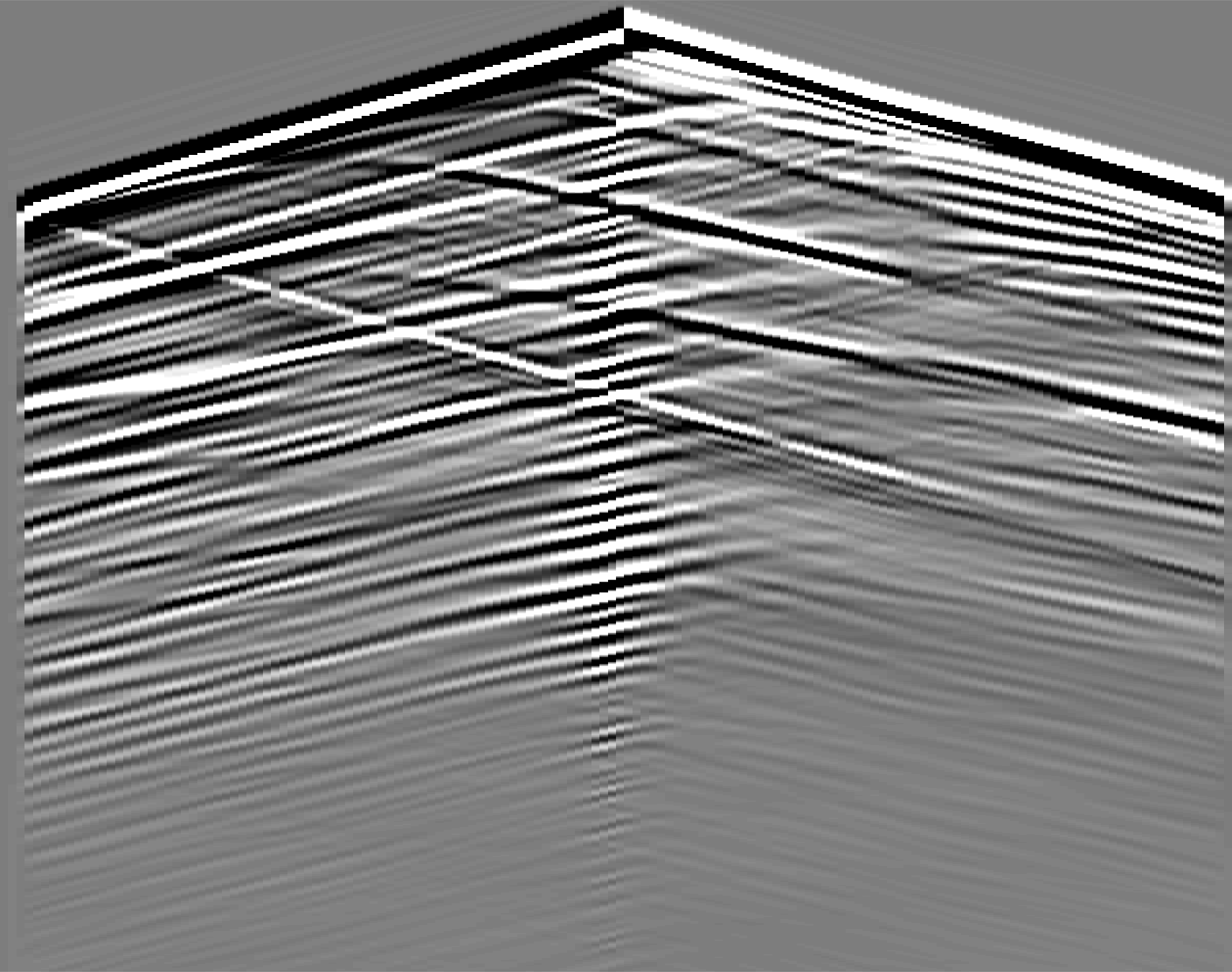}}
\hspace{0.05cm}
\subfloat{\includegraphics[width=.15\linewidth]{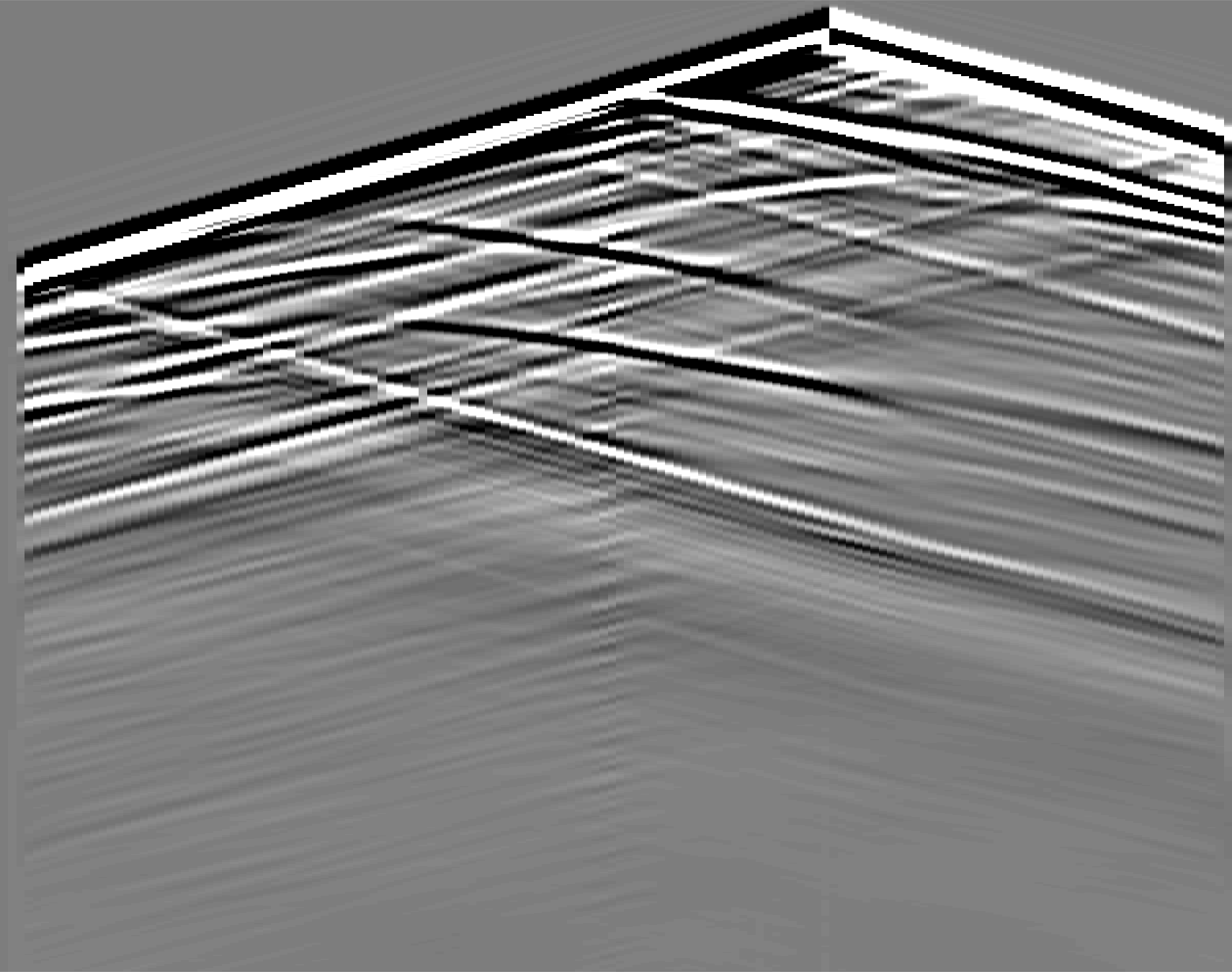}}
\hspace{0.3cm}
\subfloat{\includegraphics[width=.15\linewidth]{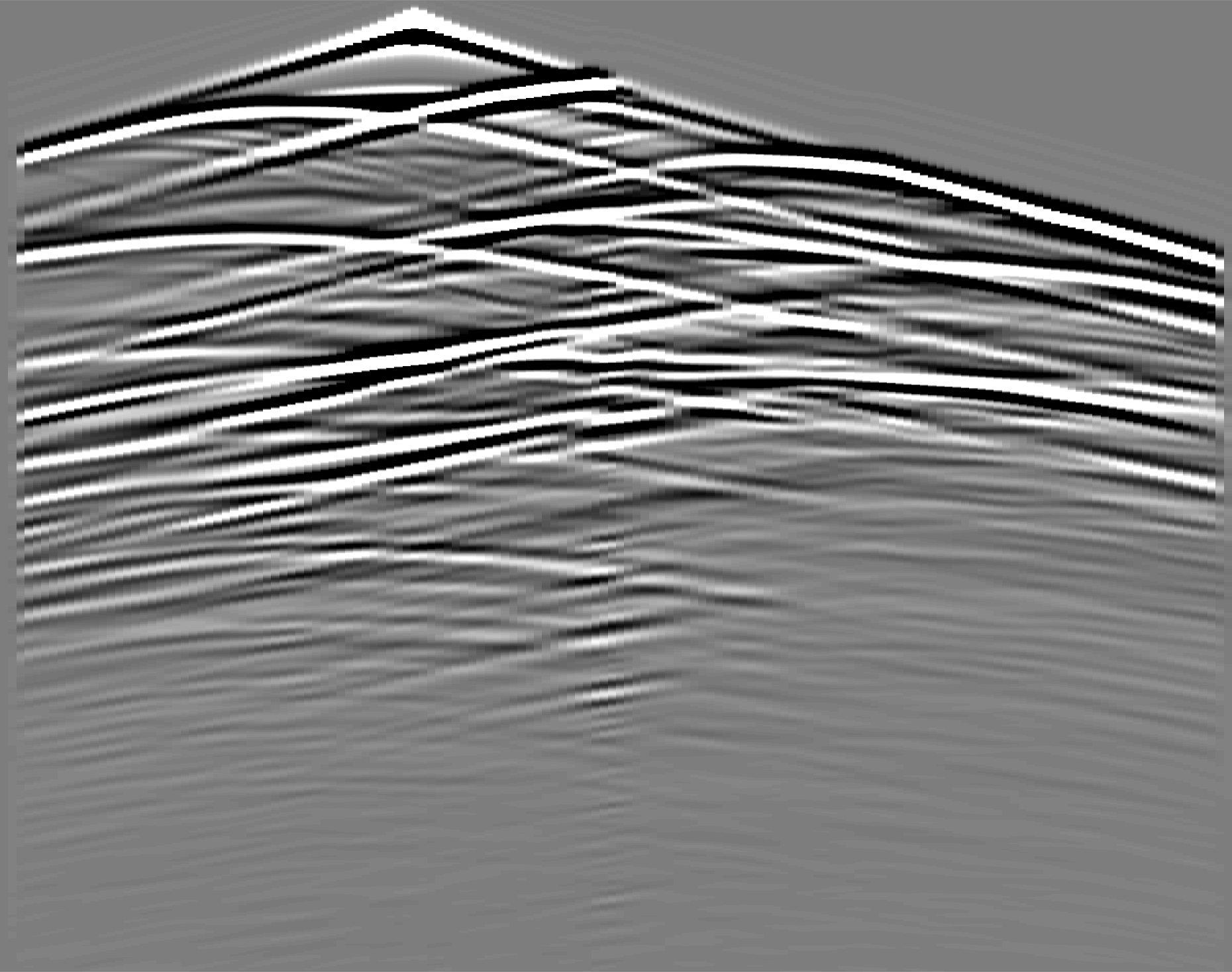}}
\hspace{0.05cm}
\subfloat{\includegraphics[width=.15\linewidth]{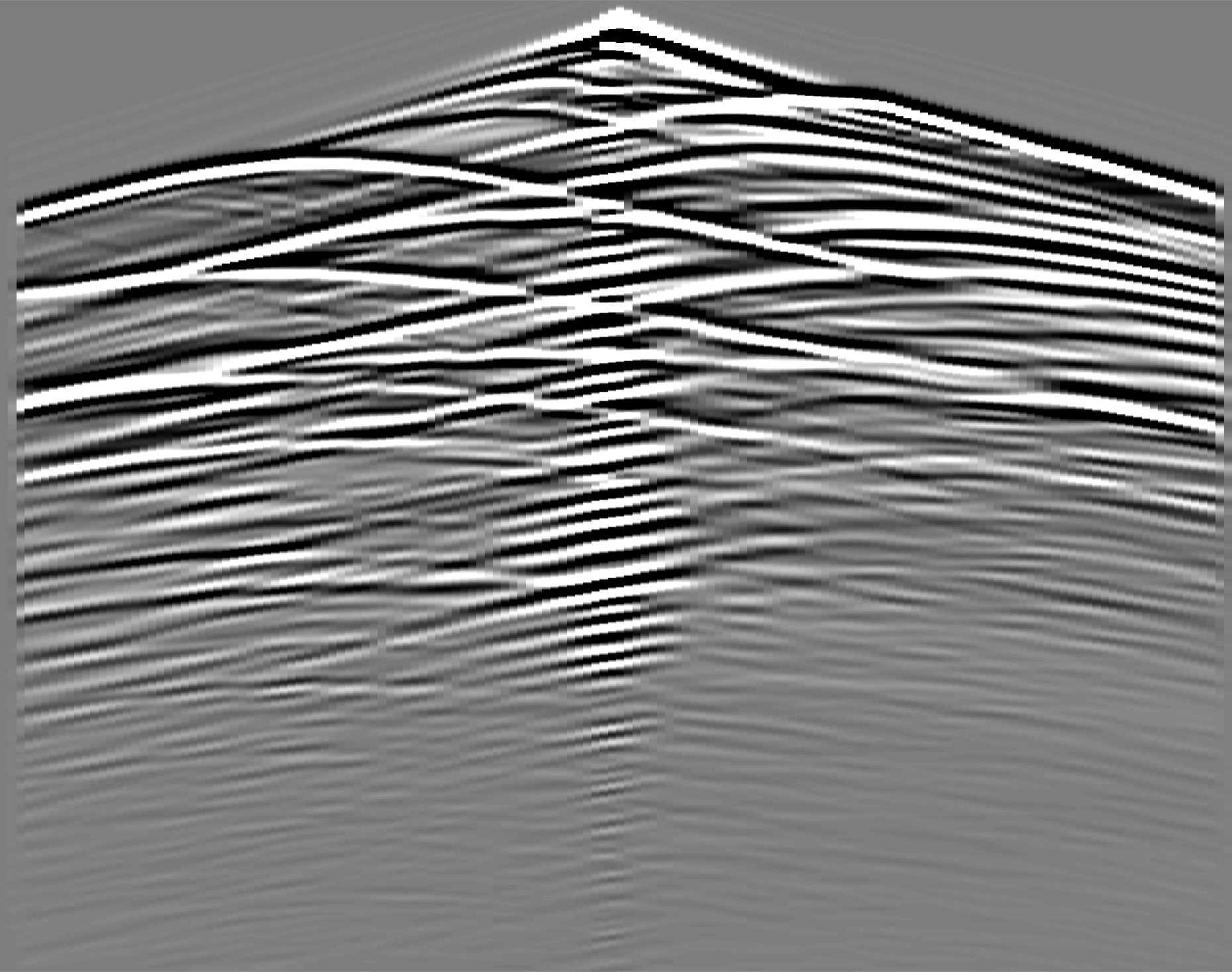}}
\hspace{0.05cm}
\subfloat{\includegraphics[width=.15\linewidth]{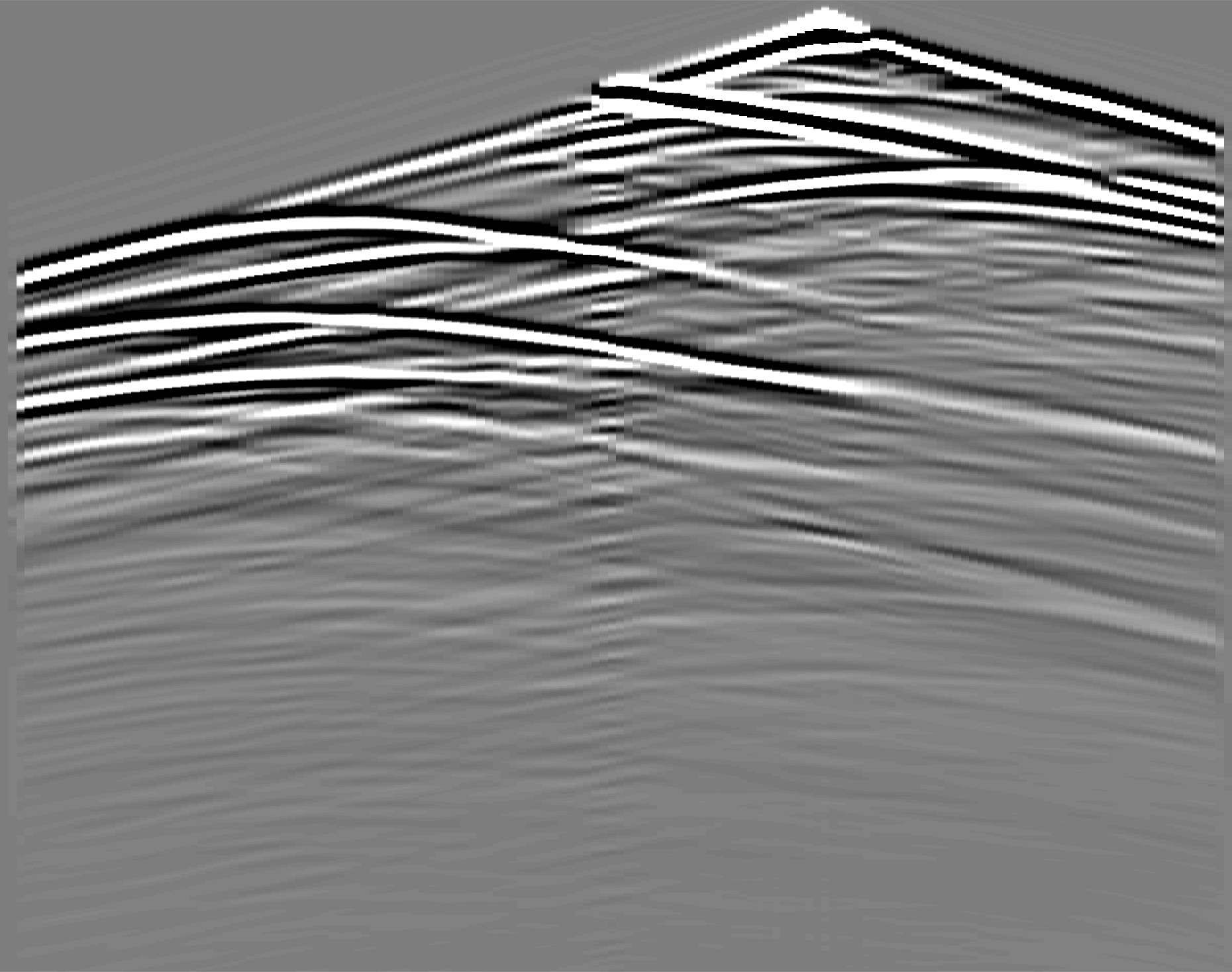}}}
\vspace{0.2cm}
\centerline{
\subfloat{\includegraphics[width=.15\linewidth]{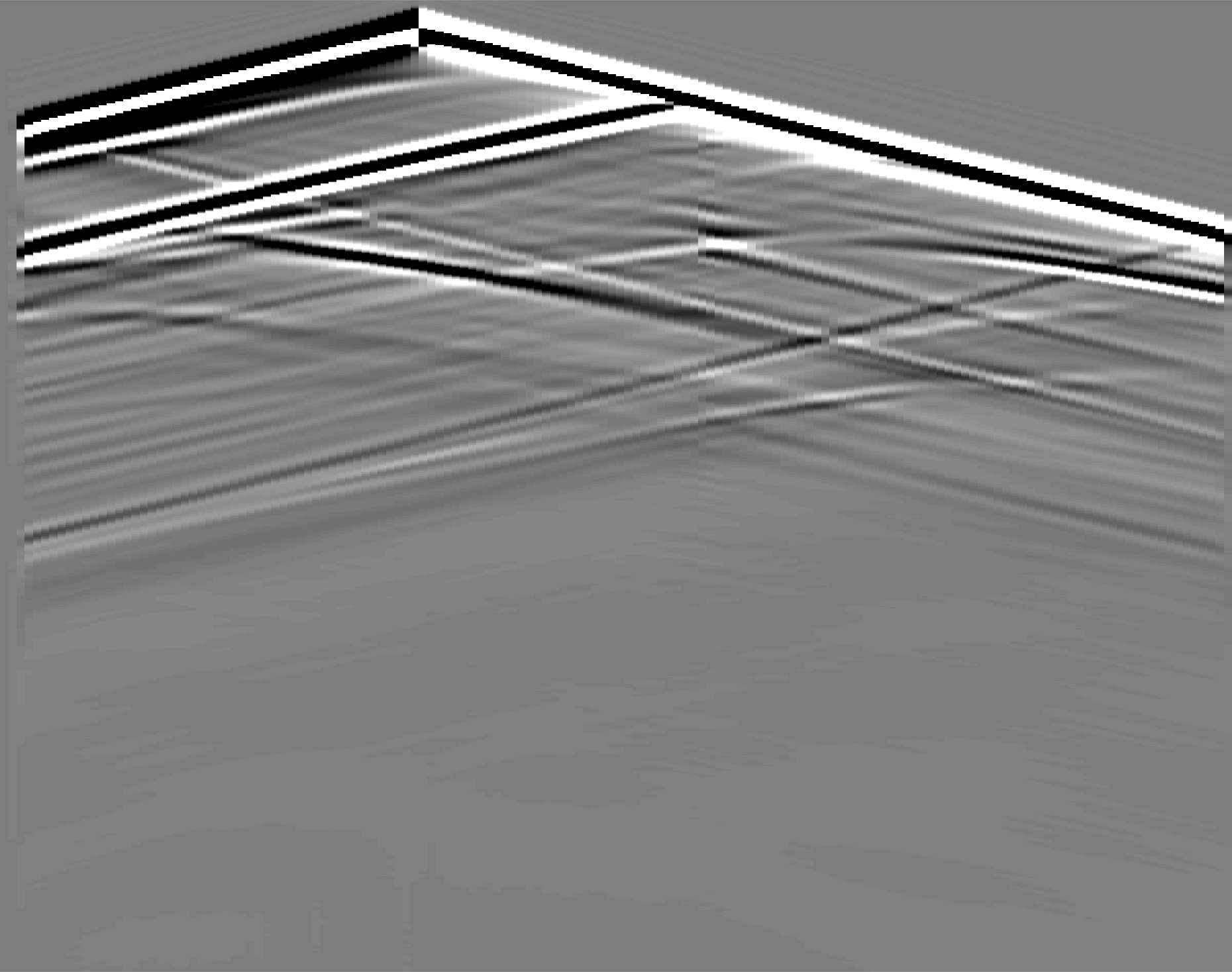}}
\hspace{0.05cm}
\subfloat{\includegraphics[width=.15\linewidth]{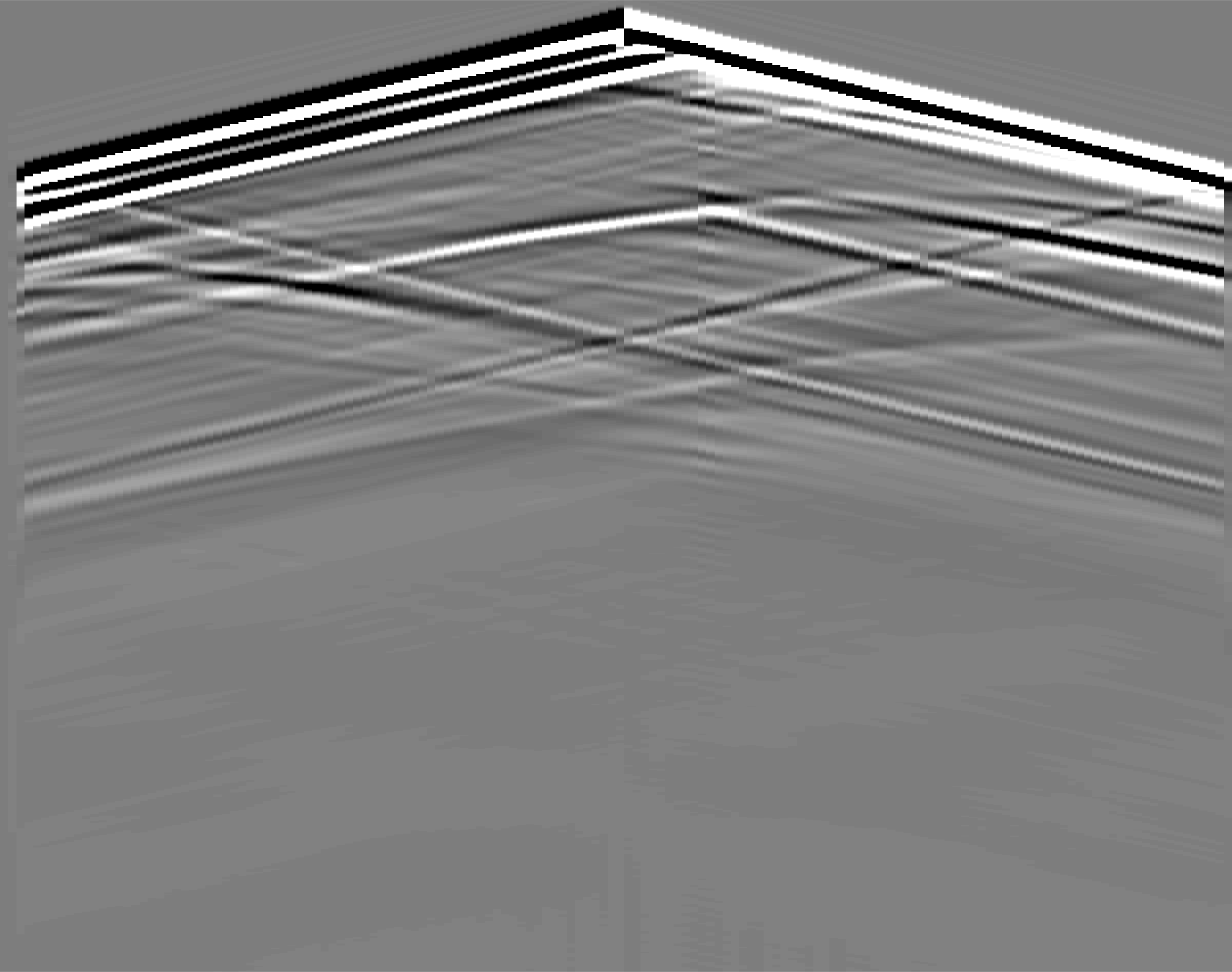}}
\hspace{0.05cm}
\subfloat{\includegraphics[width=.15\linewidth]{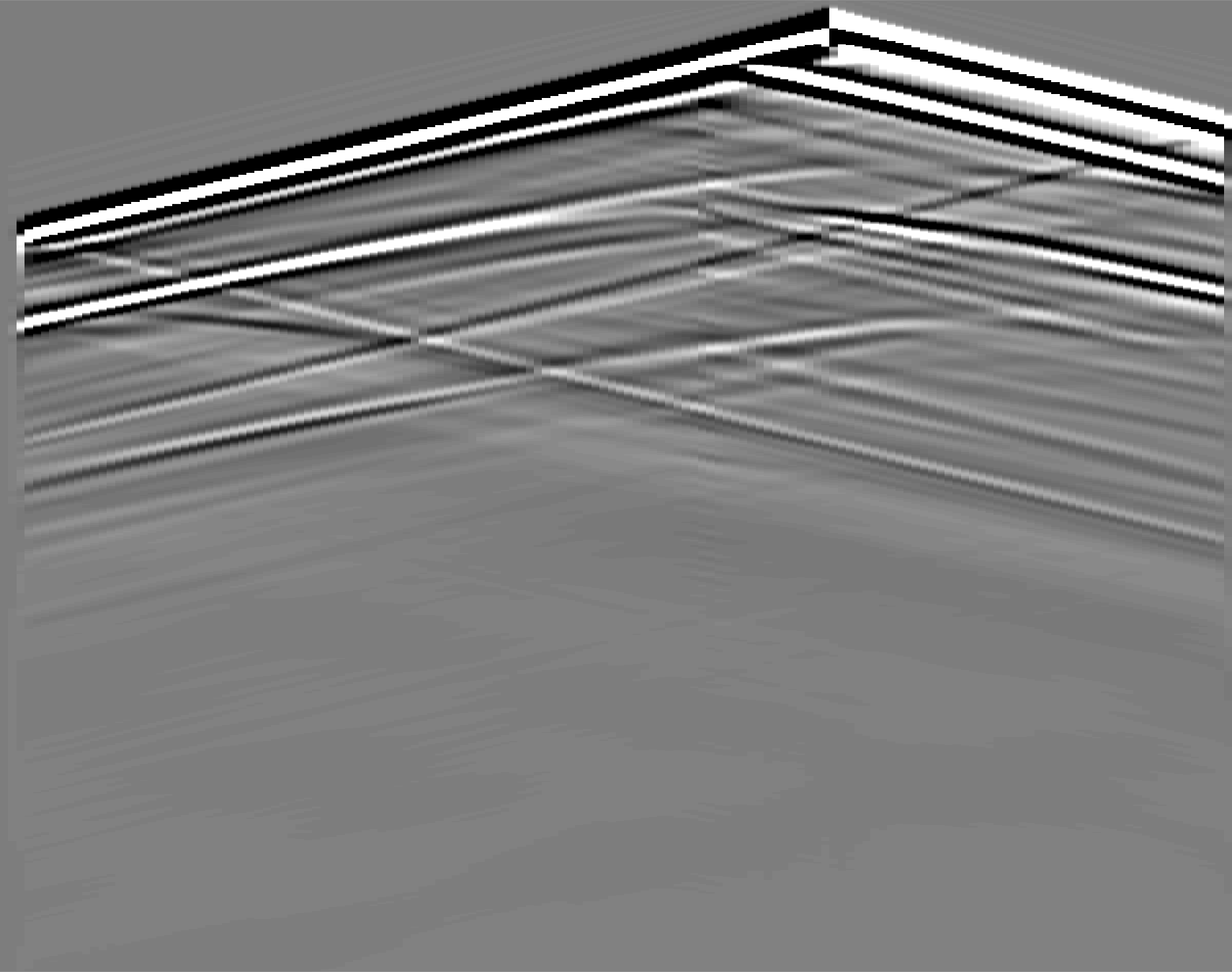}}
\hspace{0.3cm}
\subfloat{\includegraphics[width=.15\linewidth]{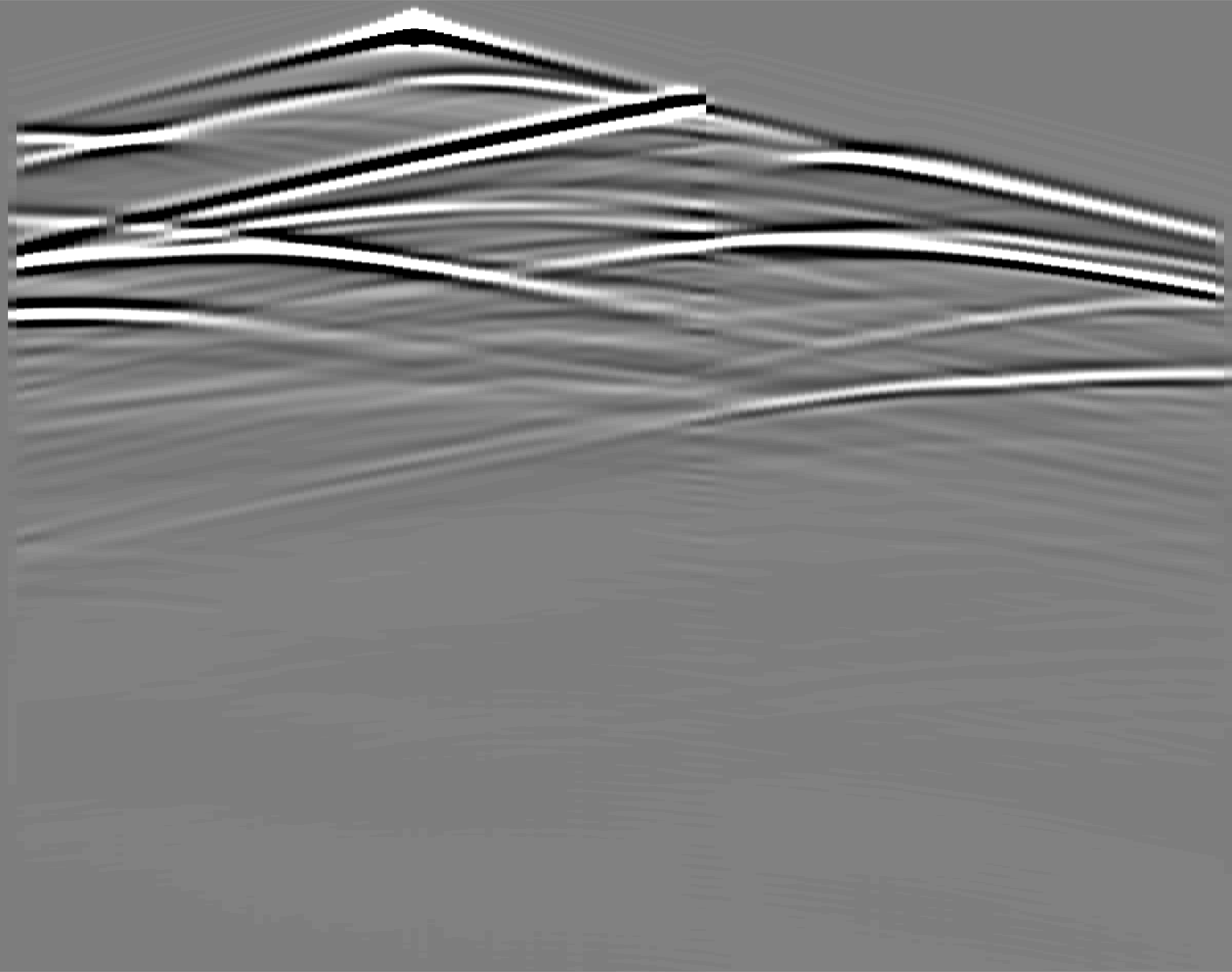}}
\hspace{0.05cm}
\subfloat{\includegraphics[width=.15\linewidth]{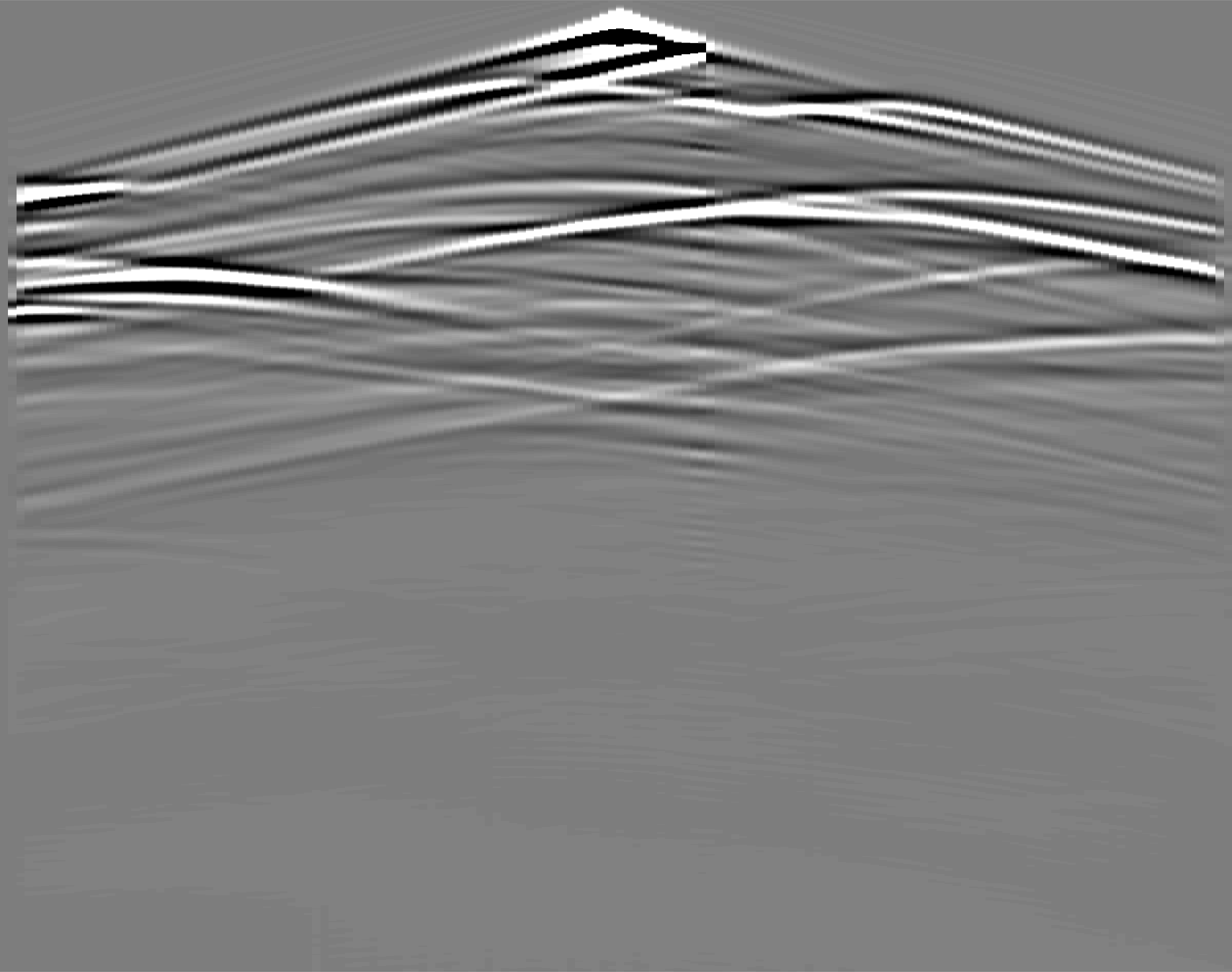}}
\hspace{0.05cm}
\subfloat{\includegraphics[width=.15\linewidth]{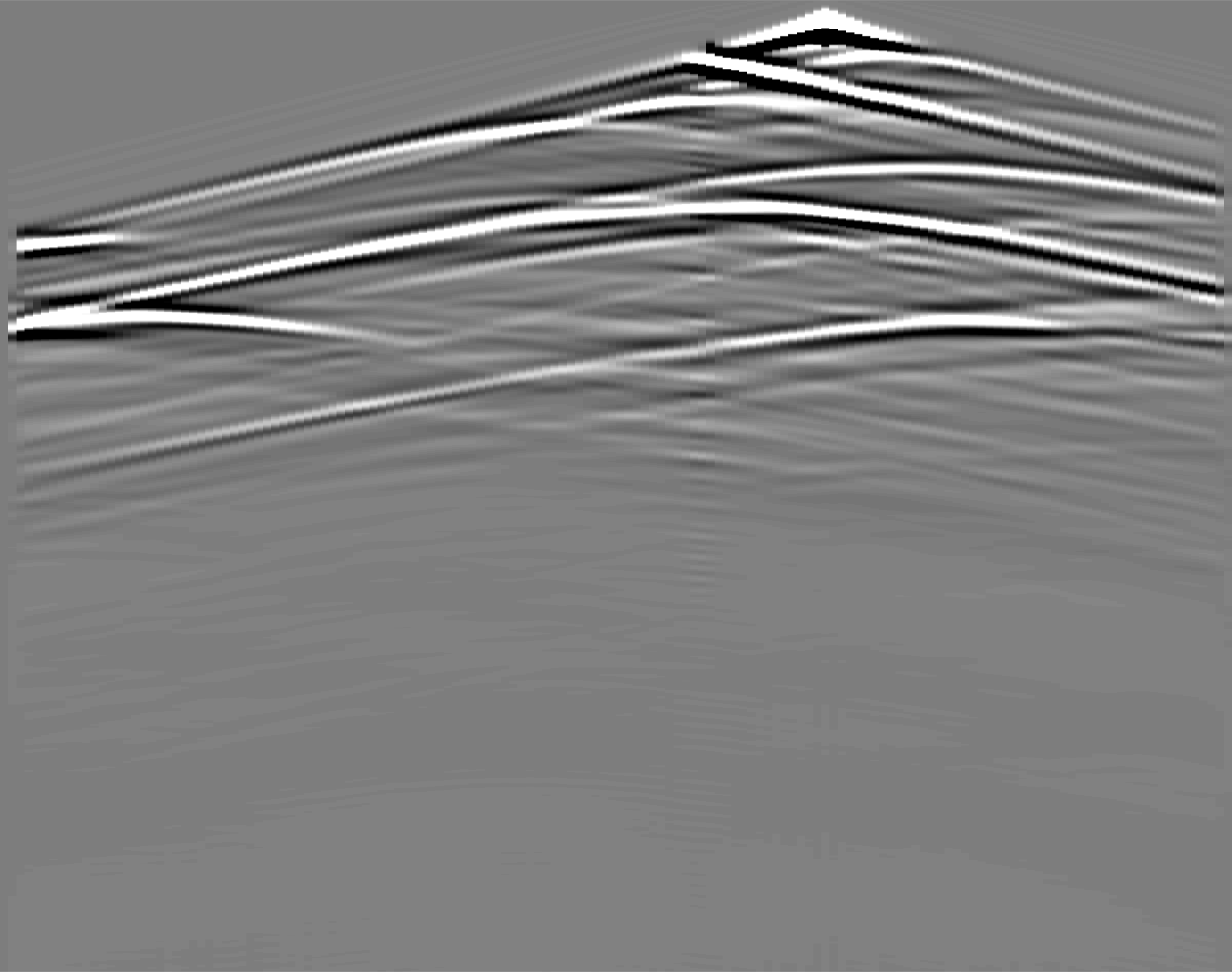}}}
\caption{Synthetic seismic data sets of CurvedVel are obtained using a staggered-grid finite-difference scheme with a perfectly matched layered absorbing boundary condition. The displacement of X direction~~(\textbf{Left} panel) and Z direction~(\textbf{Right} panel) are both used as training sets. Columns~1 and 4 are seismic data sets for first source.Columns~2 and 5 are seismic data sets for second source. Columns~3 and 6 are seismic data sets for third source. }
\label{fig:TrainingSeismicData}
\end{figure*}

\begin{figure*}[h]
\centering
\centerline{
\subfloat[]{\includegraphics[width=0.20\textwidth]{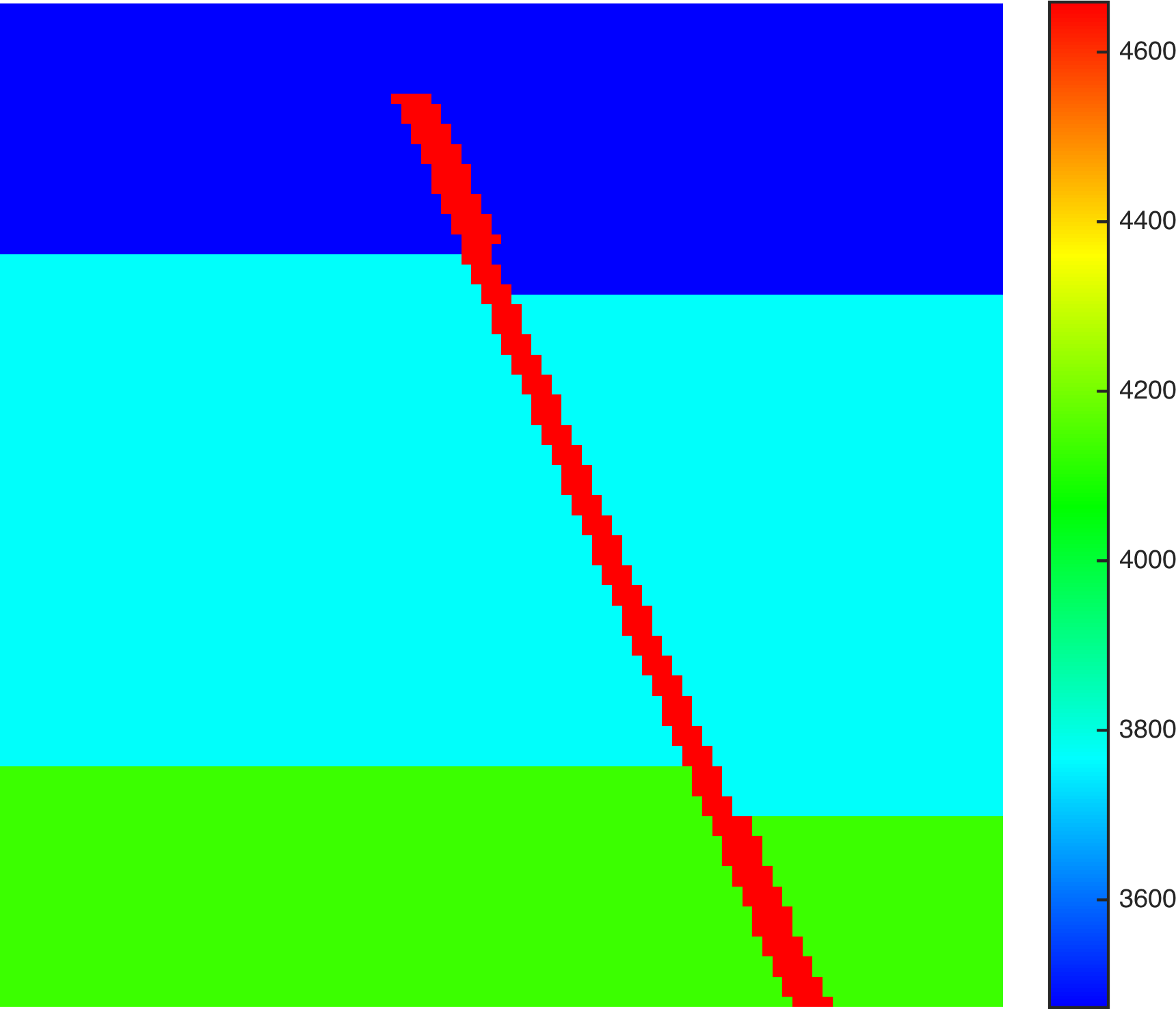}}%
\hspace{0.1cm}
\subfloat[]{\includegraphics[width=0.20\textwidth]{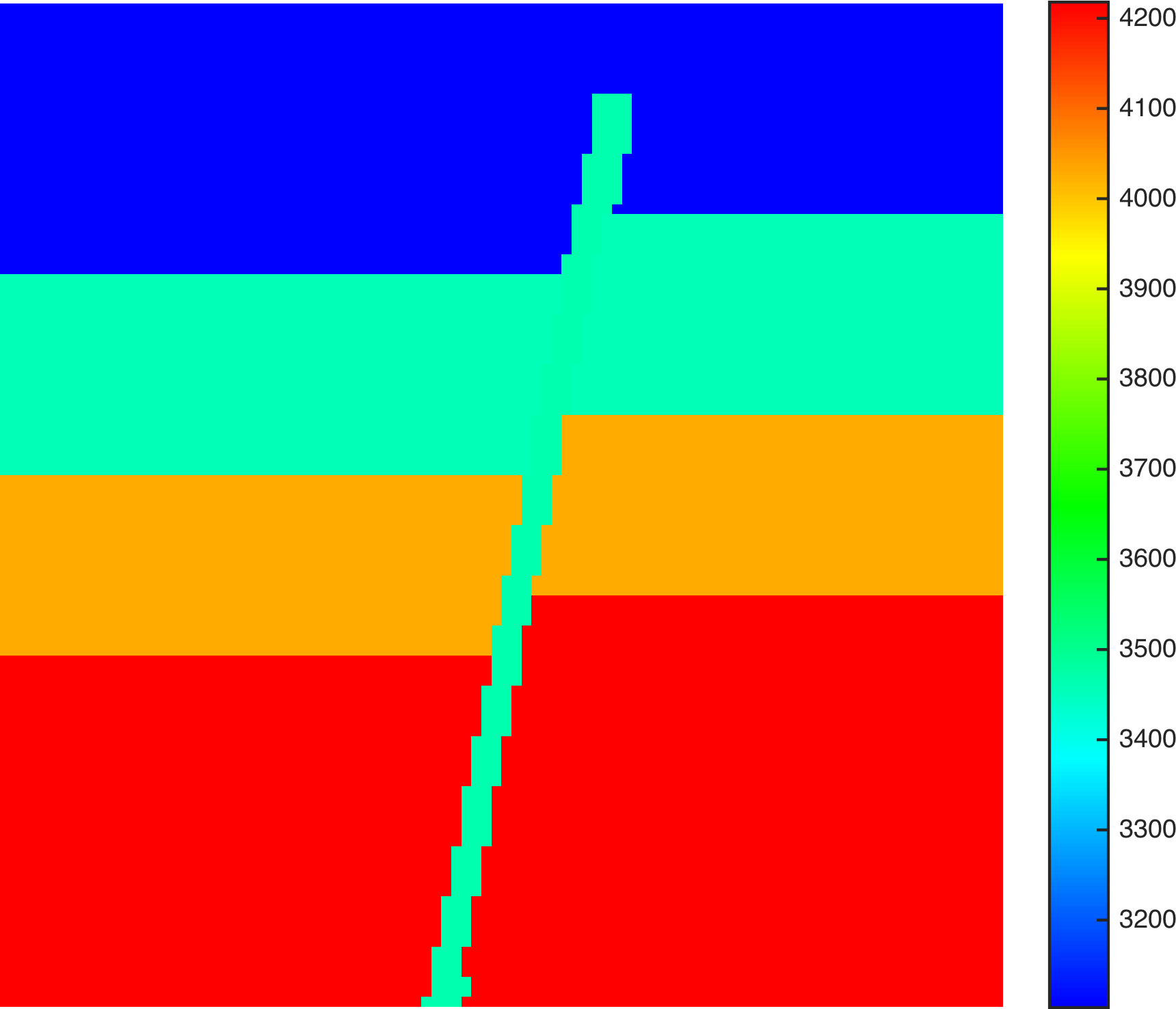}}
\hspace{0.1cm}
\subfloat[]{\includegraphics[width=0.20\textwidth]{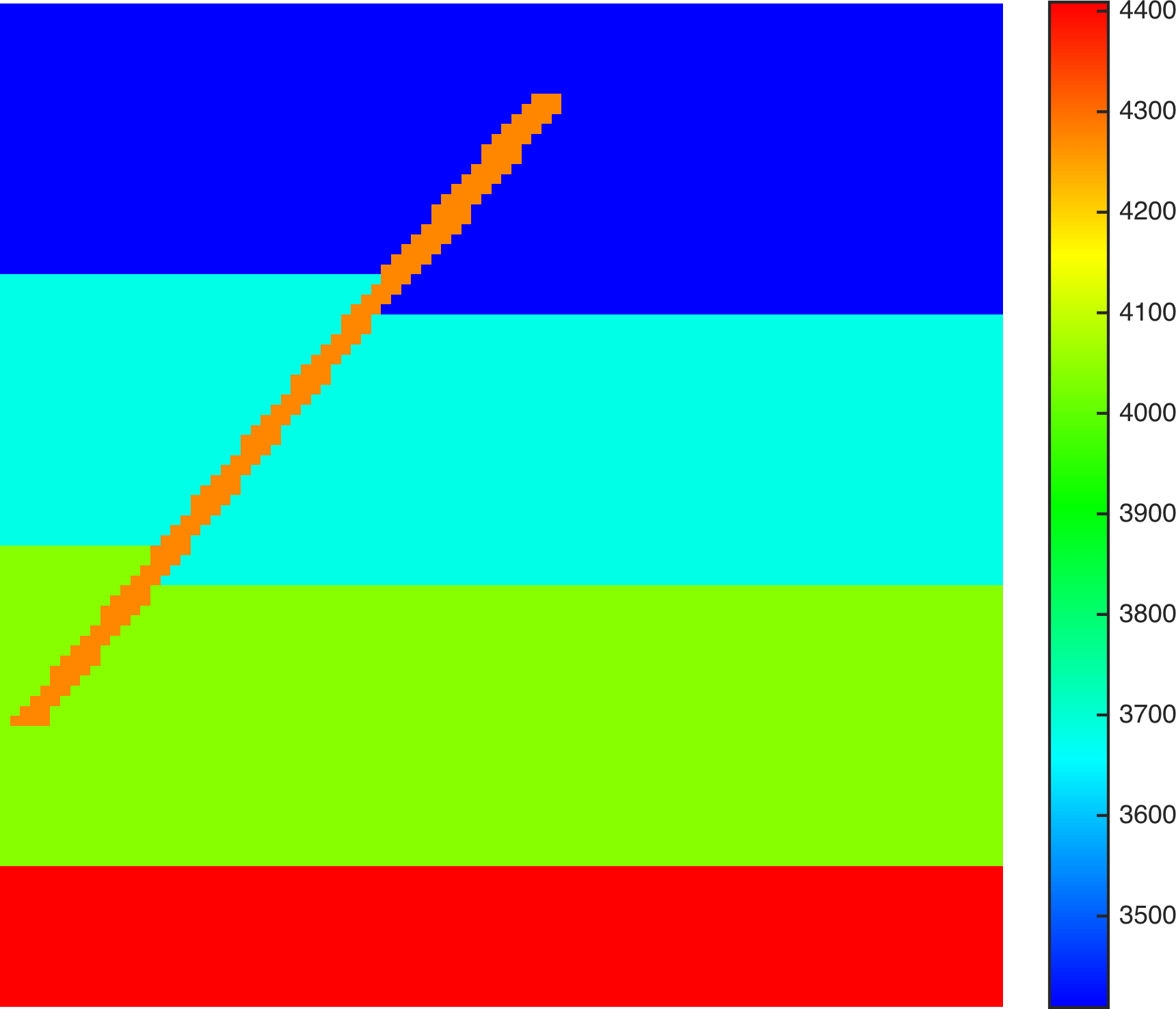}}
\hspace{0.1cm}
\subfloat[]{\includegraphics[width=0.20\textwidth]{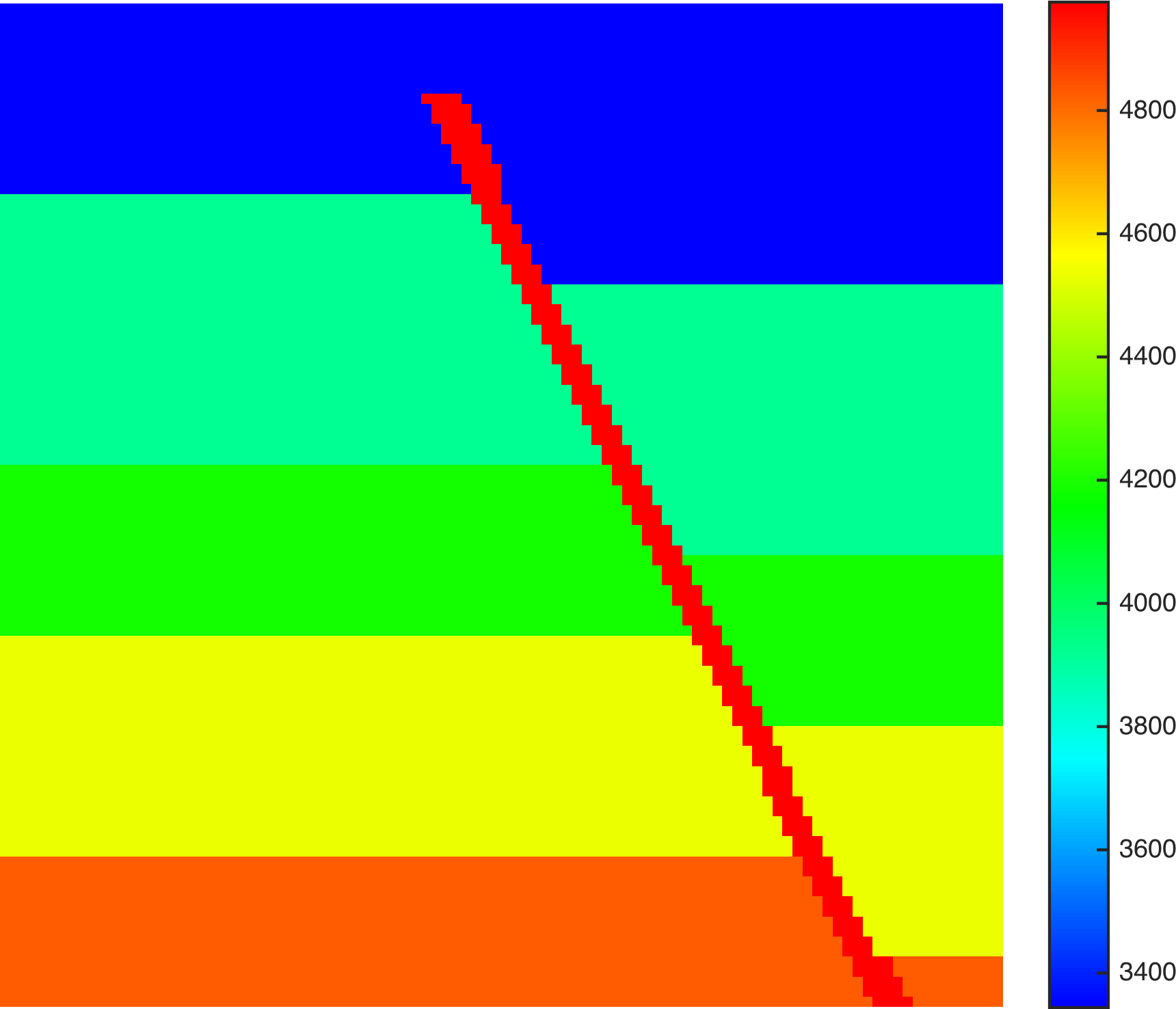}}
}
\vspace{0.1cm}
\centerline{
\subfloat[]{\includegraphics[width=0.20\textwidth]{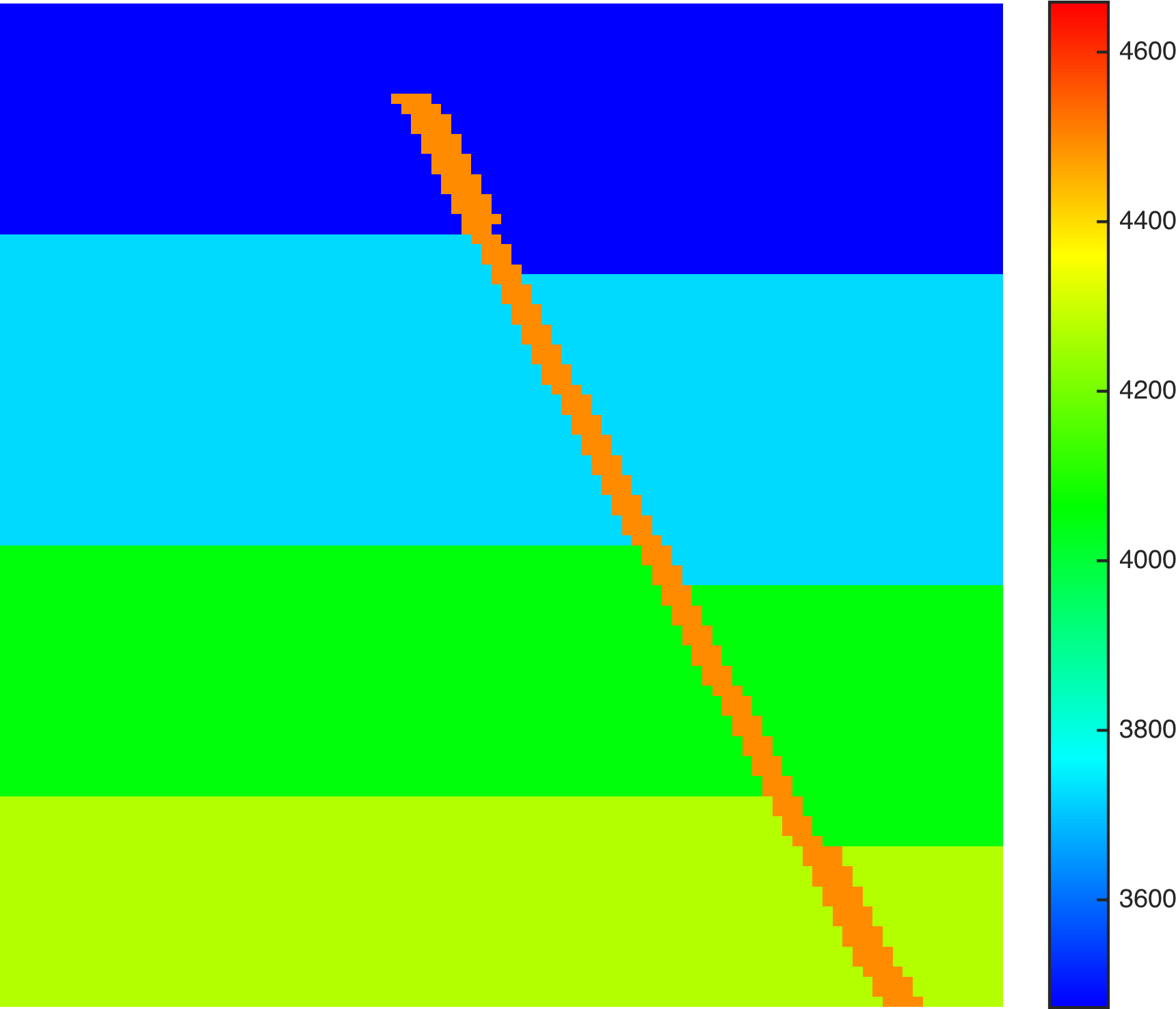}}%
\hspace{0.1cm}
\subfloat[]{\includegraphics[width=0.20\textwidth]{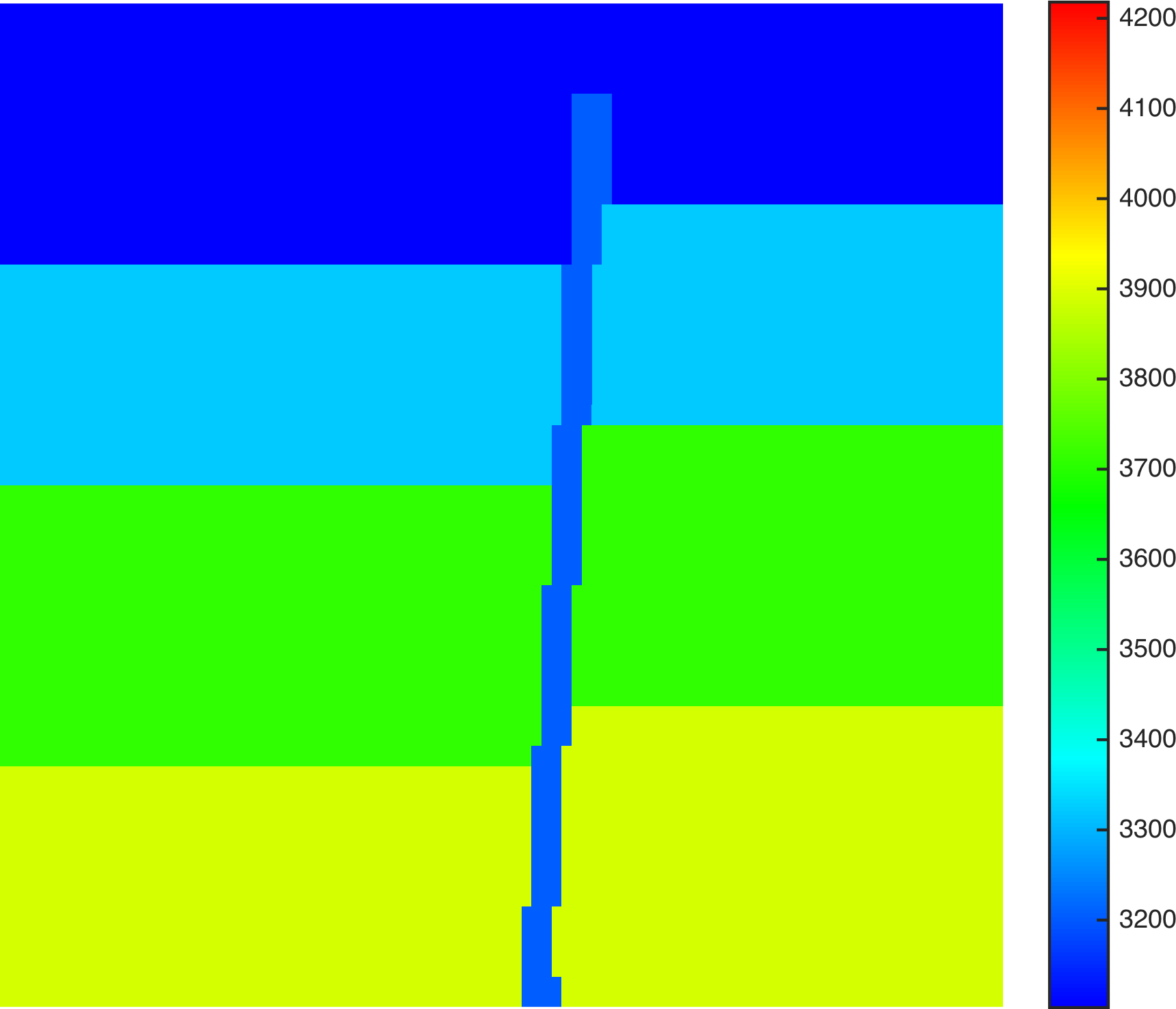}}
\hspace{0.1cm}
\subfloat[]{\includegraphics[width=0.20\textwidth]{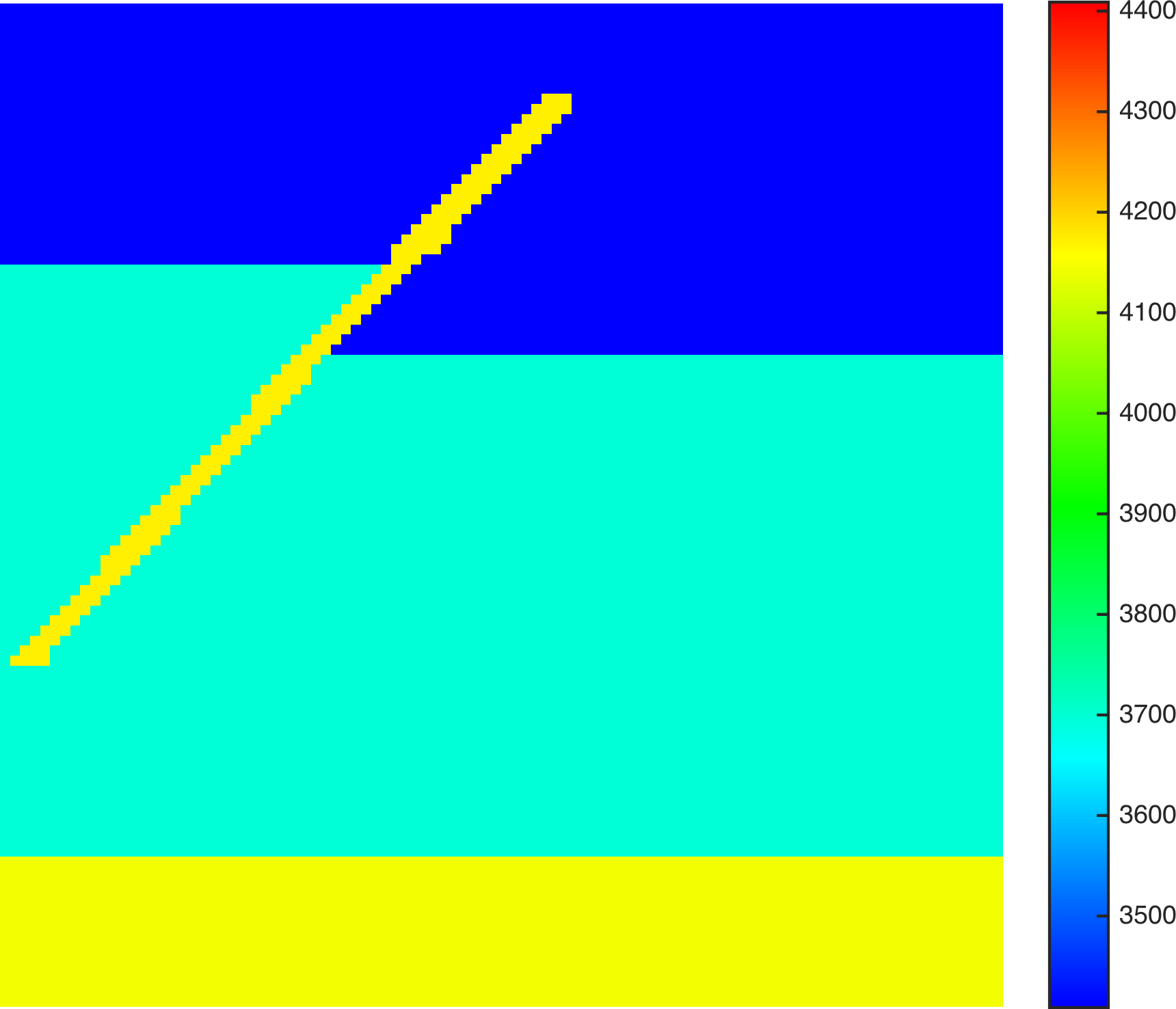}}
\hspace{0.1cm}
\subfloat[]{\includegraphics[width=0.20\textwidth]{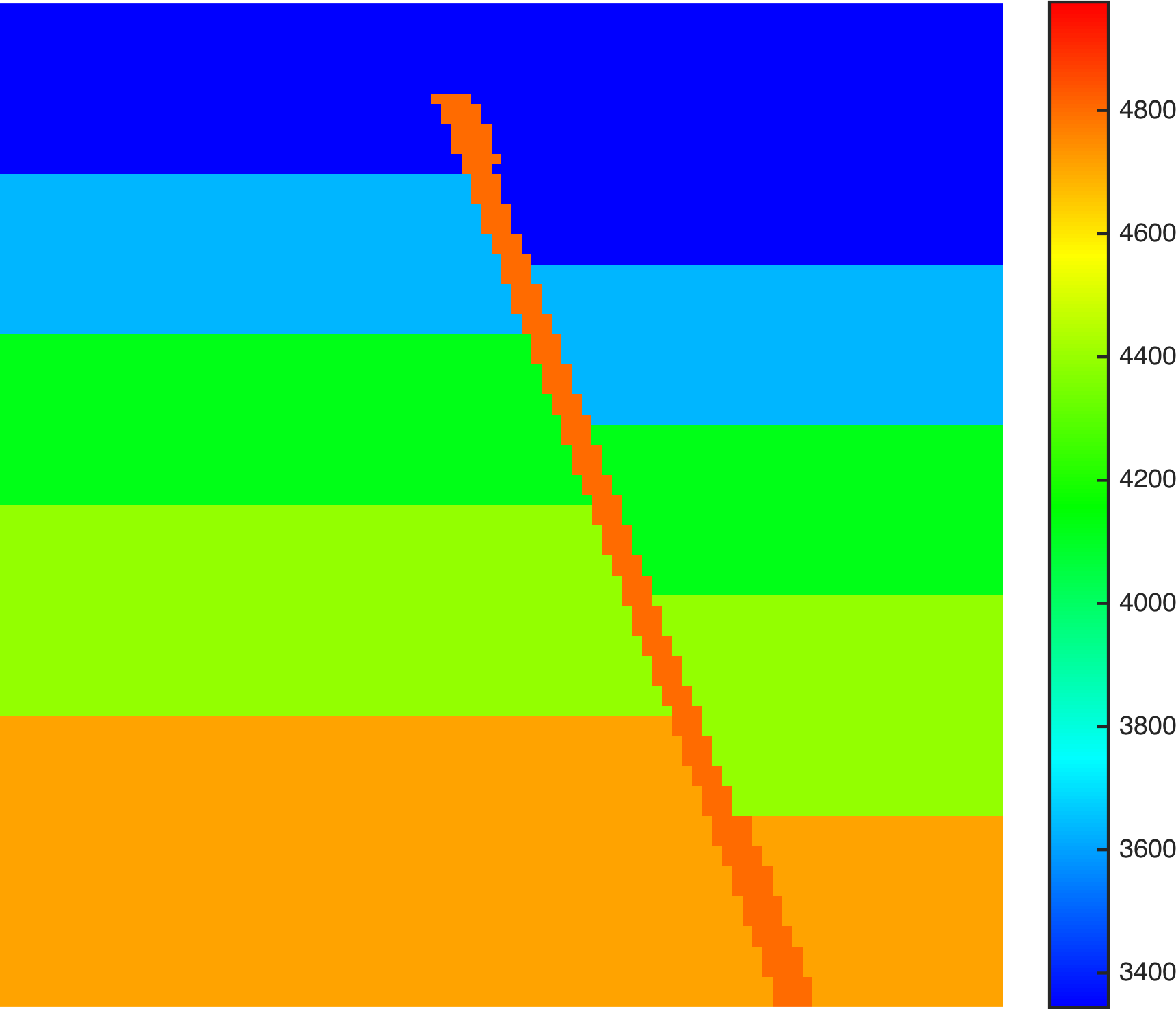}}
}
\caption{Illustration of four ground truth velocity models~(Top) randomly selected from FlatVel set. Their nearest neighbors~(Bottom) are also provided. The nearest neighbors share some similarity to their ground truth counterpart, however, the details including velocity values, fault orientation, layer location are all different.}
\label{fig:NN_Results_Flat}
\end{figure*}

\begin{figure*}[h]
\centering
\centerline{
\subfloat[]{\includegraphics[width=0.25\textwidth]{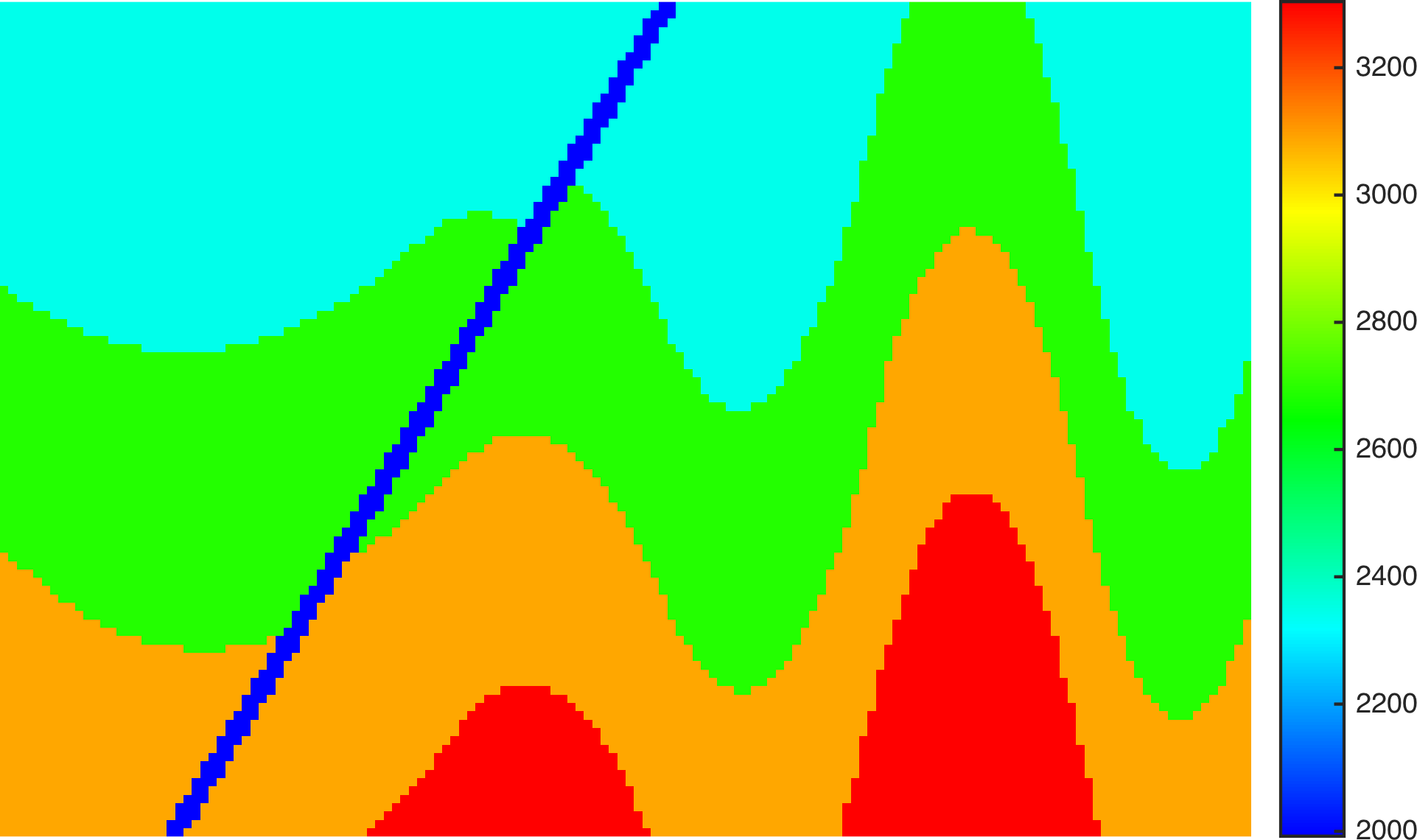}}%
\hspace{0.1cm}
\subfloat[]{\includegraphics[width=0.25\textwidth]{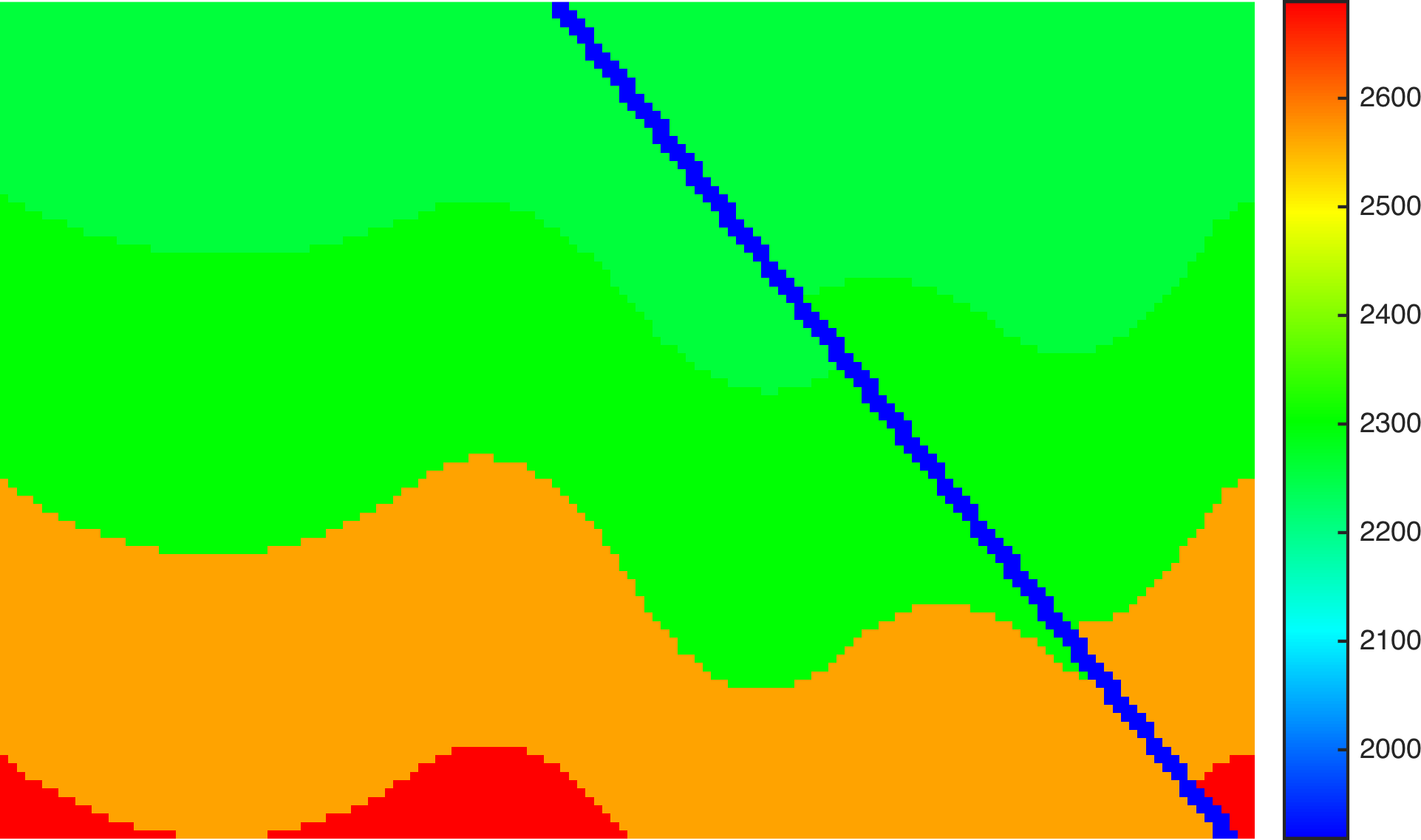}}
\hspace{0.1cm}
\subfloat[]{\includegraphics[width=0.25\textwidth]{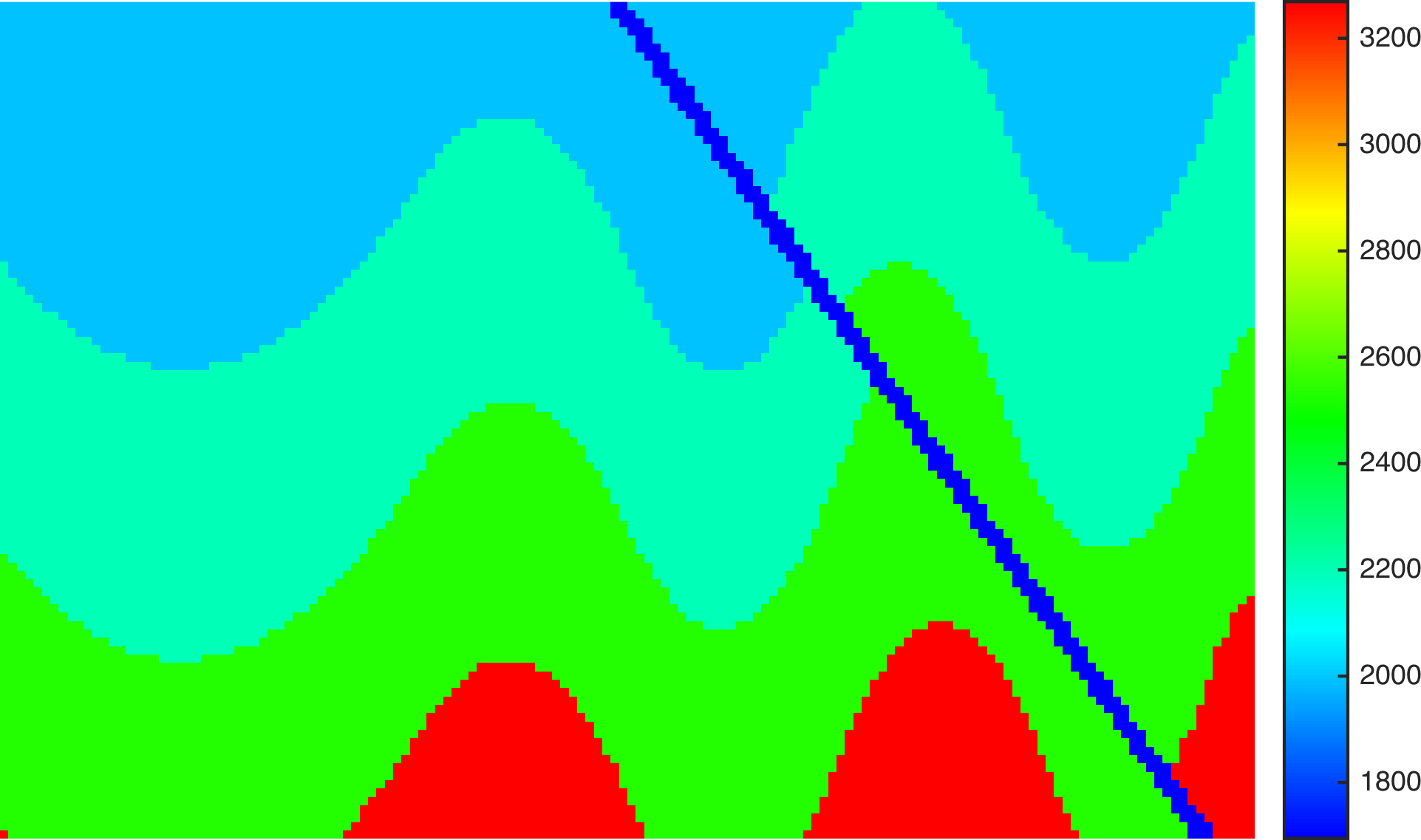}}
\hspace{0.1cm}
\subfloat[]{\includegraphics[width=0.25\textwidth]{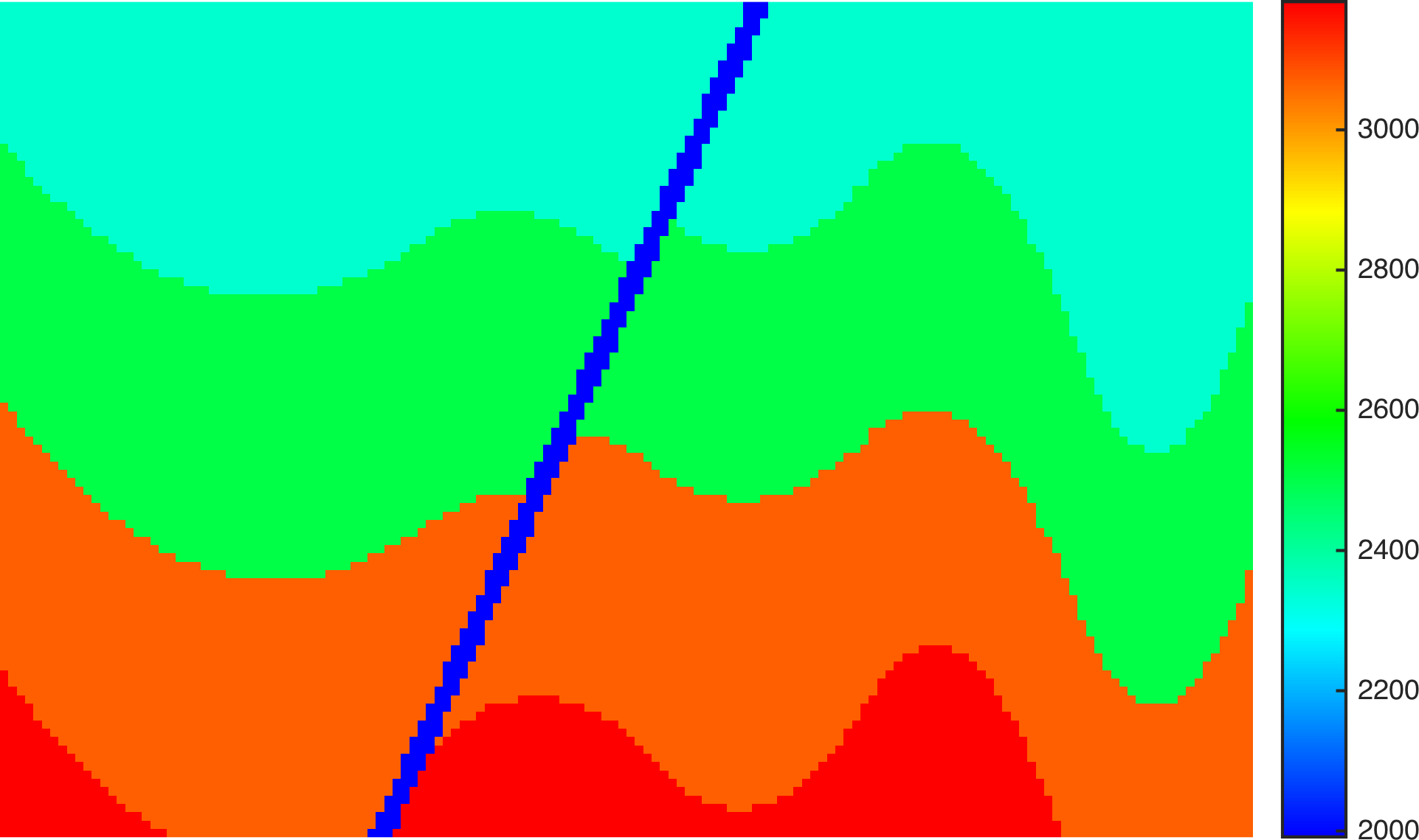}}
}
\vspace{0.1cm}
\centerline{
\subfloat[]{\includegraphics[width=0.25\textwidth]{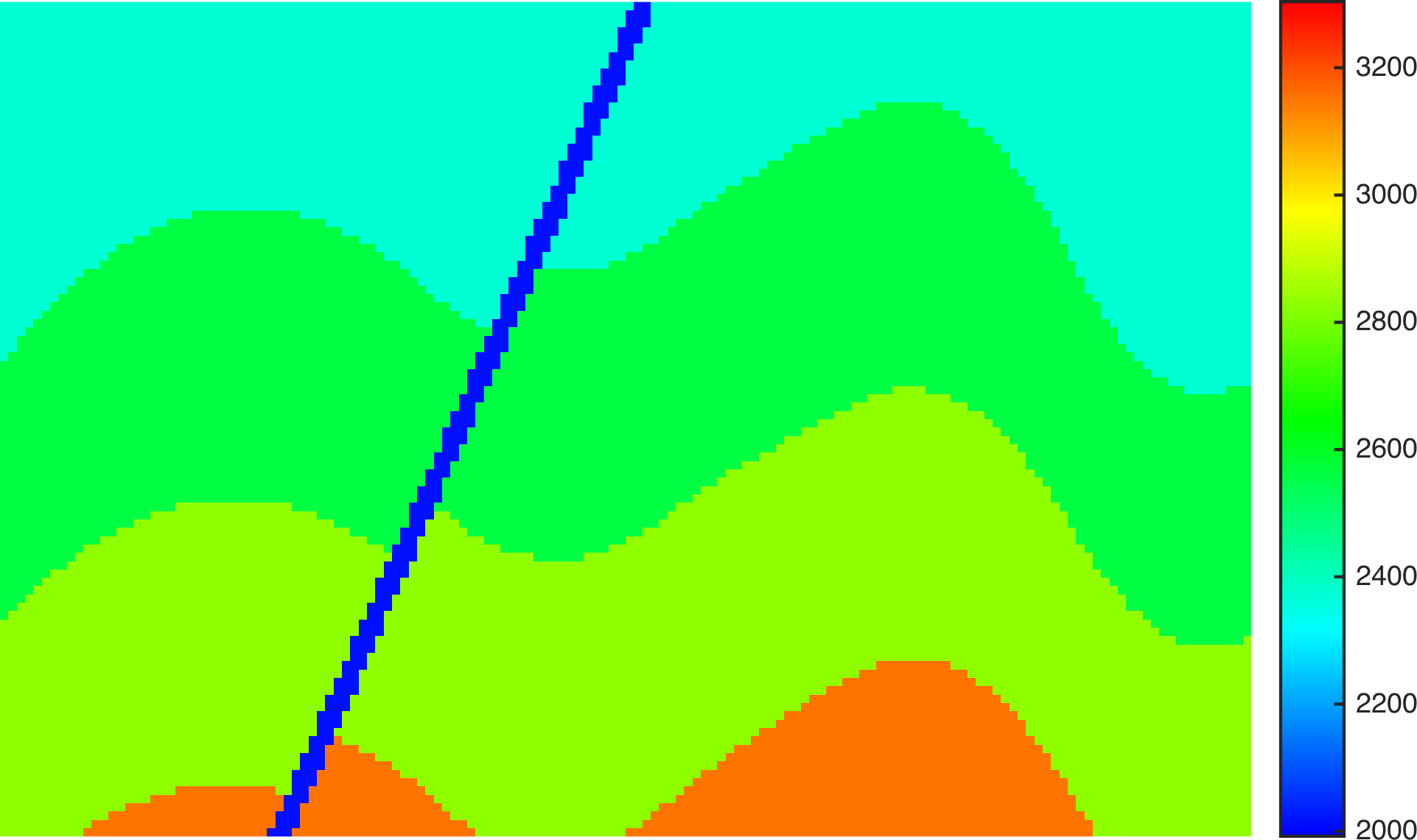}}%
\hspace{0.1cm}
\subfloat[]{\includegraphics[width=0.25\textwidth]{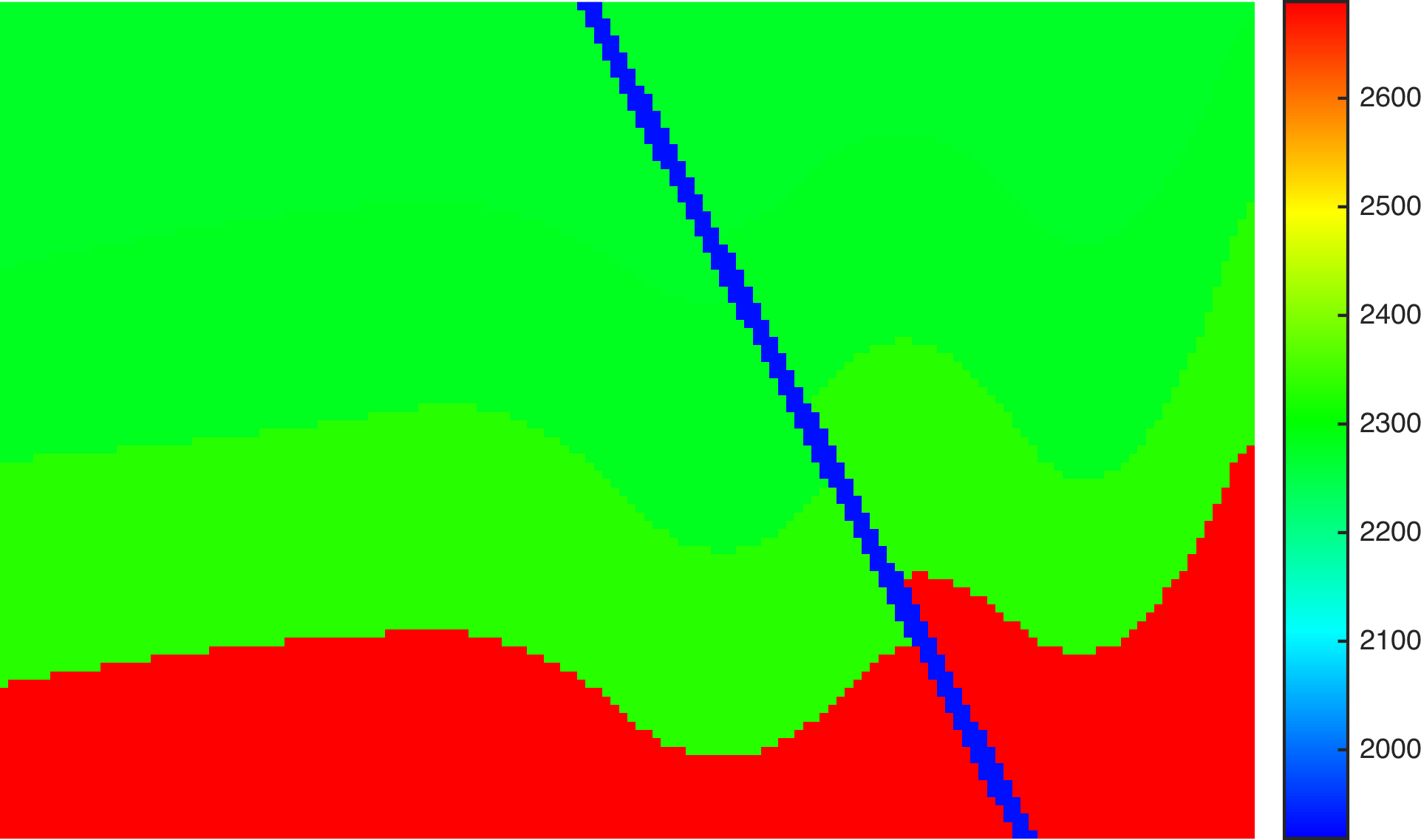}}
\hspace{0.1cm}
\subfloat[]{\includegraphics[width=0.25\textwidth]{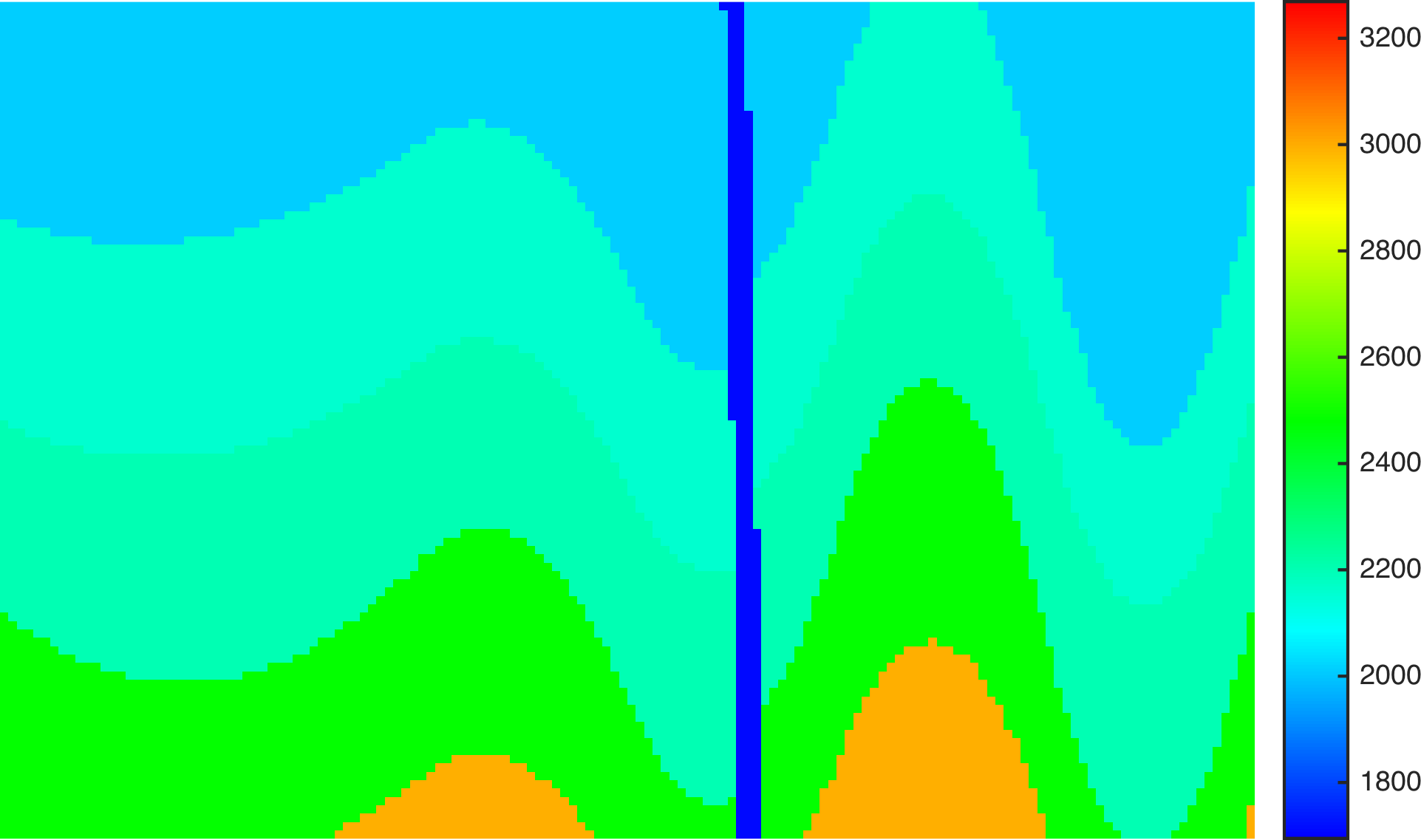}}
\hspace{0.1cm}
\subfloat[]{\includegraphics[width=0.25\textwidth]{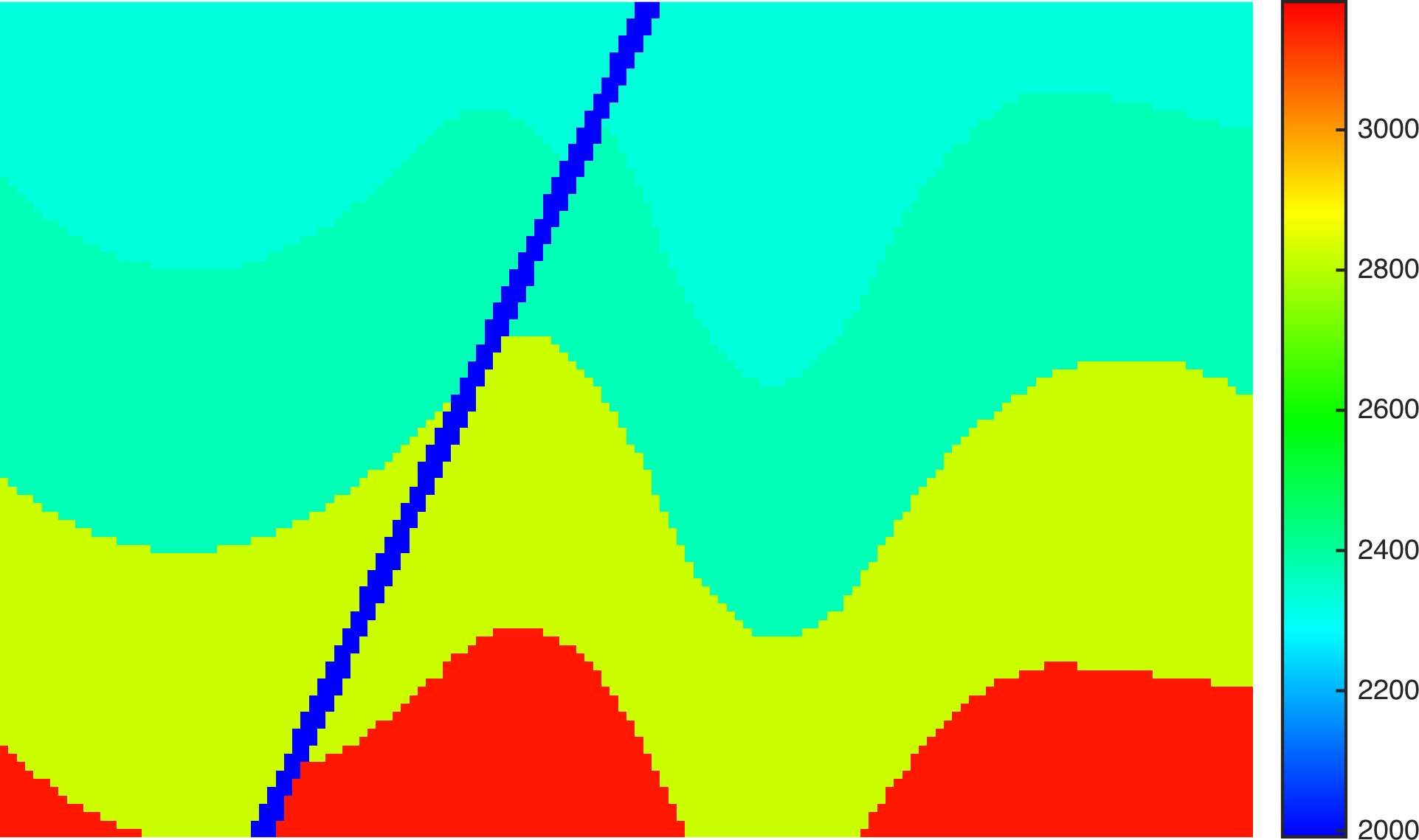}}
}
\caption{Illustration of four ground truth velocity models~(Top) randomly selected from CurvedVel set. Their nearest neighbors~(Bottom) are also provided. The nearest neighbors share some similarity to their ground truth counterpart, however, the details including velocity values, fault orientation, layer location are all different.}
\label{fig:NN_Results_Curved}
\end{figure*}

We create two datasets --- FlatVel, which is simulated with flat subsurface layers, and CurvedVel, which is simulated with curved subsurface layers. FlatVel contains $36,000$ velocity models of $100 \times 100$ grid points. The velocity models in FlatVel~(shown in Fig.~\ref{fig:TrainingSeismicModel}) are different from one another in terms of offset~(ranging from 30 grids to 70 grids), tilting angle~(ranging from $25^{\circ}$ to $165^{\circ}$), layer thickness~(ranging from 5 grids to 80 grids), and layer velocity~(ranging from 3000~m/s to 5000~m/s). CurvedVel~(shown in Fig.~\ref{fig:TrainingSeismicModel}) contains $50,000$ velocity models of $100 \times 150$ grid points.  We vary the velocity values in CurvedVel from 1,500 m/s to 3,500 m/s, the fault offset from 30 grids to 70 grids, tilting angle from 25 to 165 degrees, the number of layers from 3 to 5, and the layer thickness from 5 grids to 80 grids. CurvedVel is more challenging to reconstruct than the FlatVel model for two reasons. Firstly, CurvedVel contains much more irregular geological structures which make the inverse of the forward modeling function more difficult to approximate. Secondly, the curve model is also 1.5 times larger than the FlatVel model, which means much more velocity values need to be correctly estimated by our InversionNet.

The seismic measurements are collections of synthetic seismograms obtained by implementing forward modeling on velocity models. For CurvedVel, a total of 3 sources and 150 receivers are evenly distributed along the top boundary of the model. The source interval is $150$~m, and the receiver interval is $15$~m. We use a Ricker wavelet with a center frequency of $25$~Hz as the source time function and a staggered-grid finite-difference scheme with a perfectly matched layered absorbing boundary condition to generate synthetic seismic reflection data. The synthetic trace at each receiver is a collection of time series data of length $2,000$. In Fig.~\ref{fig:TrainingSeismicData}, we show a portion of the synthetic seismic data sets corresponding to velocity models that we generate. Specifically, the displacement in the X direction is shown in the left panel of Fig.~\ref{fig:TrainingSeismicData}, and the displacement in the Z direction is shown in the right panel of Fig.~\ref{fig:TrainingSeismicData}. Similarly, for FlatVel there are 3 sources and 32 receivers used in the curve model. Both the sources and receives are evenly distributed on the top of the model. The source interval is 125 m and the receiver interval is 5 m. Each receiver collects the time series data of length $1,000$.  We downsample the seismic measurement in the CurvedVel seismic dataset to $32 \times 1000$ to make it consistent with the FlatVel seismic meansurements.

\subsubsection{Nearest Neighbors}

Identifying the nearest neighbors to the test sets is a common approach to evaluate the quality of the training sets. It is important to create a training set that can represent the distribution of the test sets, which will be learned by the neural networks. On the other hand, we should not expect the nearest neighbors become too similar to the test sets, which will be hard to justify if our algorithms learn the true distribution from the training sets or simply memorize the training samples. We first randomly select a few ground truth for FlatVel and CurvedVel sets, then we locate the nearest neighbors from their training sets, respectively. We provide the ground truth and nearest neighbors of FlatVel and CurvedVel sets in Figs.~\ref{fig:NN_Results_Flat} and ~\ref{fig:NN_Results_Curved}. We observe from the figures that the nearest neighbors share some similarity to their ground truth counterpart, however, the details including velocity values, fault orientation, and layer location are all different.



\subsection{Implementation Details}
For FlatVel and CurvedVel, we use 30,000 / 45,000 pairs of seismic measurements and velocity models for training, respectively; and 6,000 / 5,000 pairs for testing, respectively. We adopt the piecewise training strategy to first learn a CNN backbone, then optimize the parameters in the CRF. We apply the Adam optimizer~\citep{Adam} to update the parameters of CNN. The batch size is 50. The initial learning rate is set to 0.0005, we multiply the learning rate by 0.1 after each 15 training epochs. The proposed model has approximately 30~million parameters. Our InversionNet is implemented on TensorFlow~\citep{tensorflow} with a single Nvidia GTX 1080 Ti GPU. In comparison, our physics approaches run on HPC clusters with a total of 154 nodes and each of the node is a Intel Xeon E5-2670 CPU. The learning methodology of CRF is elaborated in Sec~\ref{sec:crf_learning}. According to the dimension of velocity models, we make $h=w=7$ for the first deconvolution layer for FlatVel, and $h=7$, $w=10$ for CurvedVel. We set $\alpha = 0.2$ in Eq.~\eqref{eq:relu} for all ReLU layers. We sample the seismic measurements in CurvedVel to make the dimension $32 \times 1000$. We do not normalize the seismic measurements before feeding into the network, but we standardize the velocity models.

\begin{table*}[h]
	\centering
	\begin{tabular}{ c|ccc|cccc }
		& mae & rel ($10^{-3}$)& log10 ($10^{-3}$) & acc. (t = 1.01) & acc. (t = 1.02) & acc. (t = 1.05) & acc. (t = 1.10) \\
		\hline
		AEWI-PRE & 68.34 & 18.38 & 8.04 & 8.99\% & 14.02\% & 27.50\% & 44.88\% \\
		AEWI-MTV & 36.66 & 9.50 & 4.17 & 39.83\% & 45.28\% & 55.49\% & 63.39\% \\
		CNN & 29.59 & 7.41 & 3.22 & 81.94 \% & 95.62 \% & 98.55 \% & 99.39 \% \\
		CNN~(residual) & 34.95 & 9.10 & 3.95 & 76.53 \% & 93.48 \% & 97.83 \% & 99.09 \% \\
		CNN-CRF (d=5) & 28.28 & 7.05 & 3.06 & 83.63 \% & \textbf{96.30} \% & \textbf{98.68} \% & \textbf{99.40 \%} \\
		CNN-CRF (d=20) & \textbf{28.20} & \textbf{7.01} & \textbf{3.04} &  \textbf{83.93} \% &96.22 \% & 98.56 \% & 99.37 \% \\
		CNN-CRF (d=40) & 29.09 & 7.23 & 3.13 & 82.73 \% & 95.68\% & 98.54 \% & 99.36 \% \\
	\end{tabular}
	\caption{Quantitative results obtained on FlatVel. We compare the physics-driven models with the proposed data-driven models with different settings. The data-driven models perform significantly better under all metrics. The CRF further boosts the performance by approximately 10\%/.}
	\label{table:results_FlatVel}
\end{table*}

\begin{table*}[h]
	\centering
	\begin{tabular}{ c|ccc|cccc }
		& mae & rel ($10^{-3}$)& log10 ($10^{-3}$) & acc. (t = 1.01) & acc. (t = 1.02) & acc. (t = 1.05) & acc. (t = 1.10) \\
		\hline
		AEWI-PRE & 172.88 & 69.71 & 30.34 & 22.99 \% & 35.73 \% & 58.63 \% & 76.18 \%\\
		AEWI-MTV & 145.84 & 57.77 & 25.13 & 40.82 \% & 51.19 \% & 65.70 \% & 79.15 \%\\
		CNN & 68.70 & 29.53 & 12.57 & 39.74 \% & 62.79 \% & 87.45 \% & 94.98 \% \\
		CNN~(residual) & 73.07 & 31.51 & 13.38 & 36.30 \% & 60.05 \% & 86.24 \% & 94.71 \% \\
		CNN-CRF (d=5) & 68.29 & 29.35 & 12.48 & 40.82 \% & 63.76 \% & 87.61 \% & \textbf{97.87 \%} \\
		CNN-CRF (d=20) & 68.13 & 29.25 & 12.45 & 41.38 \% & 64.18 \% & \textbf{87.68 \%} & 94.90 \% \\
		CNN-CRF (d=40) & \textbf{68.08} & \textbf{29.23} & \textbf{12.44} & \textbf{41.39 \%} & \textbf{64.26 \%} & 87.65 \% & 94.88 \% \\
	\end{tabular}
	\caption{Quantitative results obtained on CurvedVel. The data-driven models outperform the physics-driven counterparts by a large margin and the CNN coupled with the CRF yields the best results.}
	\label{table:results_CurvedVel}
\end{table*}

\subsection{Evaluation Metrics}

Inspired by existing works on FWI and depth estimation~\citep{Lin2015, metric_1, metric_2, metric_3, crf_depth_liu, crf_multiscale_depth}, we adopt the following metrics in depth estimation: 1) mean absolute error~(mae): $\frac{1}{n}\sum_{i}|\mathbf{m_{i}} - \mathbf{m_{i}}^{\star}|$; 2) mean relative error~(rel): $\frac{1}{n}\sum_{i}\frac{|\mathbf{m_{i}} - \mathbf{m_{i}}^{\star}|}{\mathbf{m_{i}}^{\star}}$; 3) mean $\log10$ error~($\log10$): $\frac{1}{n}|\log_{10}\mathbf{m_{i}} - \log_{10}\mathbf{m_{i}}^{\star}|$; 4) The percentage of $\mathbf{m_{i}}$ s.t. $\max (\frac{\mathbf{m_{i}}^{\star}}{\mathbf{m_{i}}}, \frac{\mathbf{m_{i}}}{\mathbf{m_{i}}^{\star}}) < t$.

\begin{figure*}[h]
\centering
\centerline{
\subfloat{\includegraphics[width=0.16\textwidth]{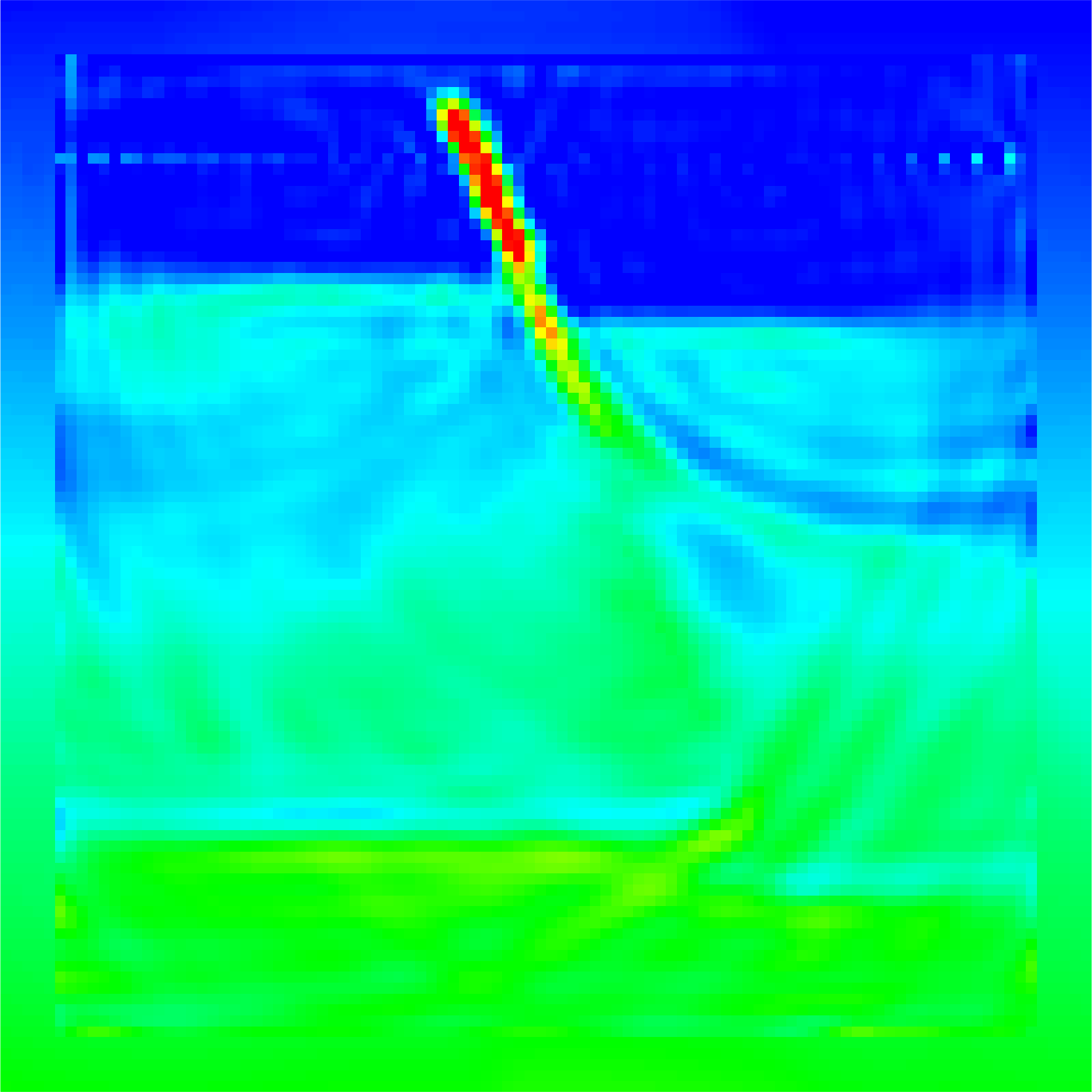}}%
\hspace{0.05cm}
\subfloat{\includegraphics[width=0.16\textwidth]{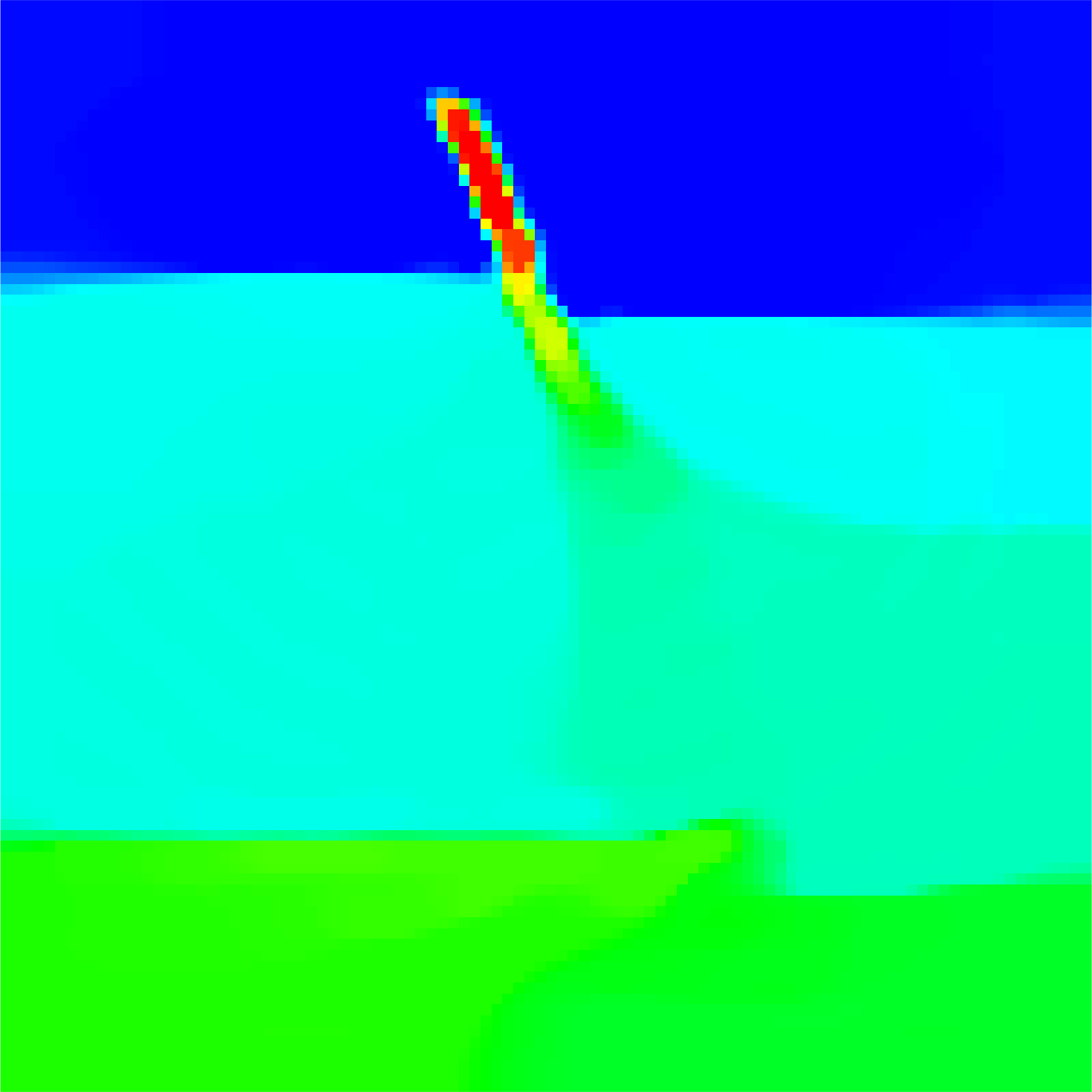}}
\hspace{0.05cm}
\subfloat{\includegraphics[width=0.16\textwidth]{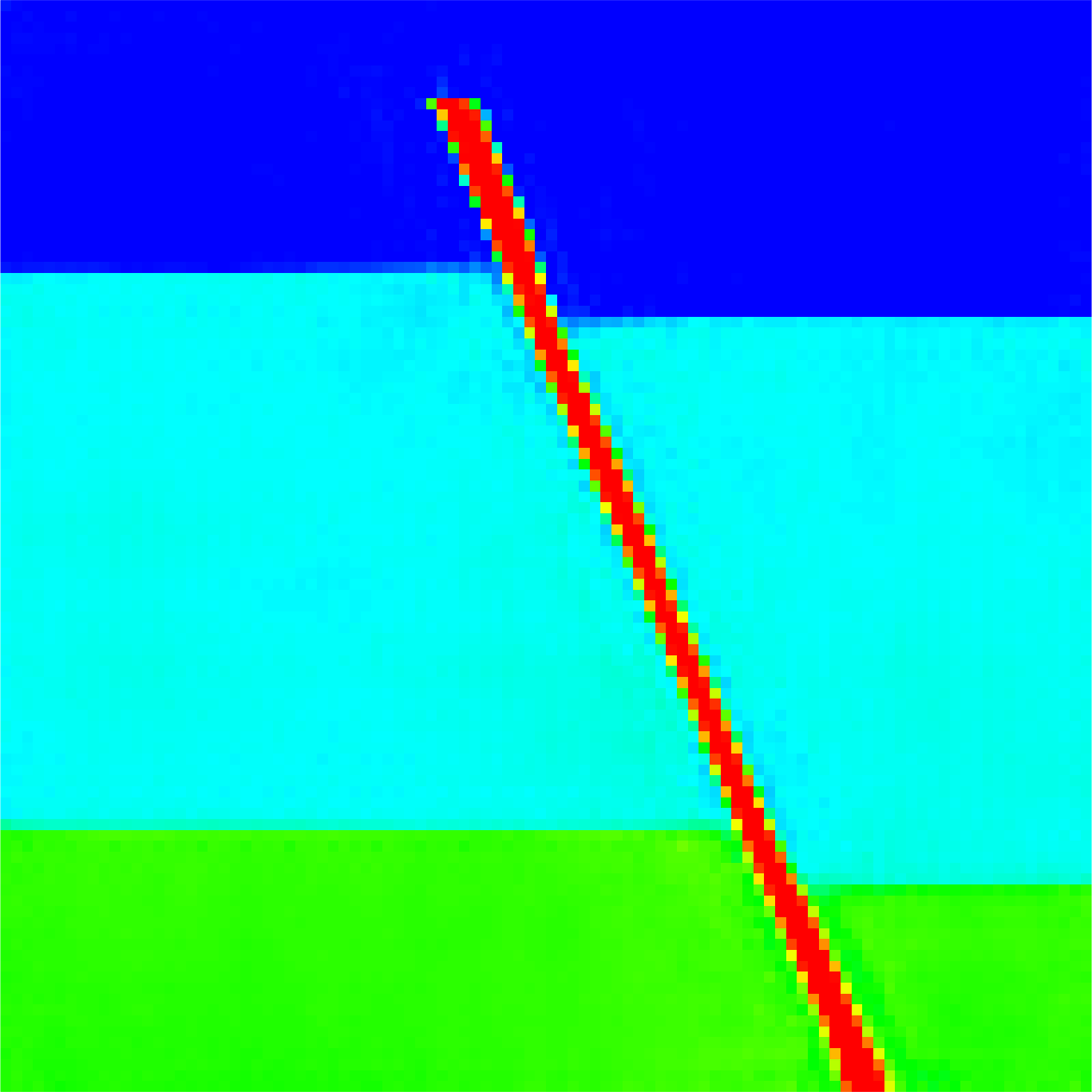}}
\hspace{0.05cm}
\subfloat{\includegraphics[width=0.16\textwidth]{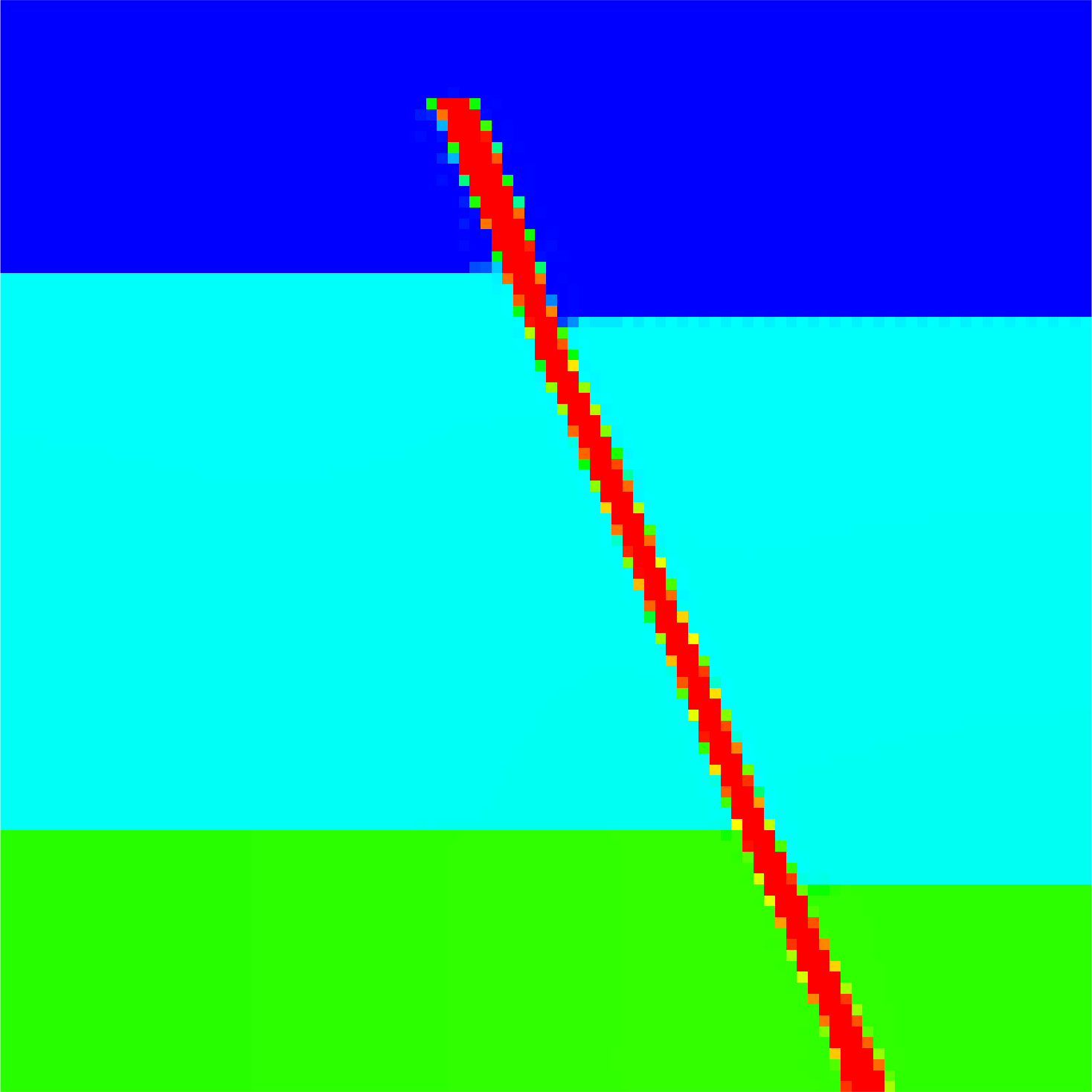}}
\hspace{0.05cm}
\subfloat{\includegraphics[width=0.16\textwidth]{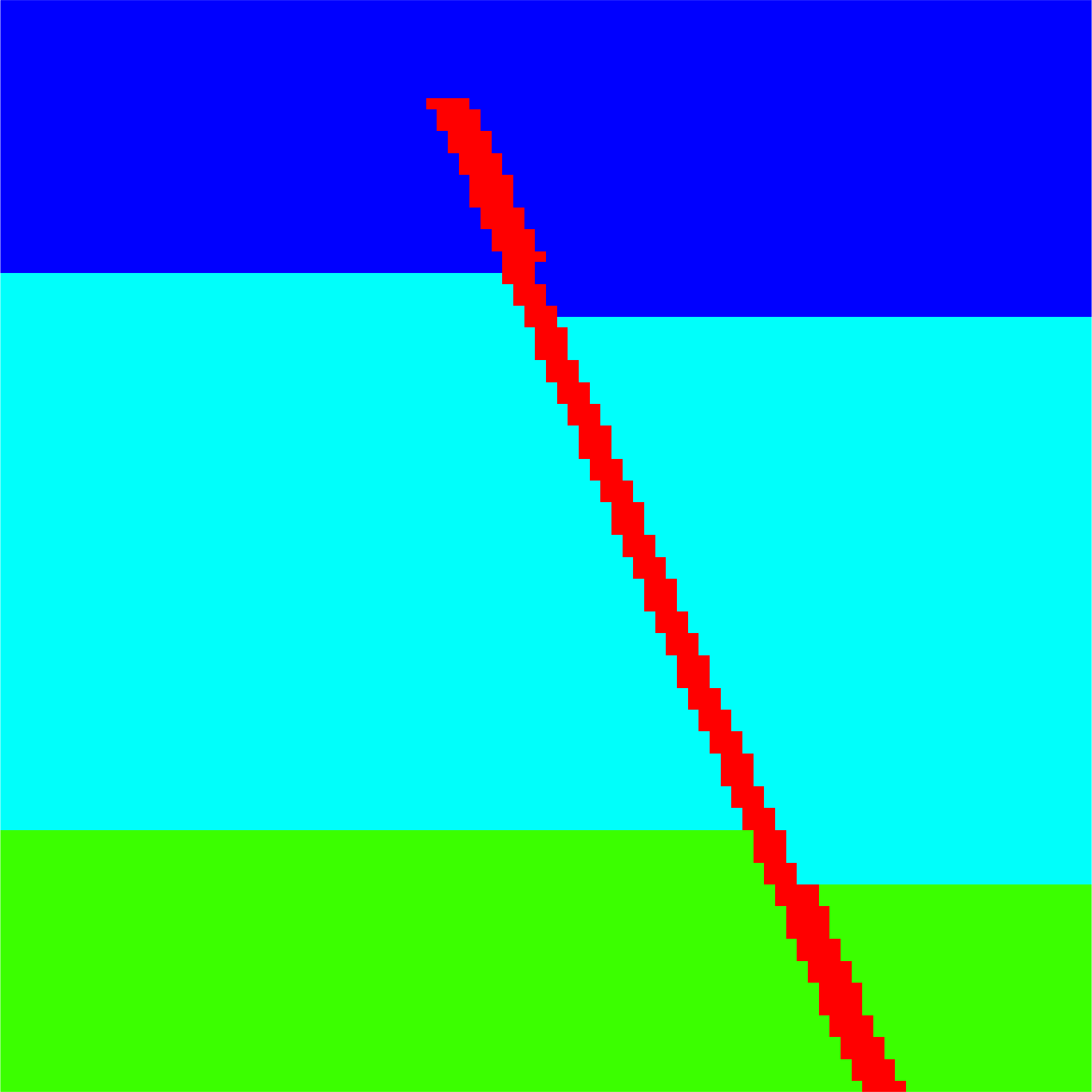}}
}
\vspace{0.1cm}
\centerline{
\subfloat{\includegraphics[width=0.16\textwidth]{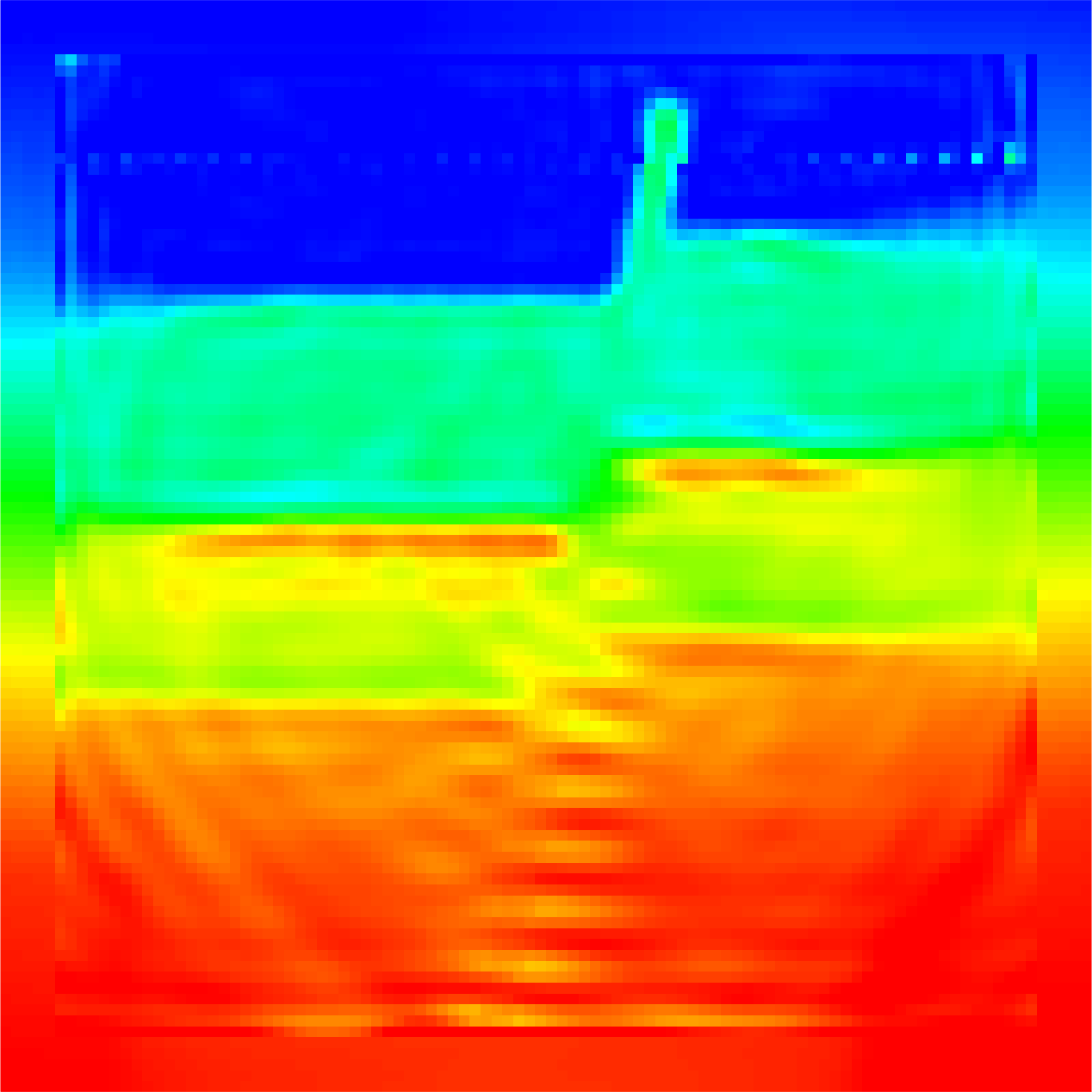}}%
\hspace{0.05cm}
\subfloat{\includegraphics[width=0.16\textwidth]{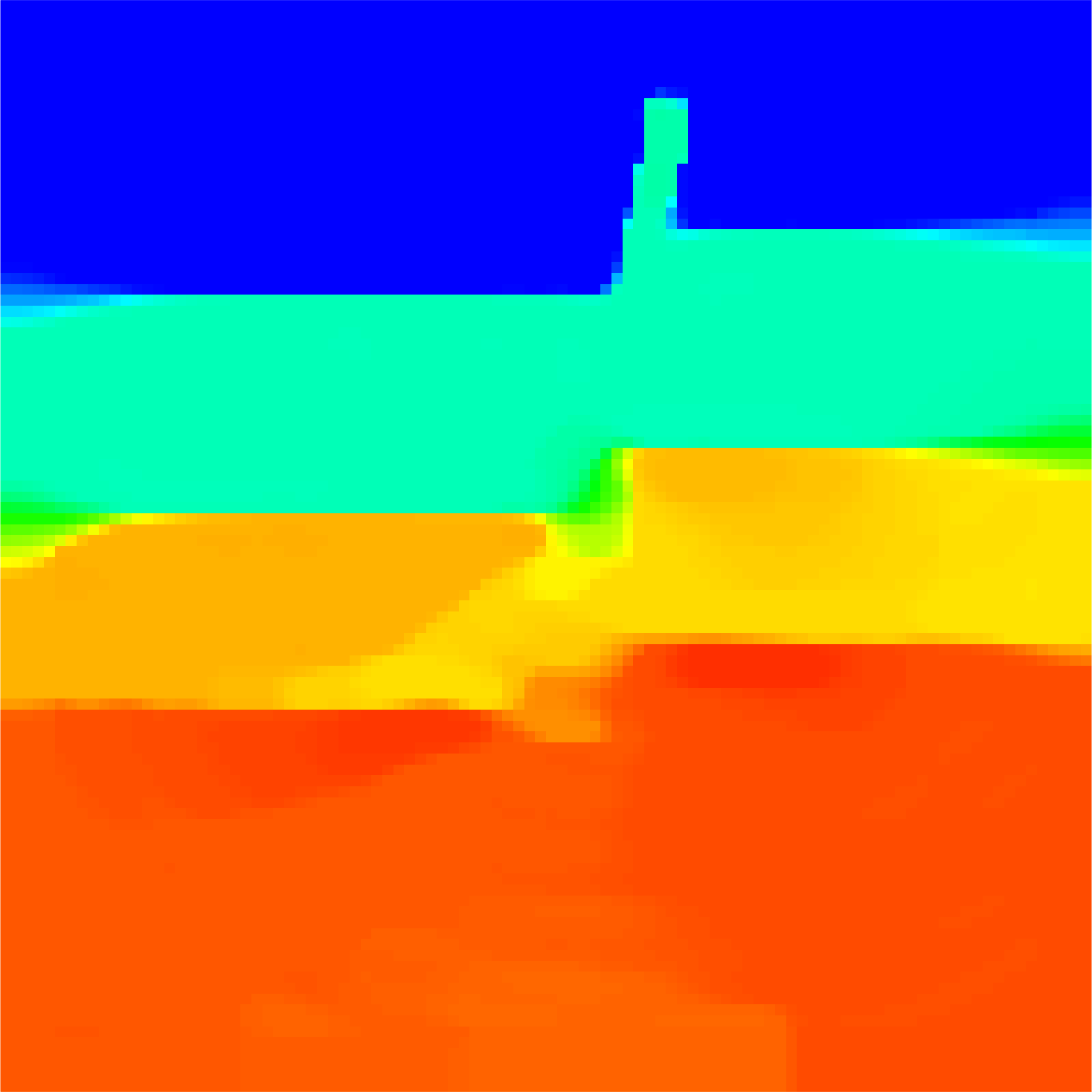}}
\hspace{0.05cm}
\subfloat{\includegraphics[width=0.16\textwidth]{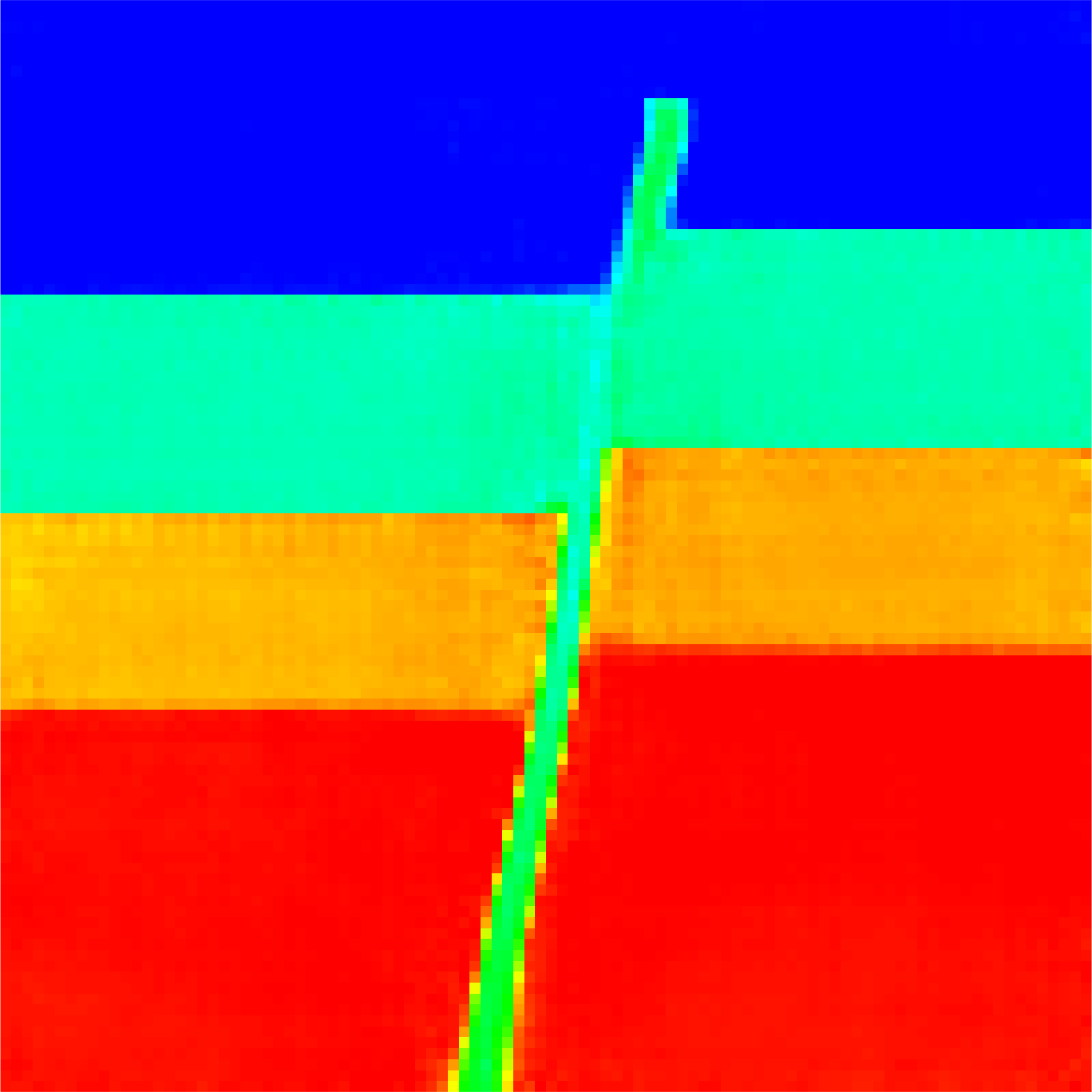}}
\hspace{0.05cm}
\subfloat{\includegraphics[width=0.16\textwidth]{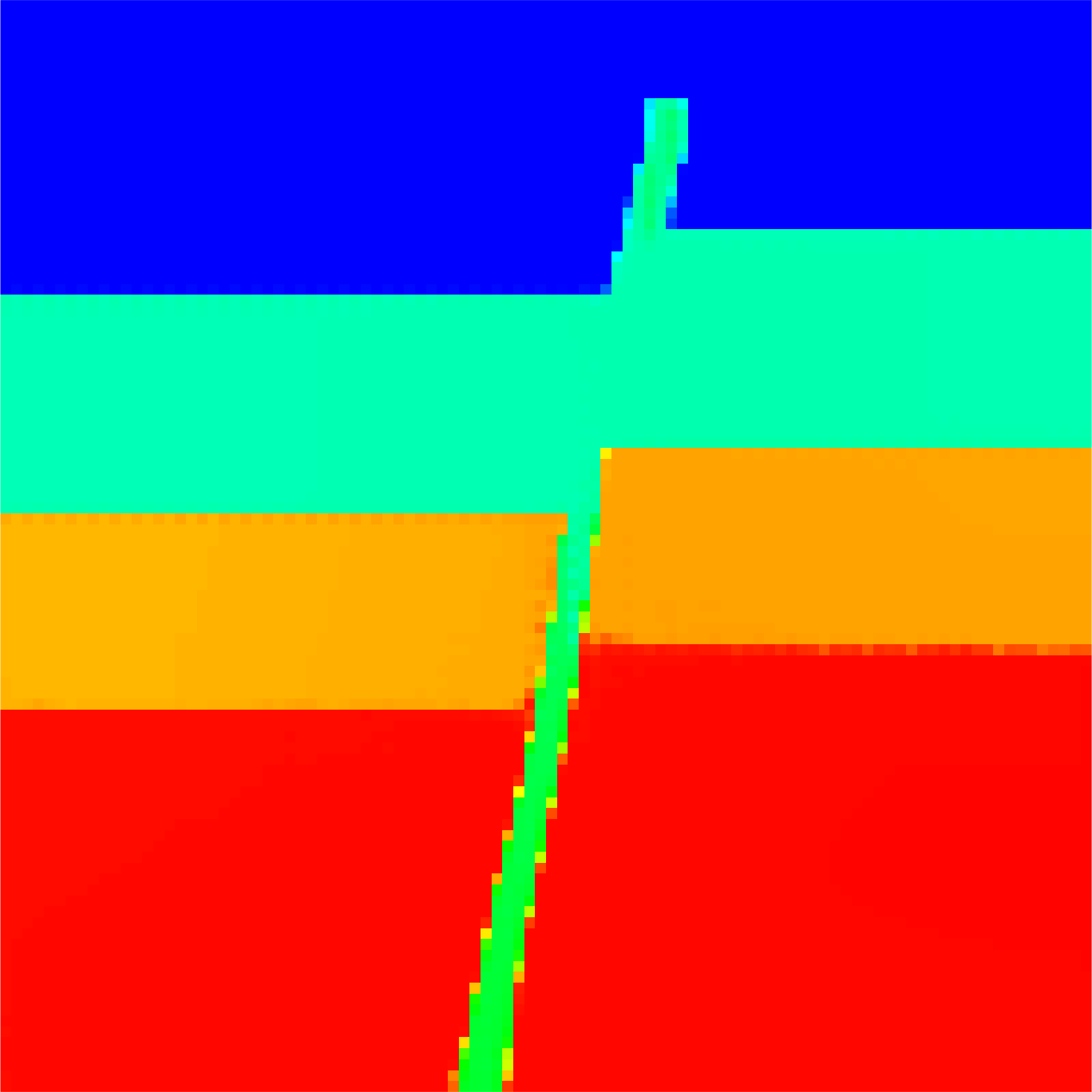}}
\hspace{0.05cm}
\subfloat{\includegraphics[width=0.16\textwidth]{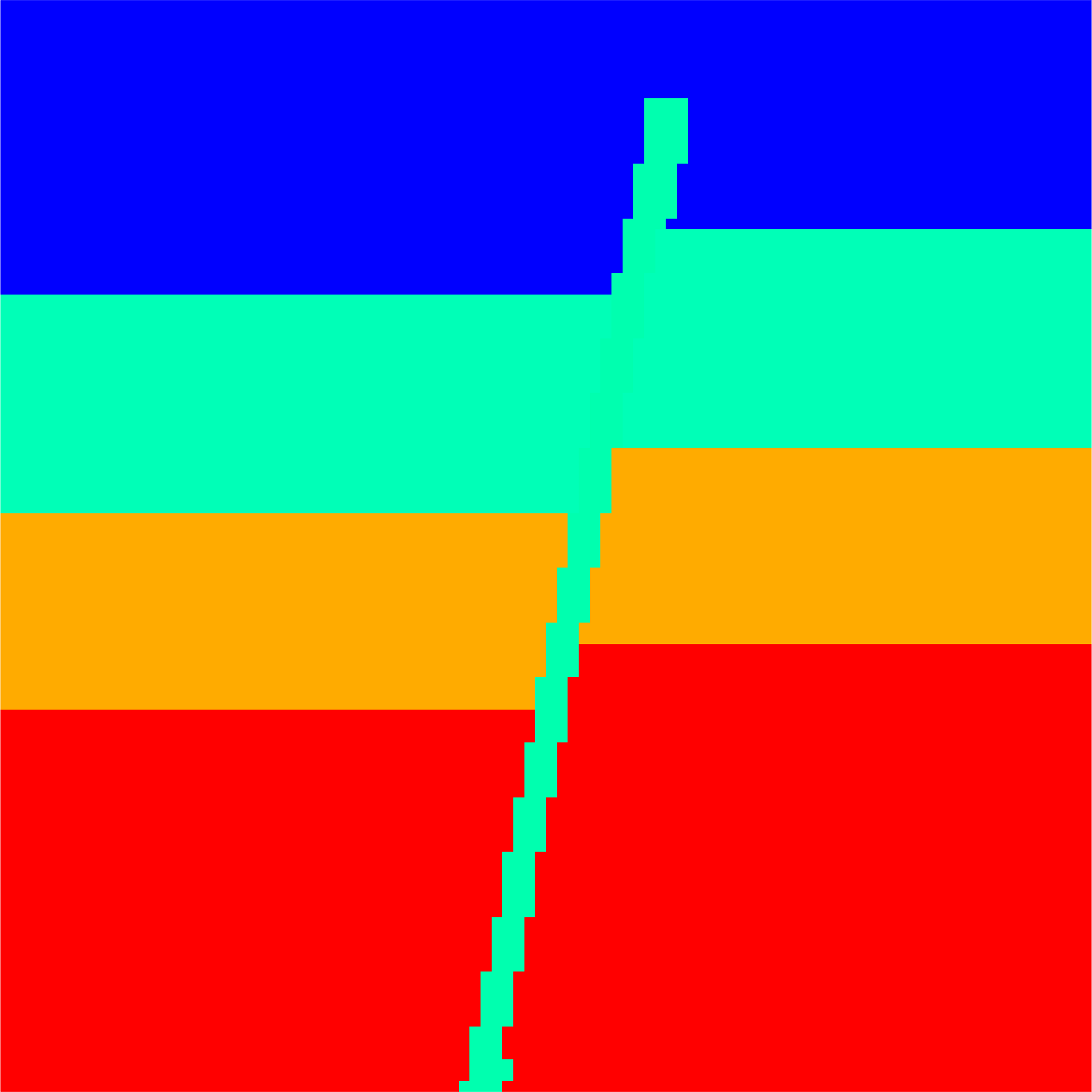}}
}
\vspace{0.1cm}
\centerline{
\subfloat{\includegraphics[width=0.16\textwidth]{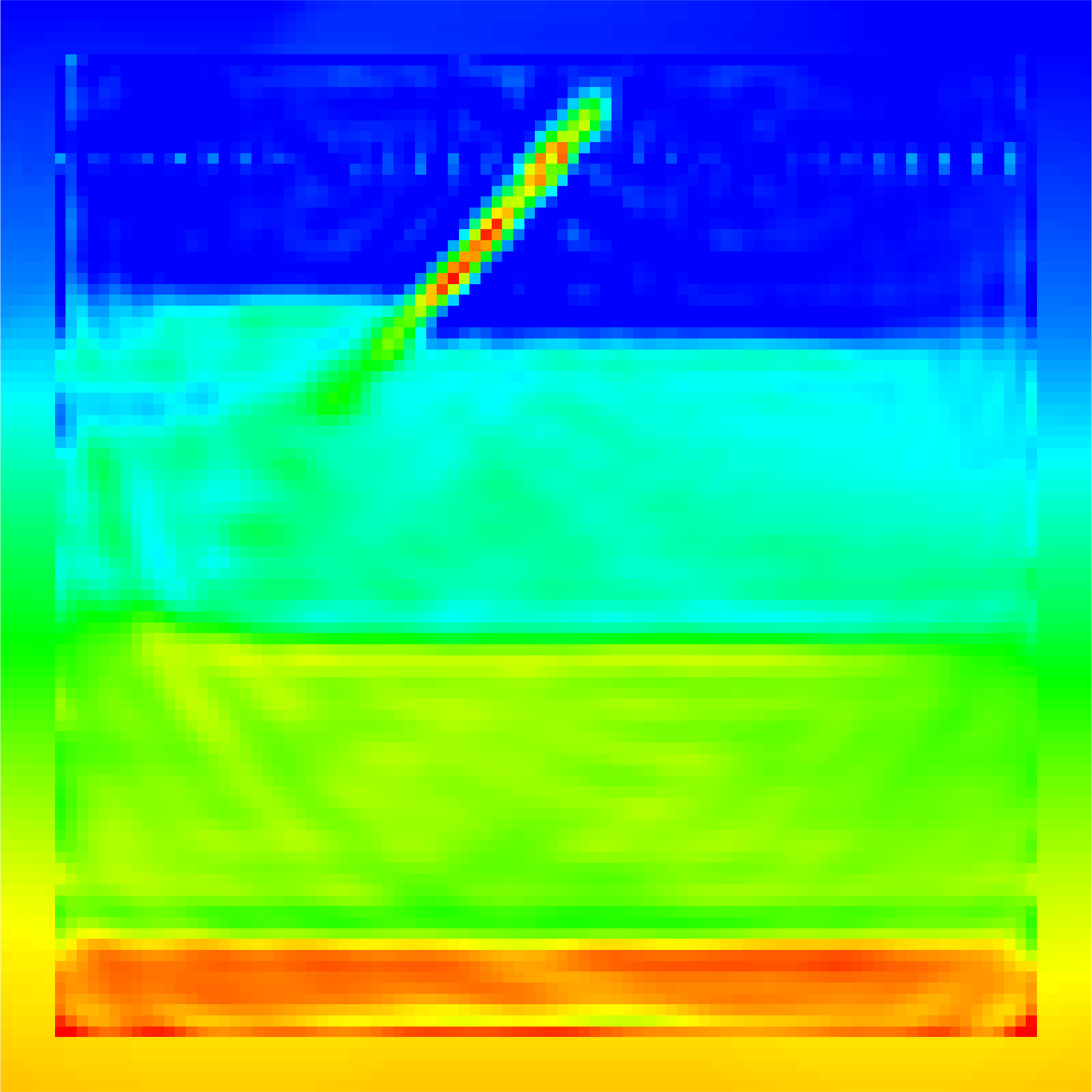}}%
\hspace{0.05cm}
\subfloat{\includegraphics[width=0.16\textwidth]{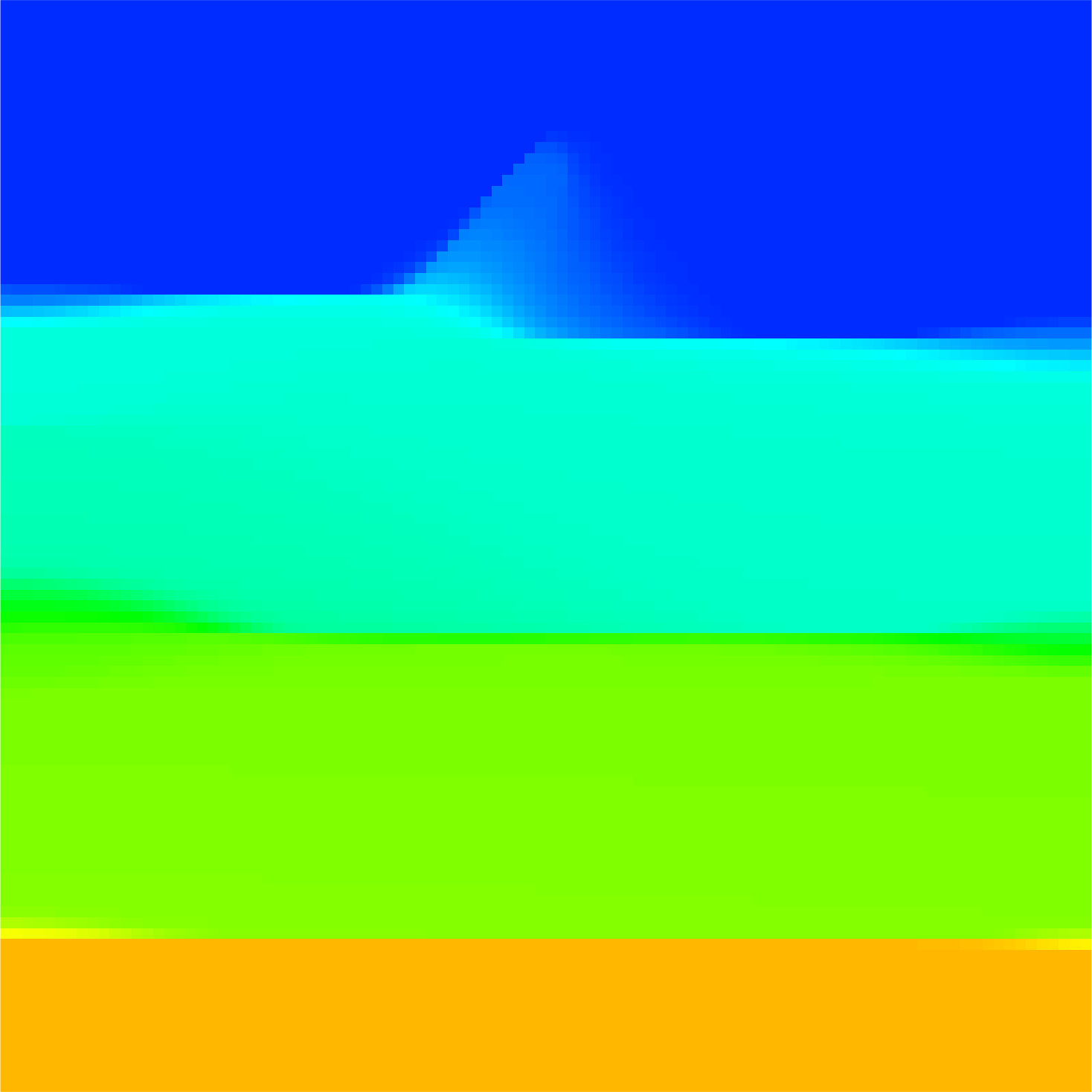}}
\hspace{0.05cm}
\subfloat{\includegraphics[width=0.16\textwidth]{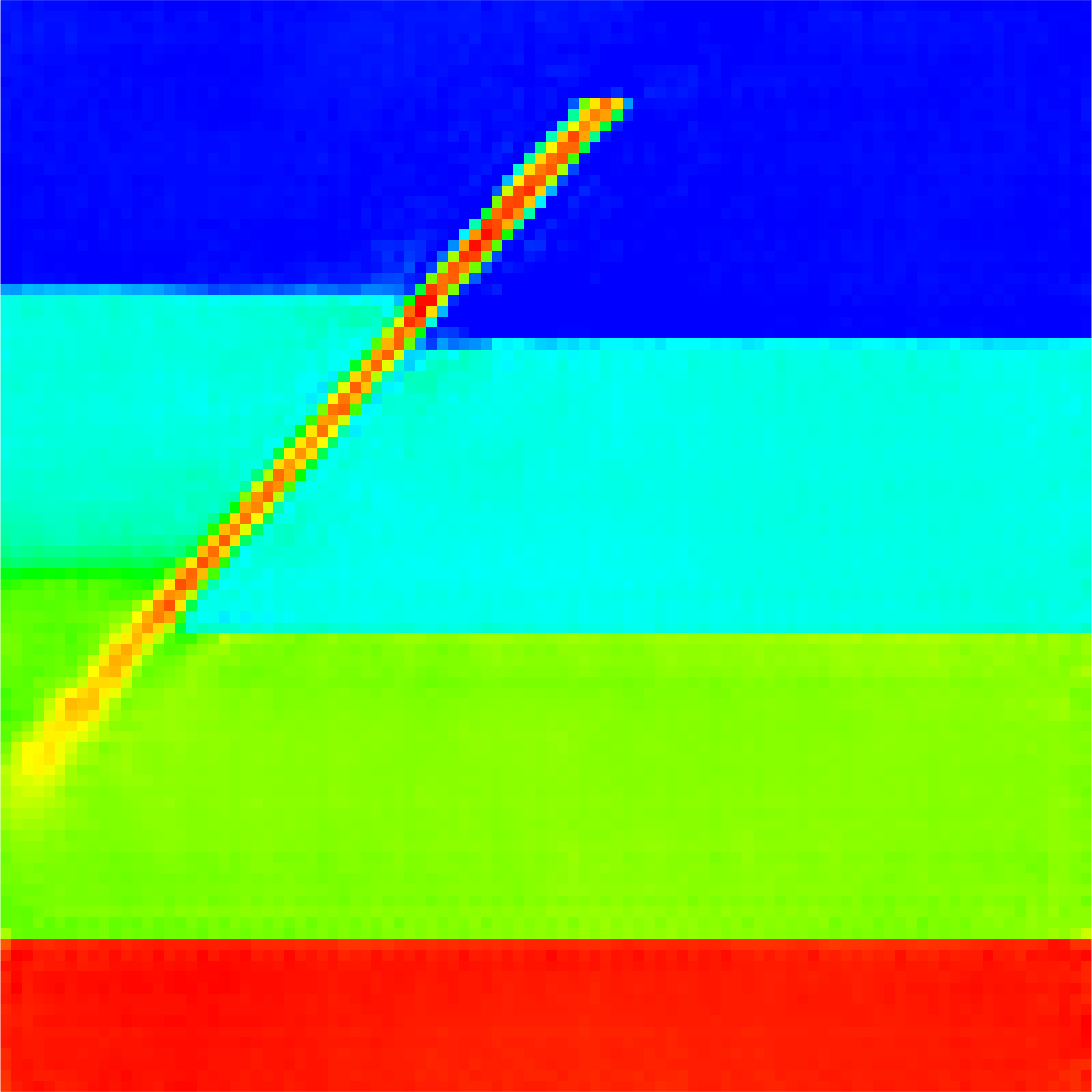}}
\hspace{0.05cm}
\subfloat{\includegraphics[width=0.16\textwidth]{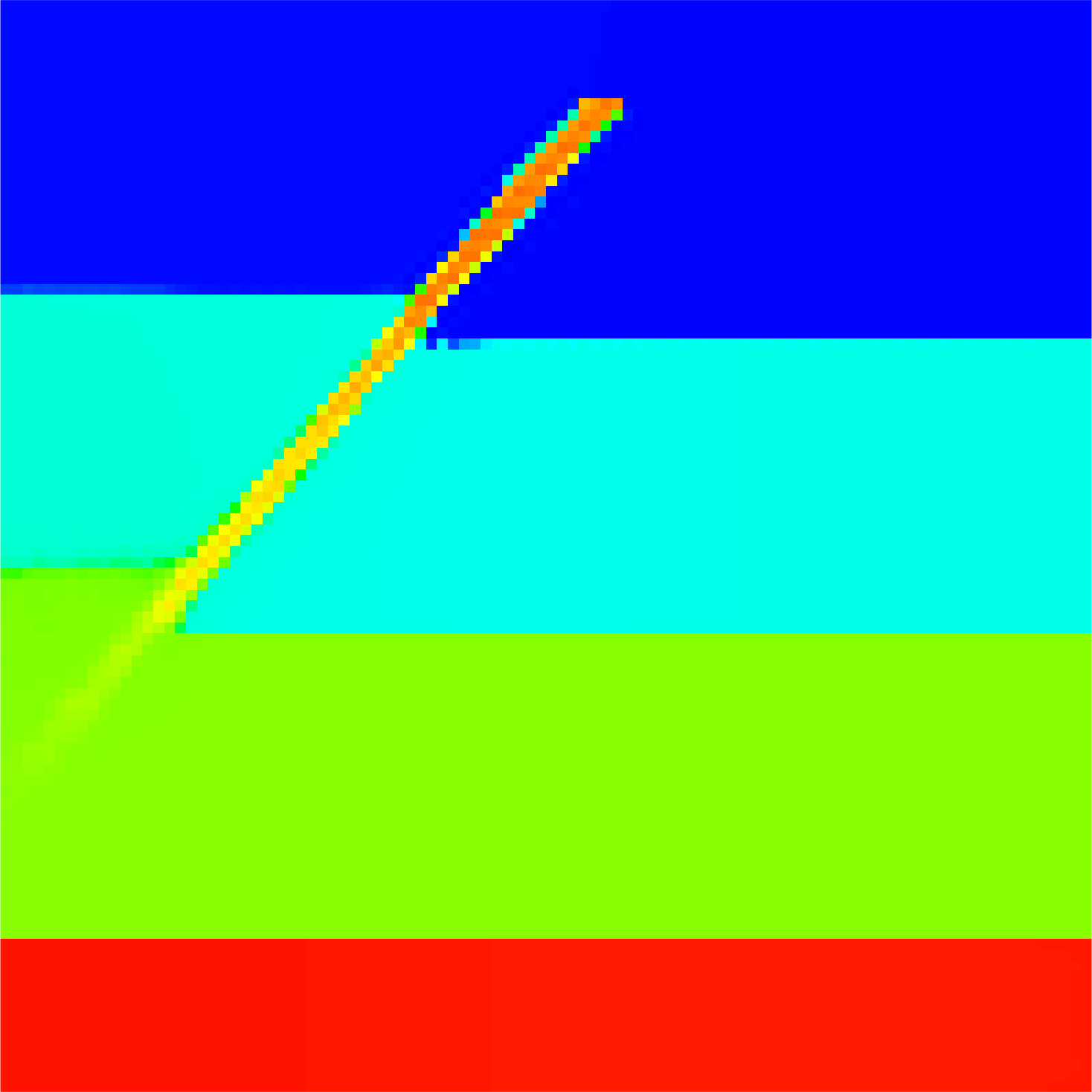}}
\hspace{0.05cm}
\subfloat{\includegraphics[width=0.16\textwidth]{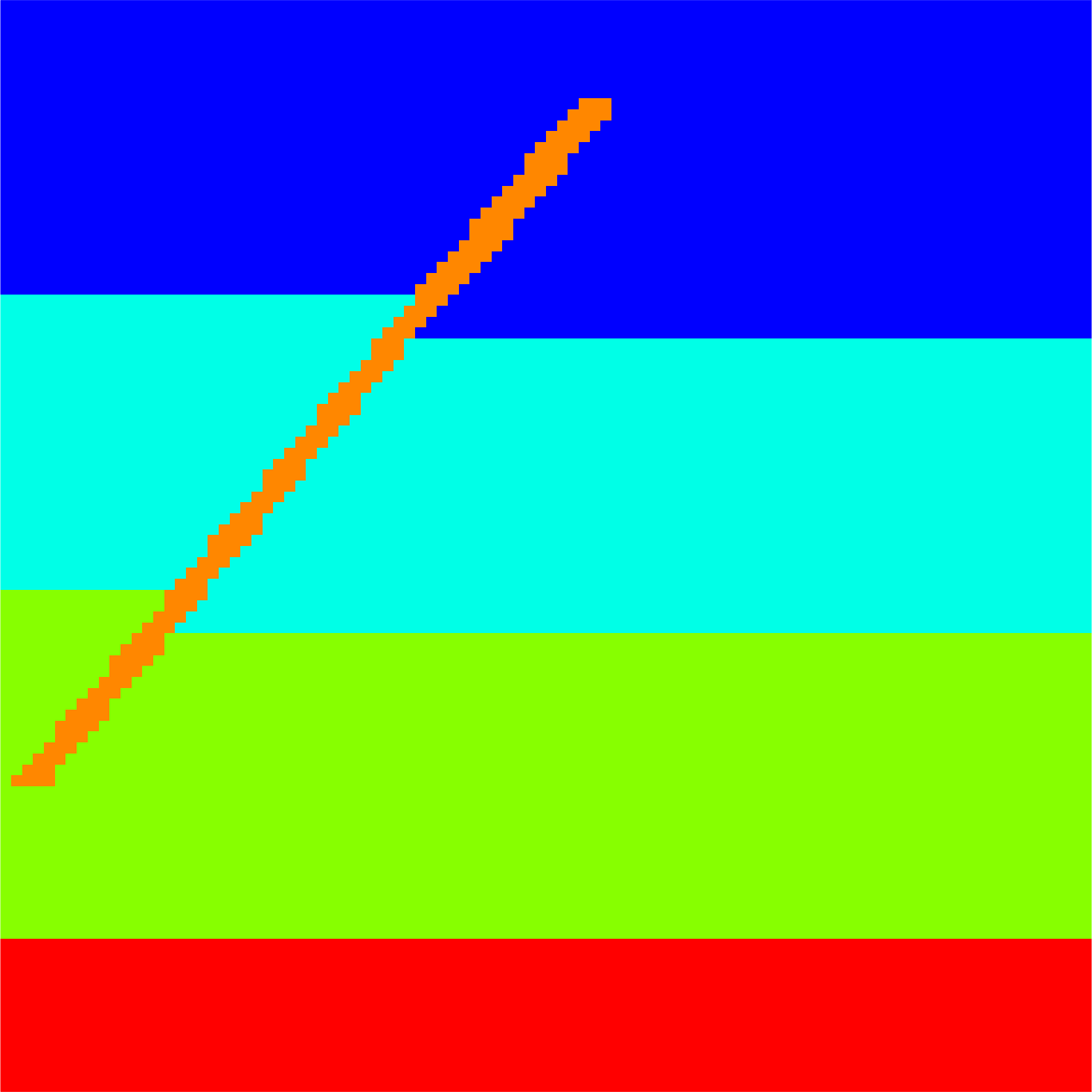}}
}
\vspace{0.1cm}
\centerline{
\subfloat{\includegraphics[width=0.16\textwidth]{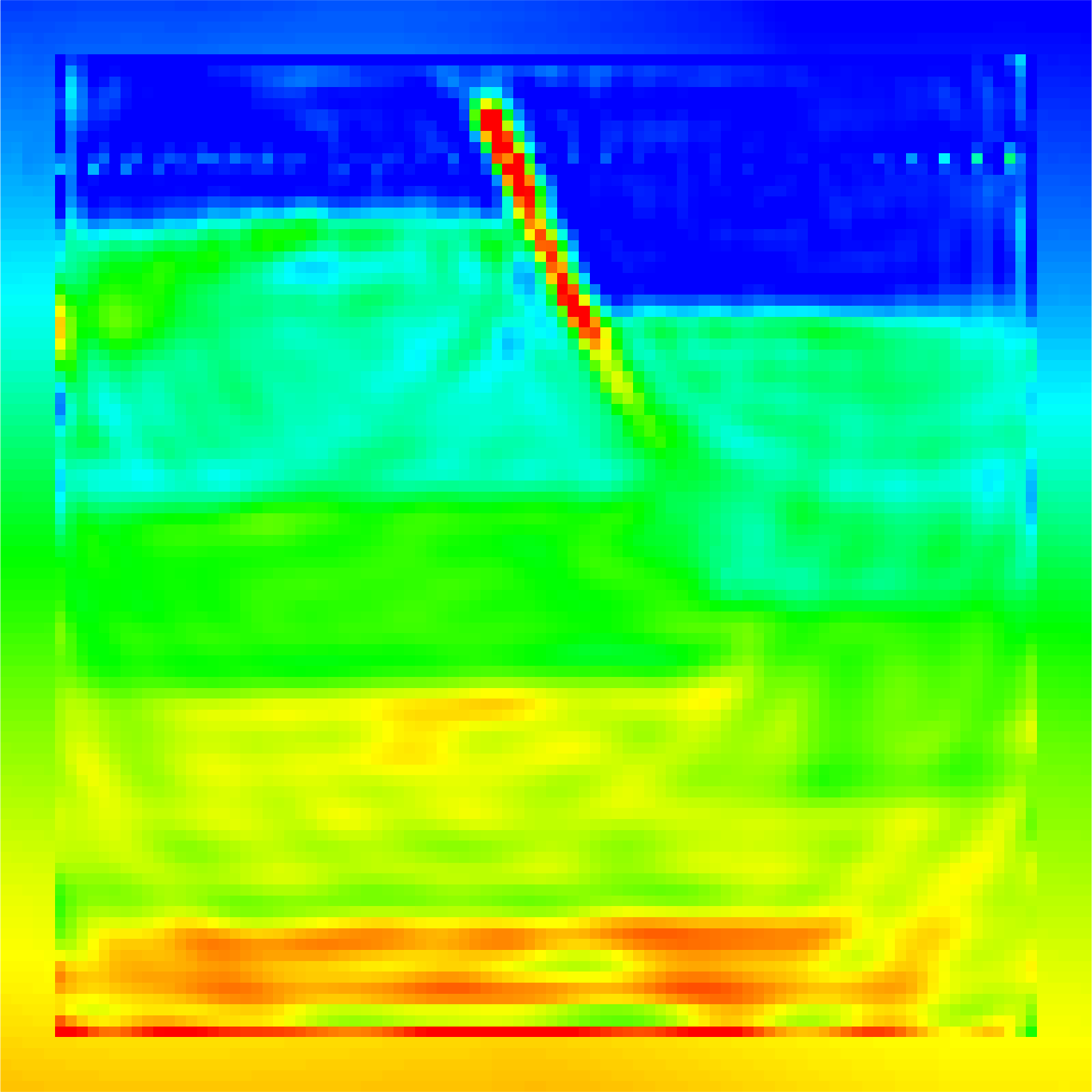}}%
\hspace{0.05cm}
\subfloat{\includegraphics[width=0.16\textwidth]{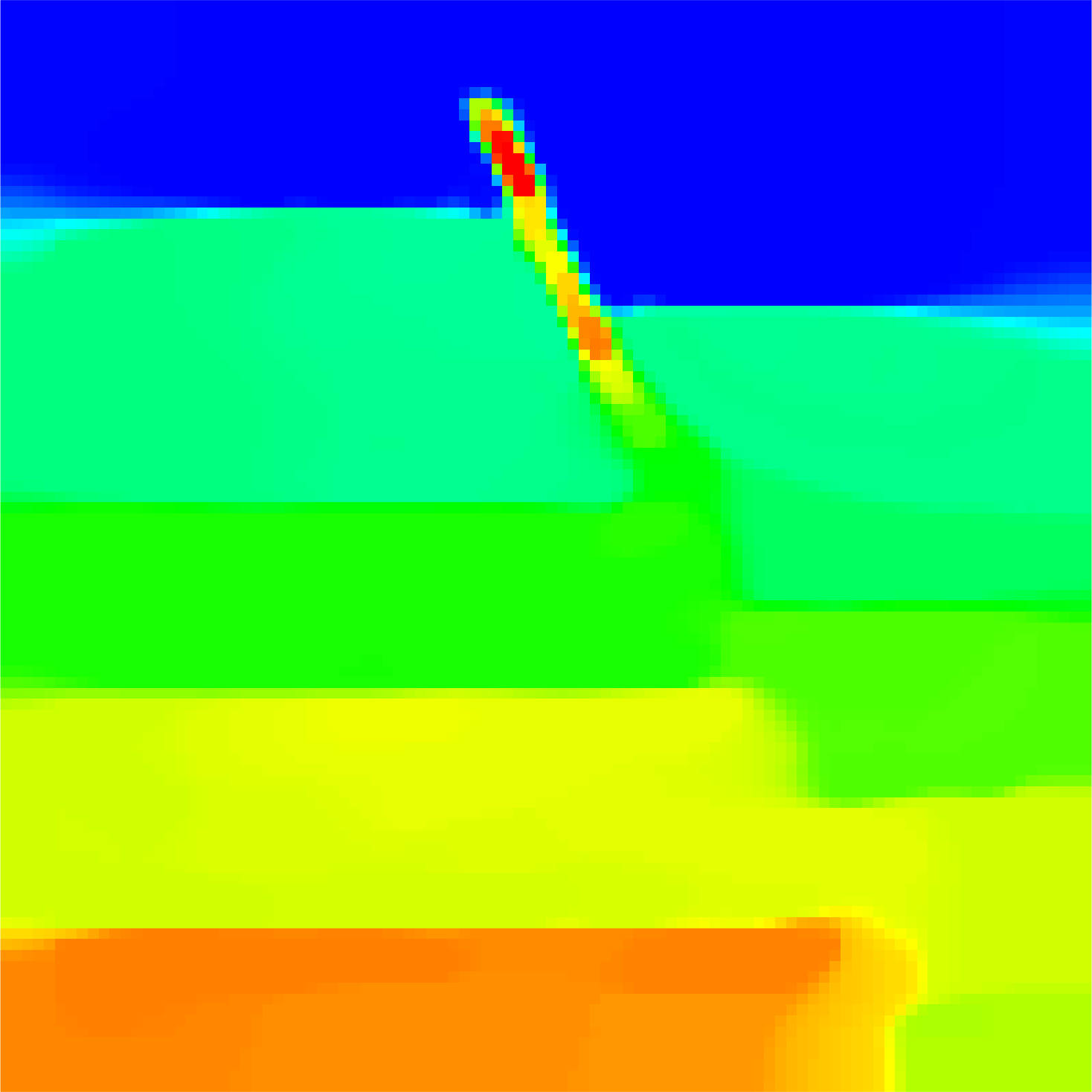}}
\hspace{0.05cm}
\subfloat{\includegraphics[width=0.16\textwidth]{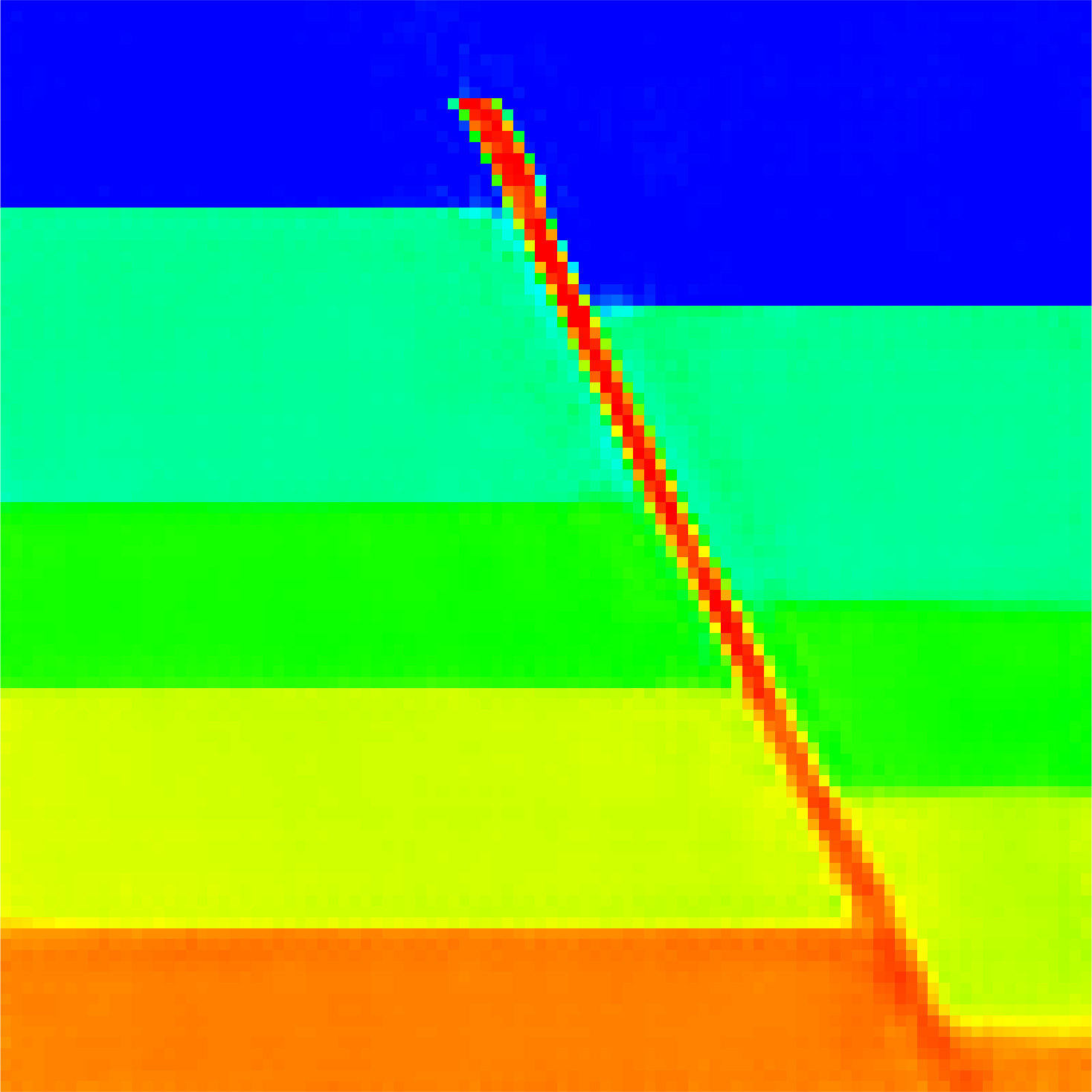}}
\hspace{0.05cm}
\subfloat{\includegraphics[width=0.16\textwidth]{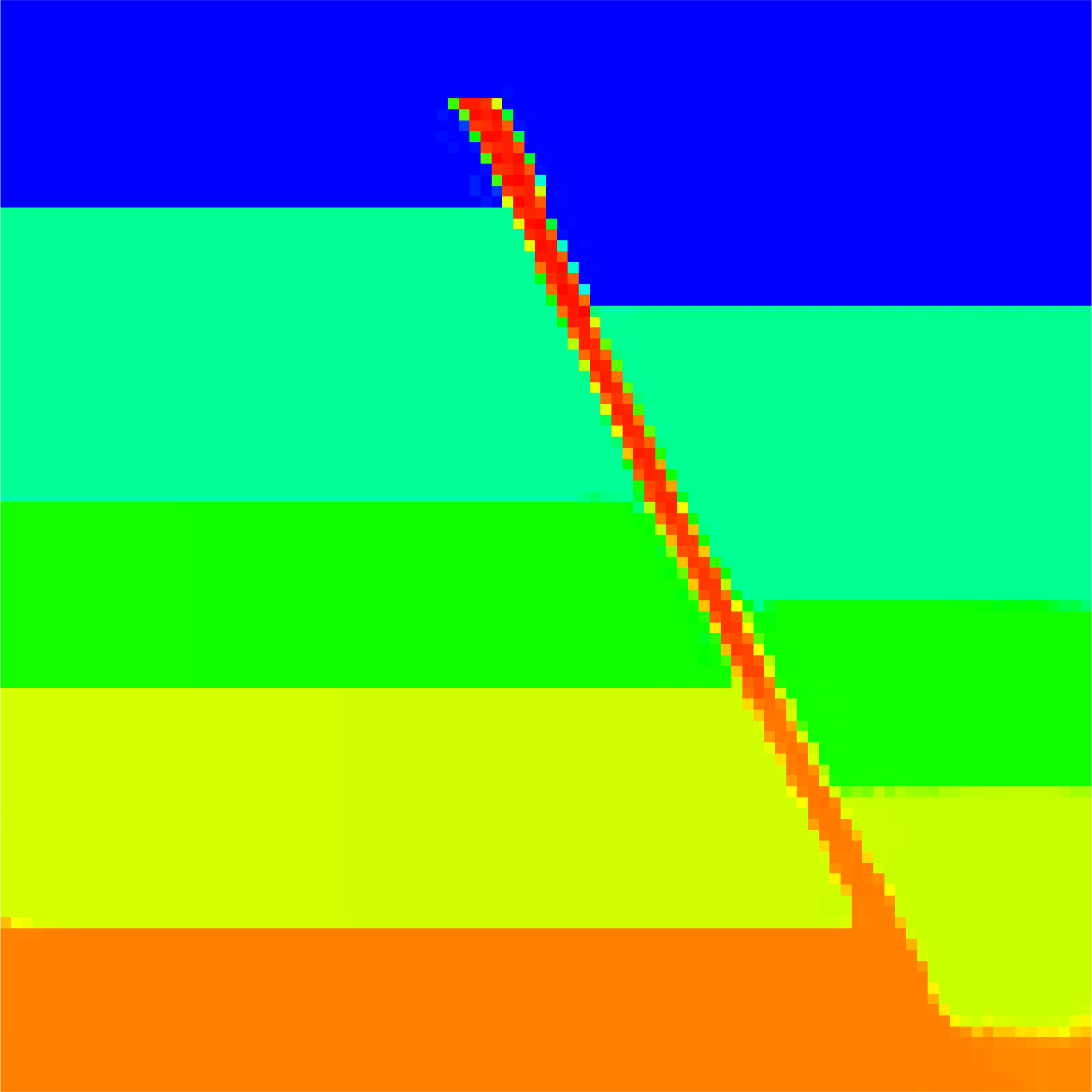}}
\hspace{0.05cm}
\subfloat{\includegraphics[width=0.16\textwidth]{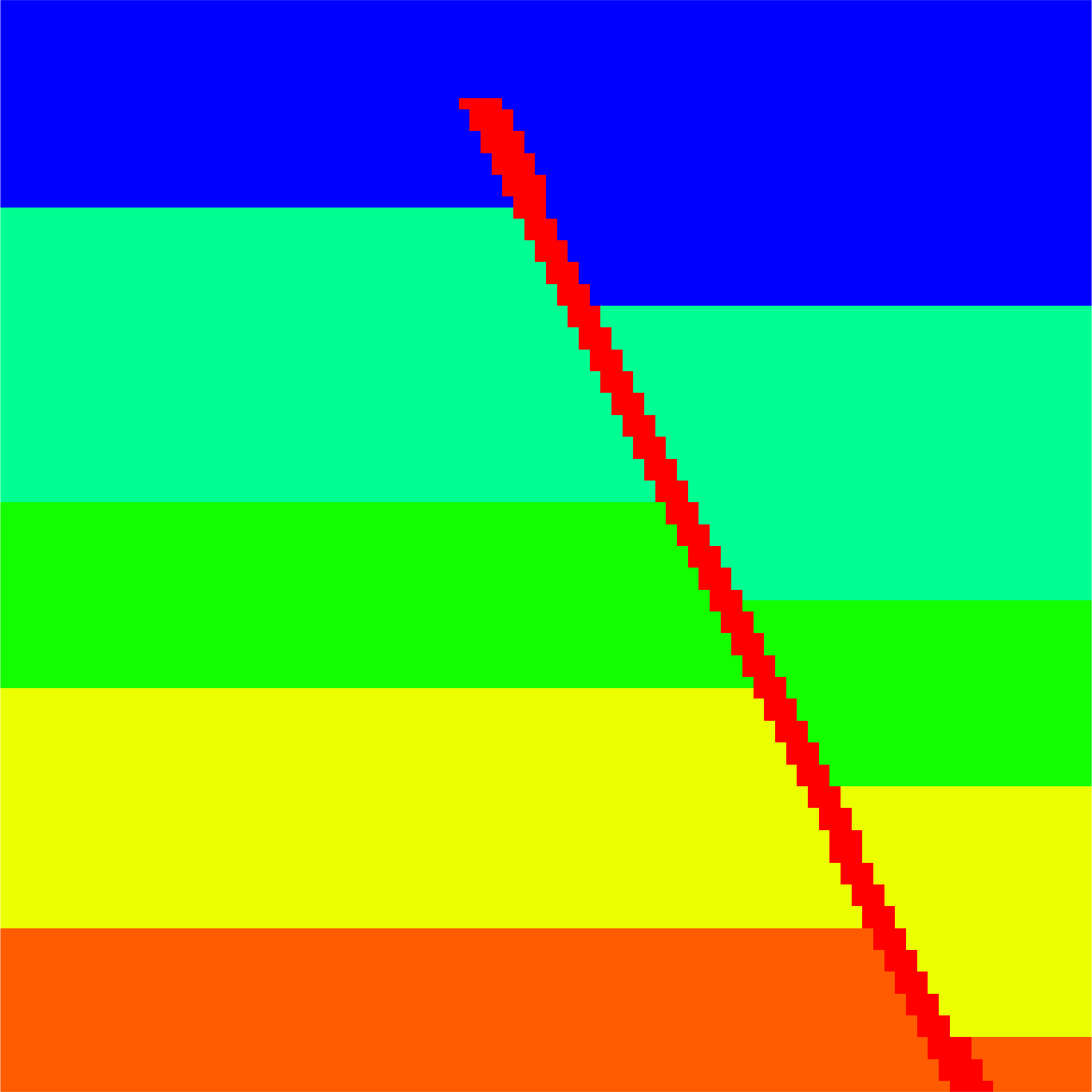}}
}
\caption{We juxtapose four inverted velocity models~(Col~1 to 4) and the ground-truth~(Col~5) on FlatVel. The physics-driven models~(Cols~1~(PRE) and 2~(MTV)) produce significant defects. The CNN model~(Col~3) yields much better results and CNN-CRF model~(Col~4) generates the most accurate velocity estimation and captures the subsurface structure.}
\label{fig:flat_visualize}
\end{figure*}

\begin{figure*}[h]
\centering
\centerline{
\subfloat{\includegraphics[width=0.23\textwidth]{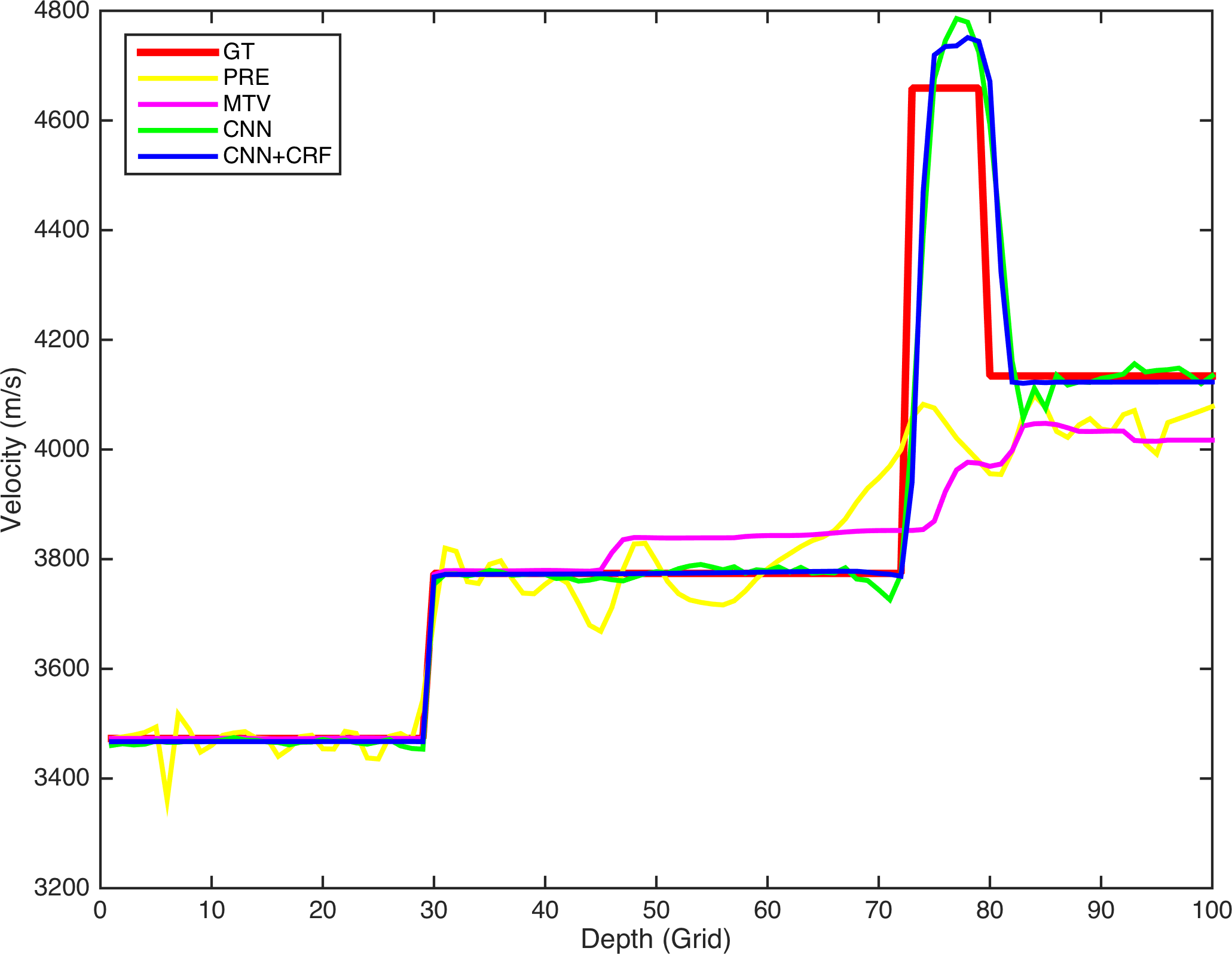}}%
\hspace{0.1cm}
\subfloat{\includegraphics[width=0.23\textwidth]{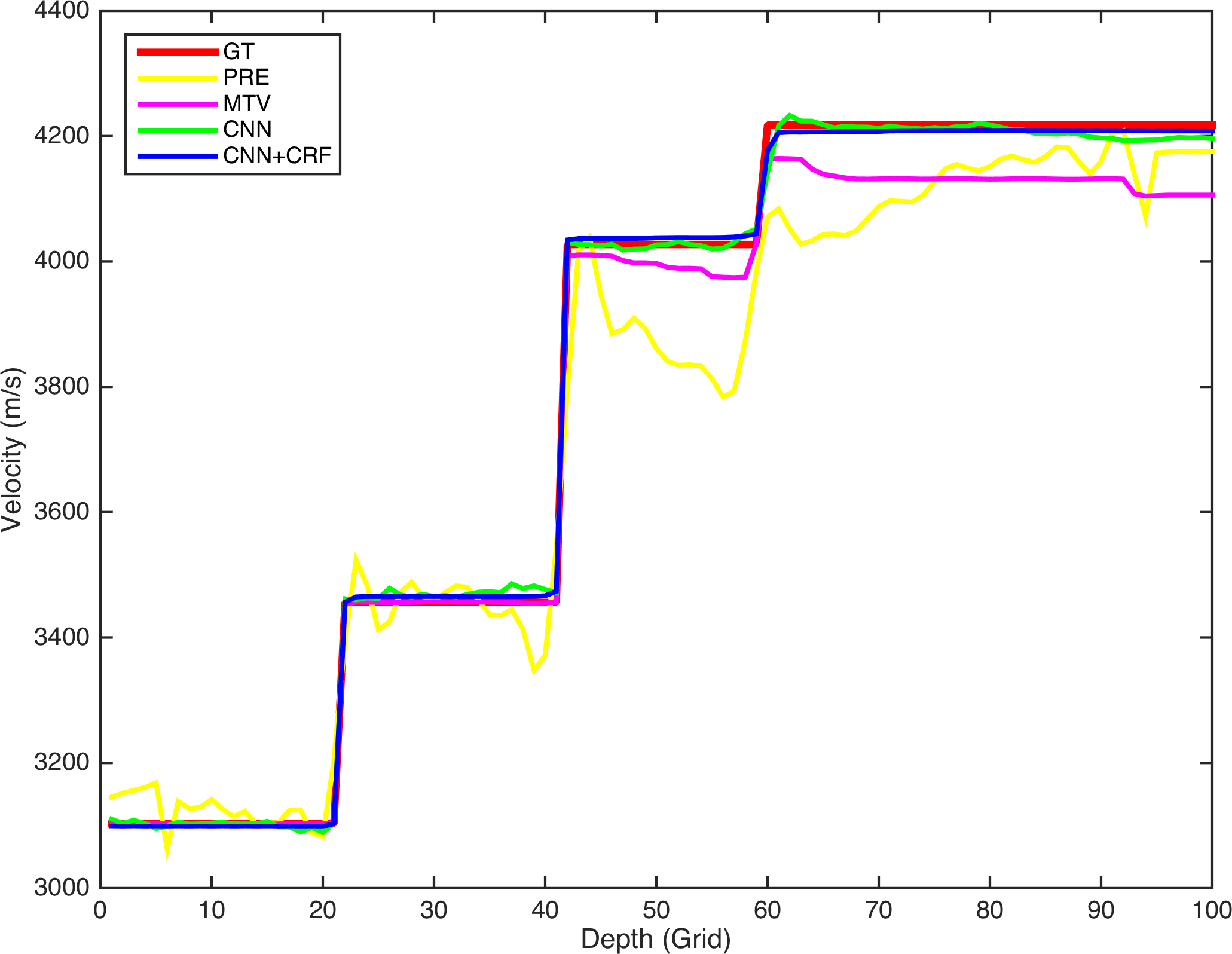}}
\hspace{0.1cm}
\subfloat{\includegraphics[width=0.23\textwidth]{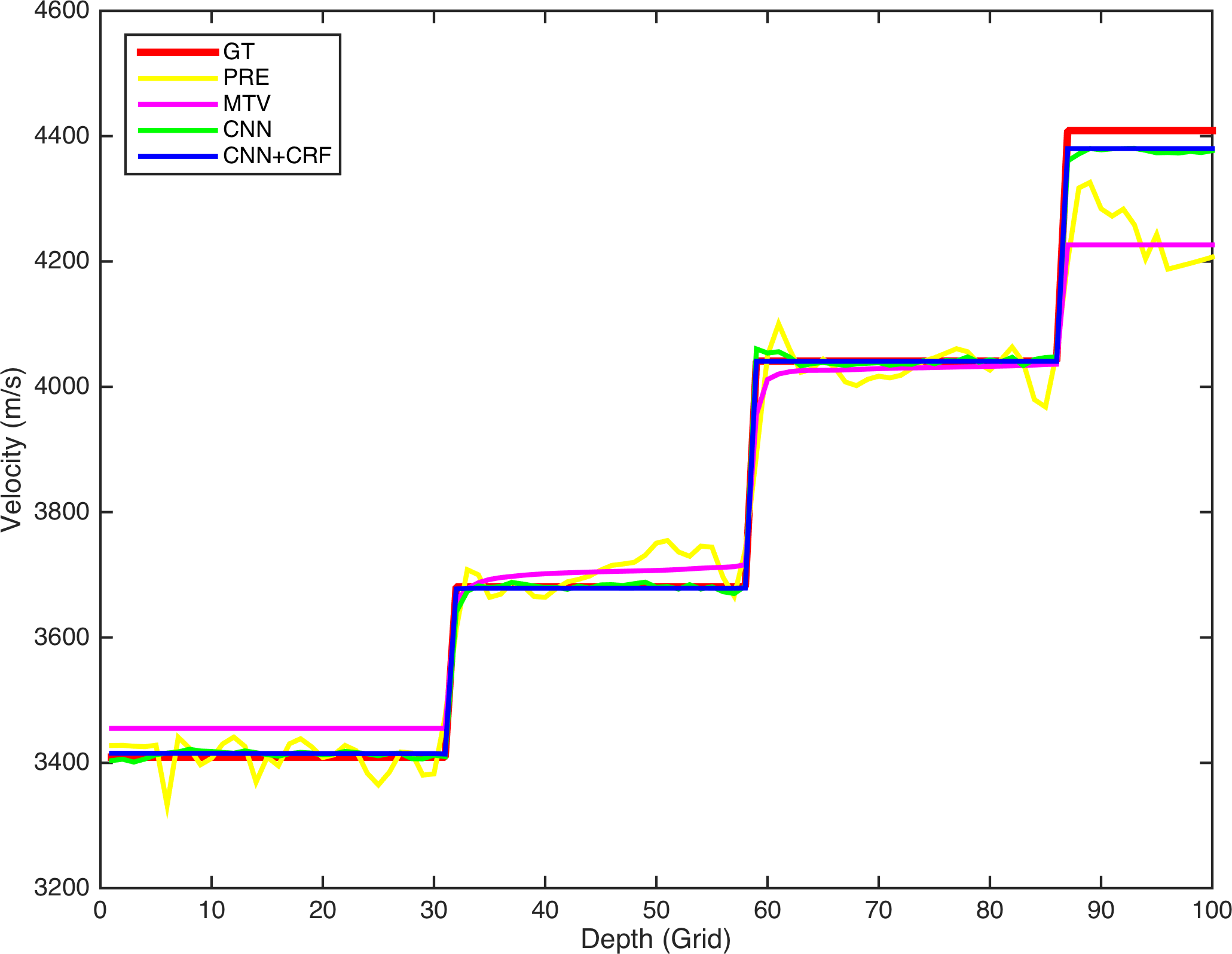}}
\hspace{0.1cm}
\subfloat{\includegraphics[width=0.23\textwidth]{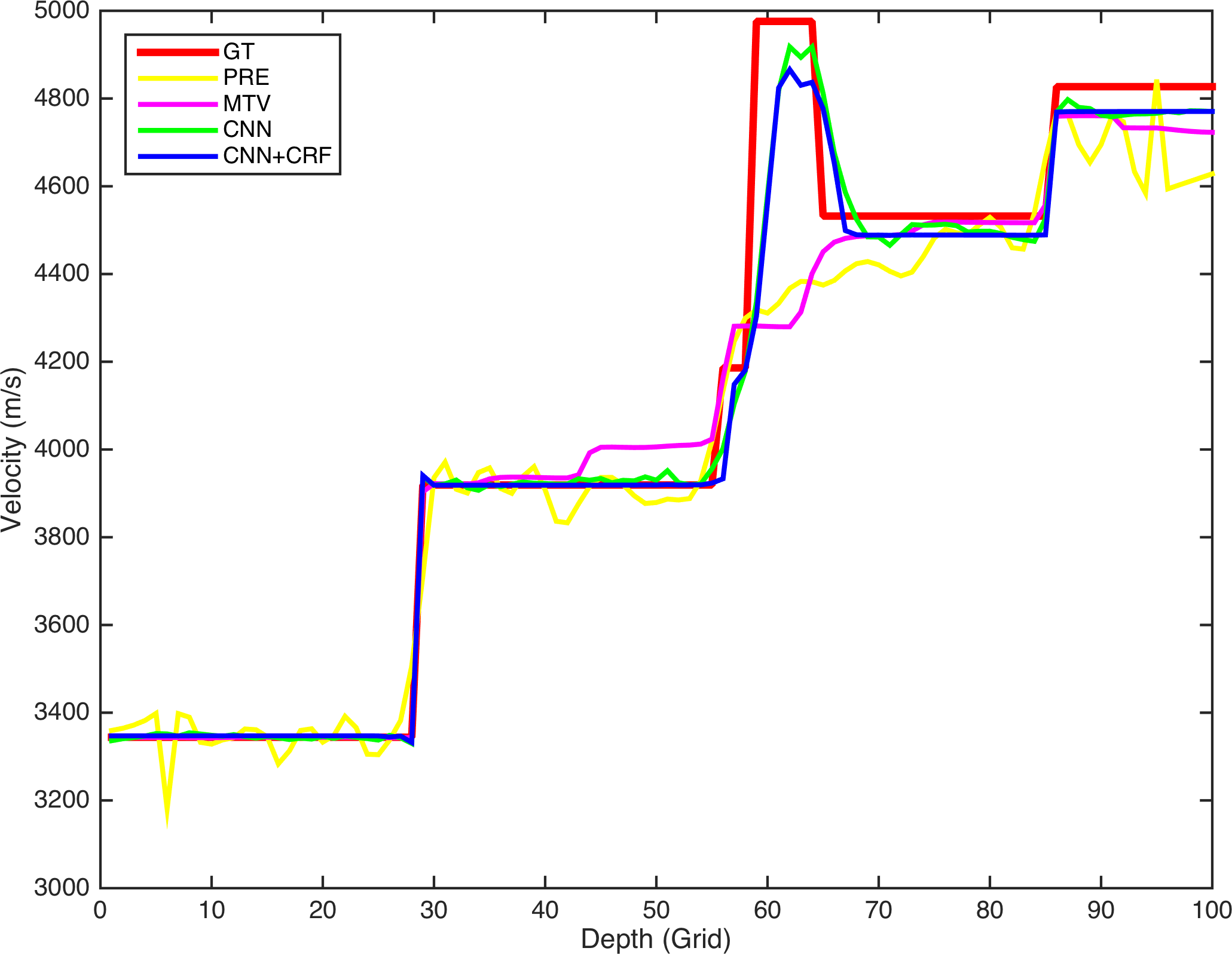}}
}
\caption{We plot the  profile comparison of the four cases in Fig.~\ref{fig:flat_visualize} to ground-truth (in red). The physics-driven methods~(in yellow and magenta) produce oscillated velocity values whereas the velocity reconstruction given by the data-driven methods~(in green and blue) essentially match the ground-truth.}
\label{fig:flat_profile}
\end{figure*}



\begin{figure*}[t]
\begin{center}
\includegraphics[width=0.9\textwidth]{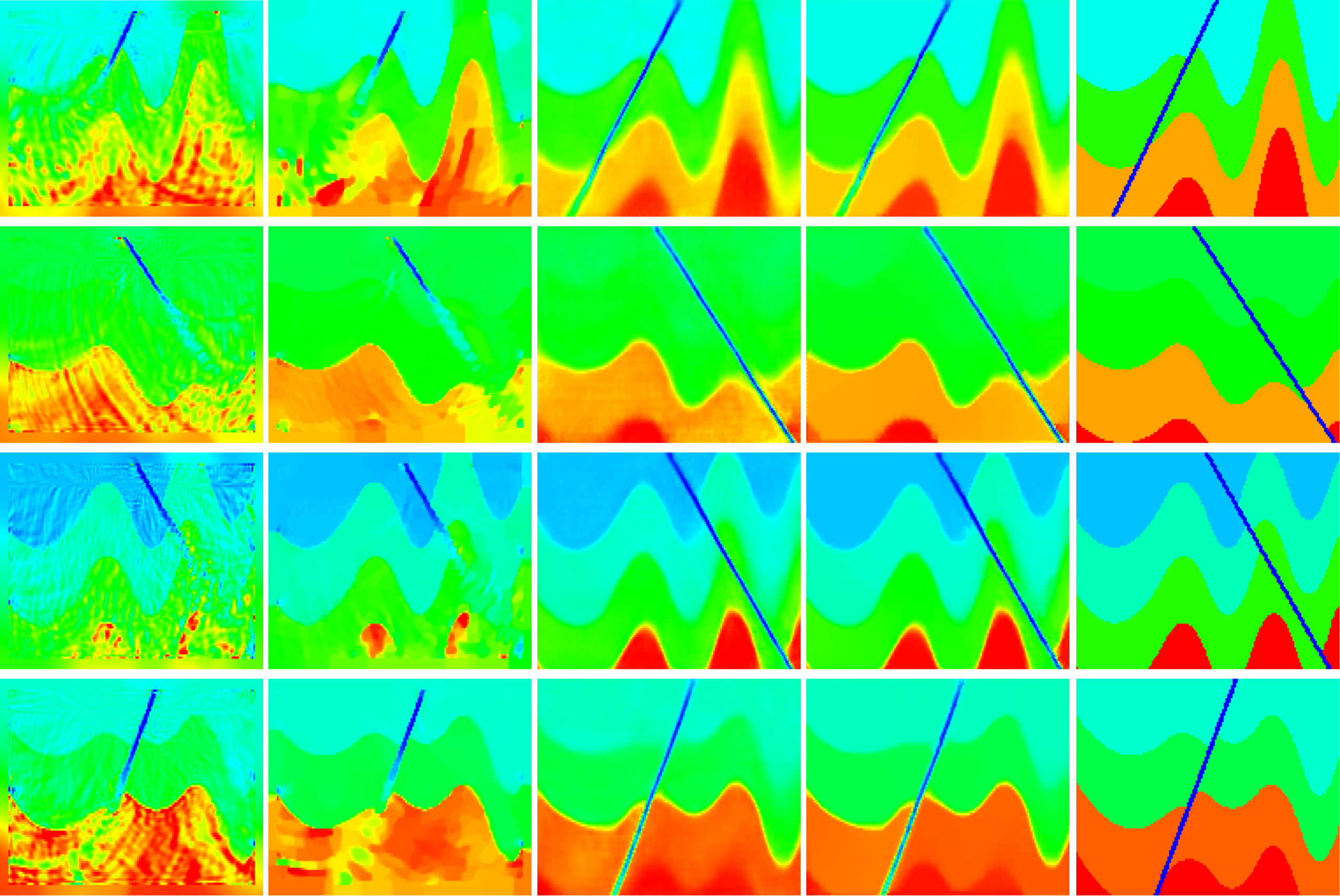}
\caption{We juxtapose four inverted velocity models~(Col~1 to 4) and the ground-truth~(Col~5) on CurvedVel. The physics-driven models~(Cols~1 and 2) yield inversion results with large amount of artifacts. The CNN-CRF model~(Col~4) generate more accurate velocity reconstruction with only some artifacts on boundaries.}
\label{fig:curved_visualize}
\end{center}
\end{figure*}

\begin{figure*}[h]
\begin{center}
\includegraphics[width=\textwidth]{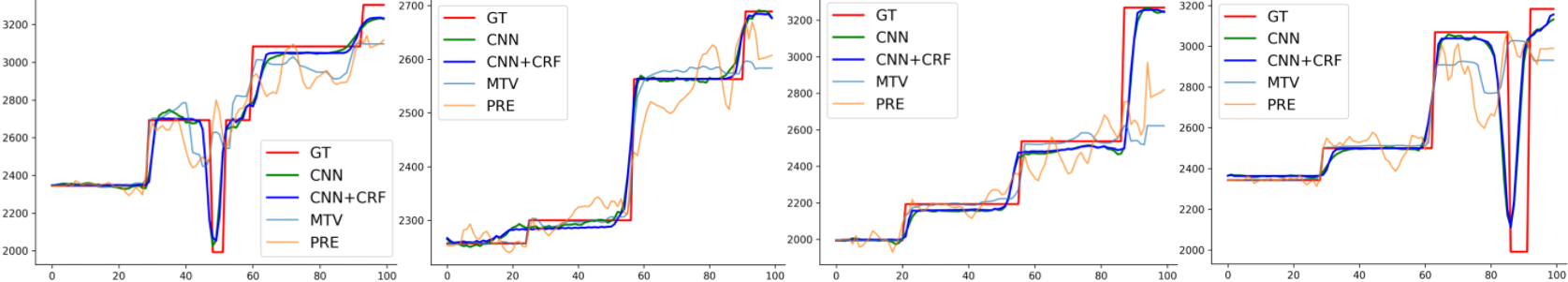}
\caption{We plot the  profile comparison of the four cases in Fig.~\ref{fig:curved_visualize} to ground-truth (in red). The physics-driven methods~(in light blue and orange) produce oscillated velocity values whereas the velocity reconstruction given by the data-driven methods~(in green and blue) yield much more accurate velocity values.}
\label{fig:curved_profile}
\end{center}
\end{figure*}

\begin{figure*}[h]
\centering
\centerline{
\subfloat{\includegraphics[width=0.17\textwidth]{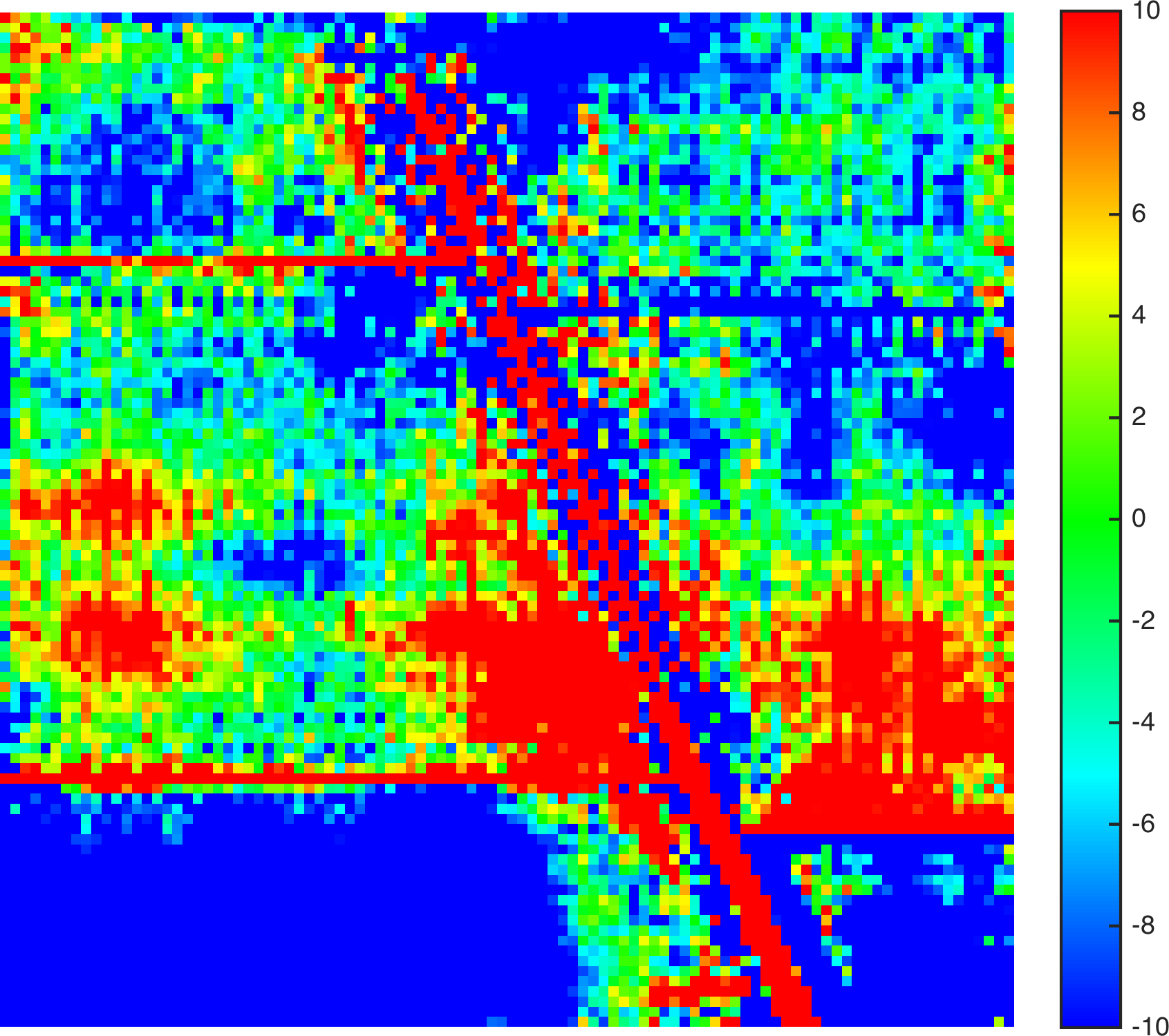}}%
\hspace{0.1cm}
\subfloat{\includegraphics[width=0.17\textwidth]{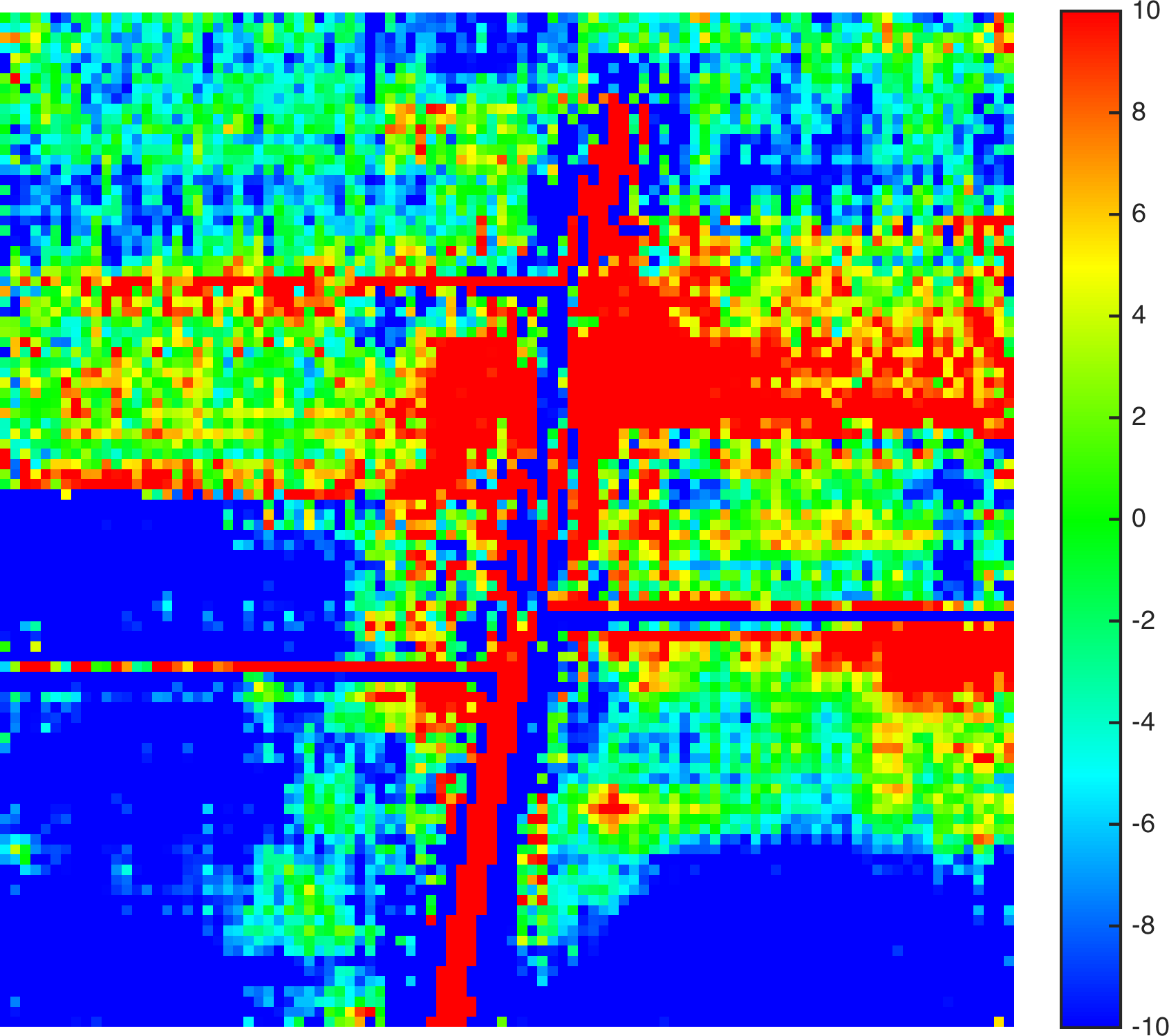}}
\hspace{0.5cm}
\subfloat{\includegraphics[width=0.25\textwidth]{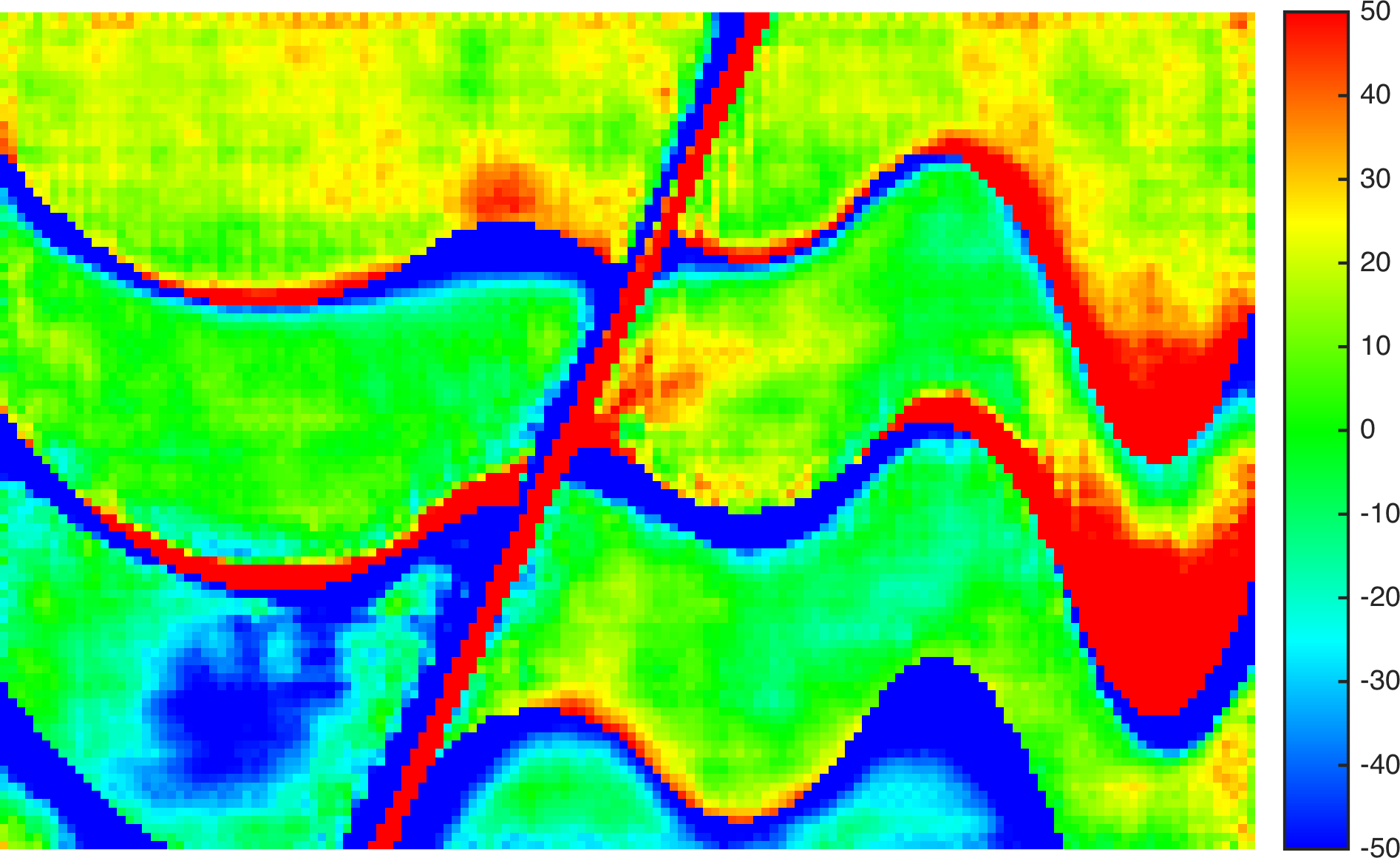}}
\hspace{0.1cm}
\subfloat{\includegraphics[width=0.25\textwidth]{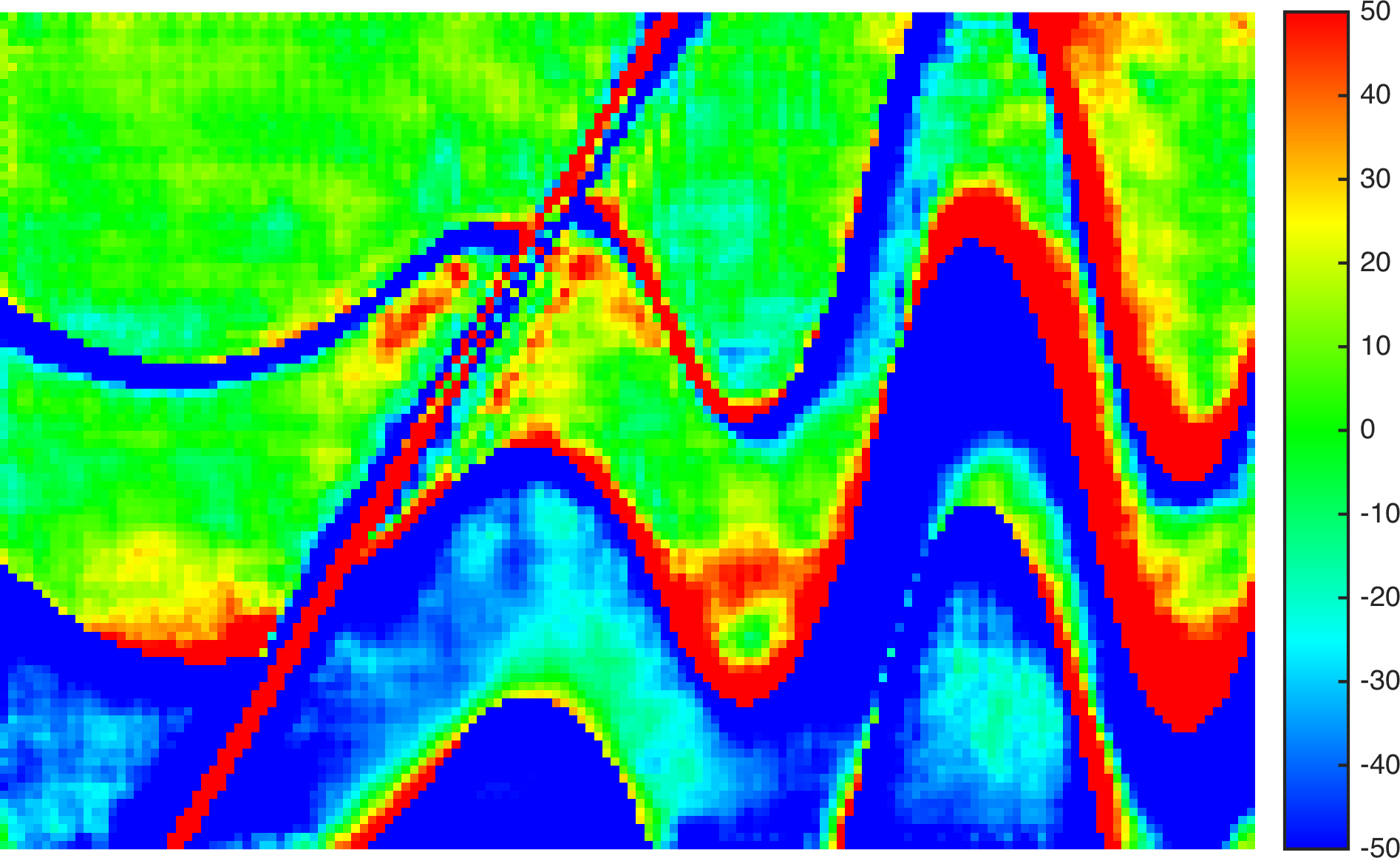}}
}
\vspace{0.1cm}
\centerline{
\subfloat{\includegraphics[width=0.17\textwidth]{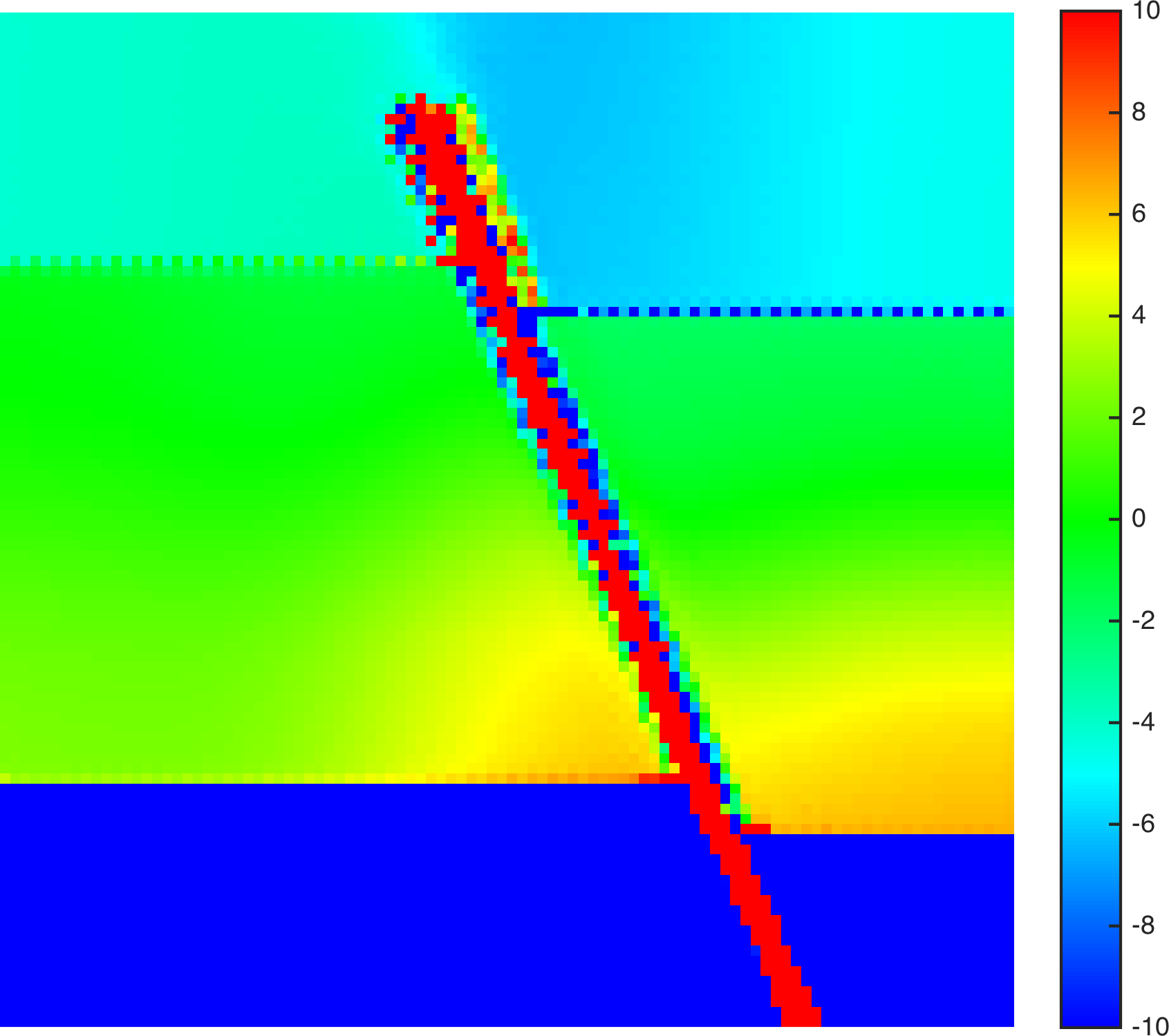}}%
\hspace{0.1cm}
\subfloat{\includegraphics[width=0.17\textwidth]{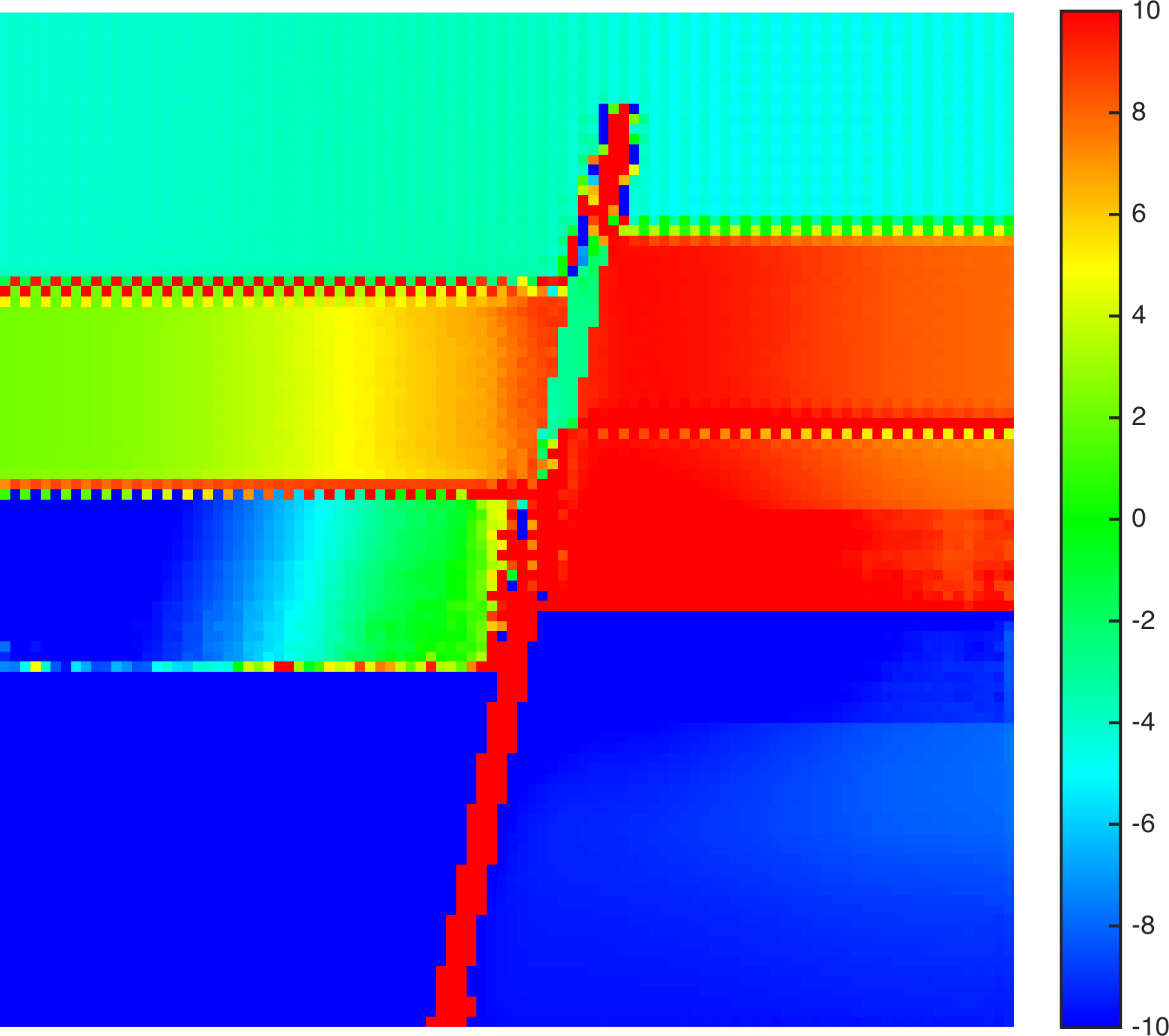}}
\hspace{0.5cm}
\subfloat{\includegraphics[width=0.25\textwidth]{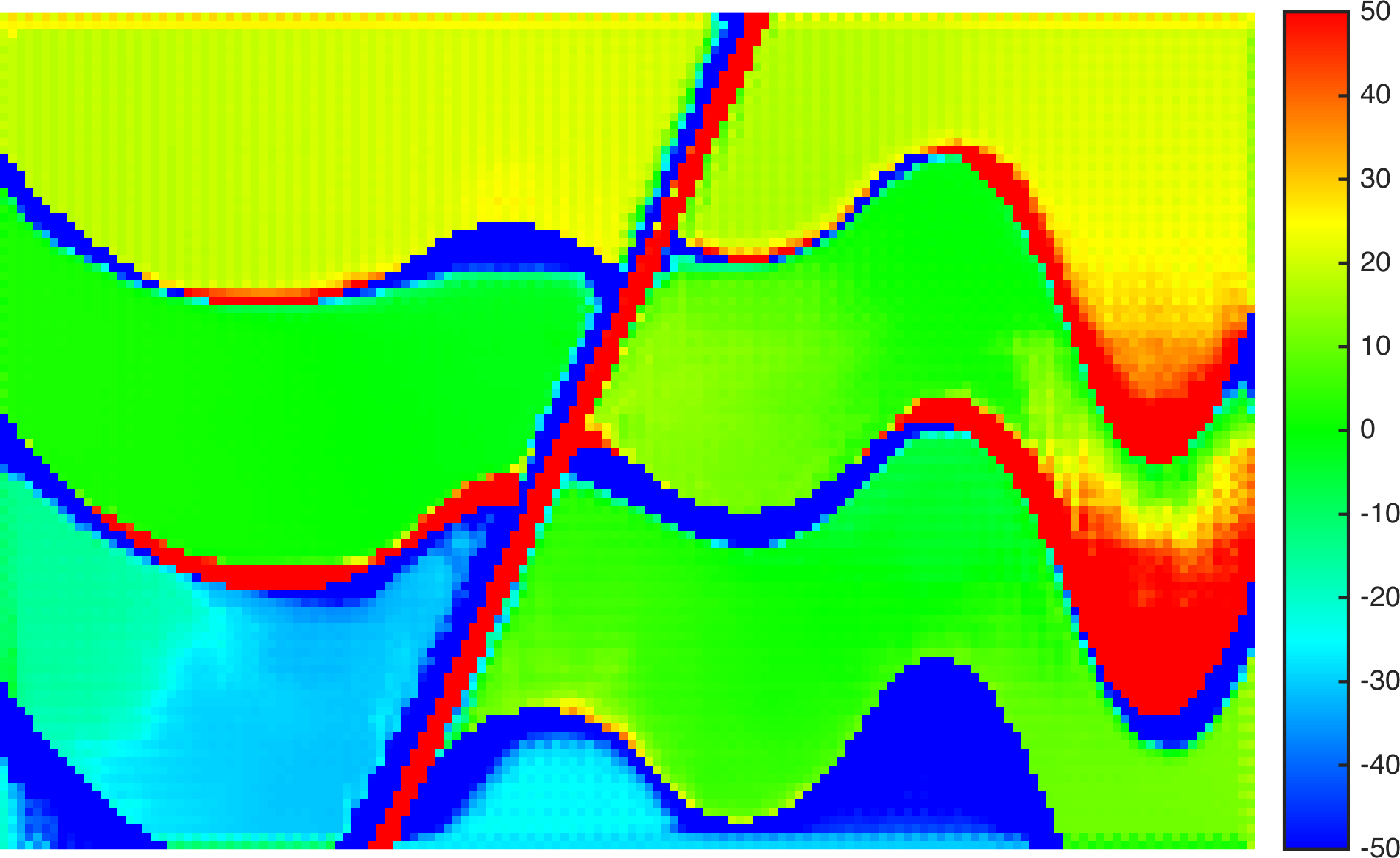}}
\hspace{0.1cm}
\subfloat{\includegraphics[width=0.25\textwidth]{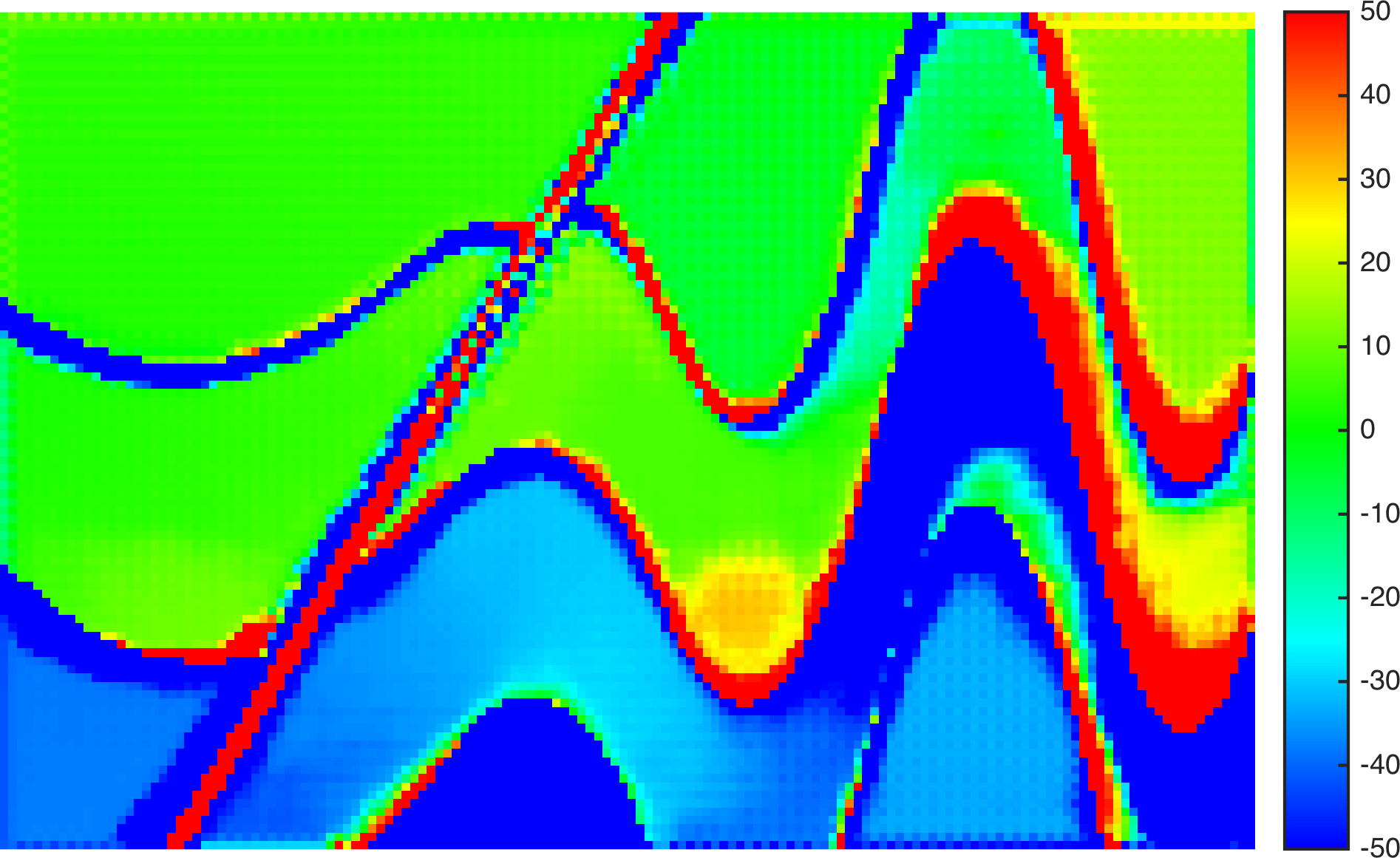}}
}
\caption{We show four cases in this figure to demonstrate that inversion artifacts are alleviated using our CNN-CRF model~(bottom) by comparing with the inversion results given by the CNN model~(top).}
\label{fig:cnn_crf_diff}
\end{figure*}

\begin{figure*}[ht]
	\centering
	\subfloat[]{\includegraphics[width=.45\linewidth]{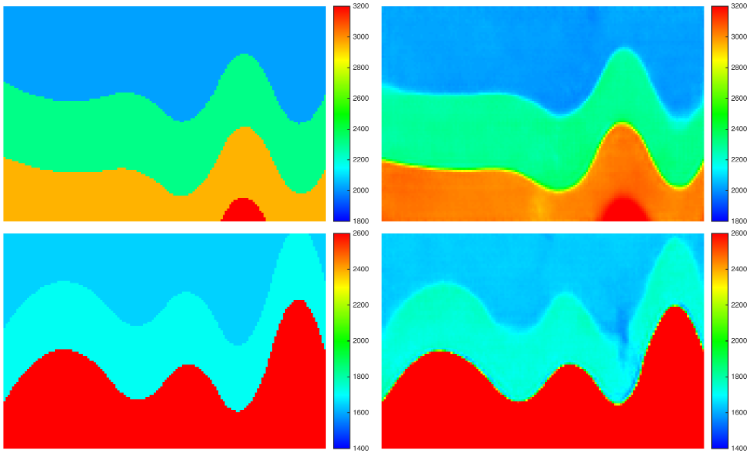}}
	\quad
	\quad
	\quad
	\subfloat[]{\includegraphics[width=.46\linewidth]{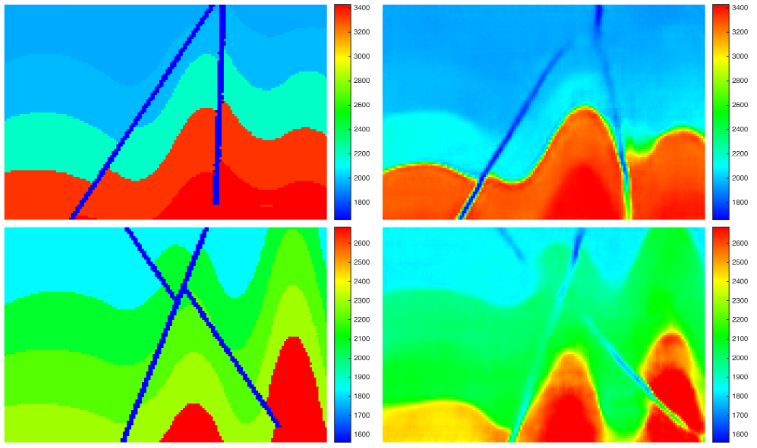}}
	\caption{Two velocity models without geologic faults~(a) and two velocity models with two geologic faults~(b) are used to demonstrate the generalization of our InversionNet. The results demonstrate that our InversionNet has the generalization ability.}
	\label{fig:Robustness_Tests}
\end{figure*}

\begin{figure*}[ht]
	\centering
	\centerline{
		\subfloat[]{\includegraphics[width=.30\linewidth]{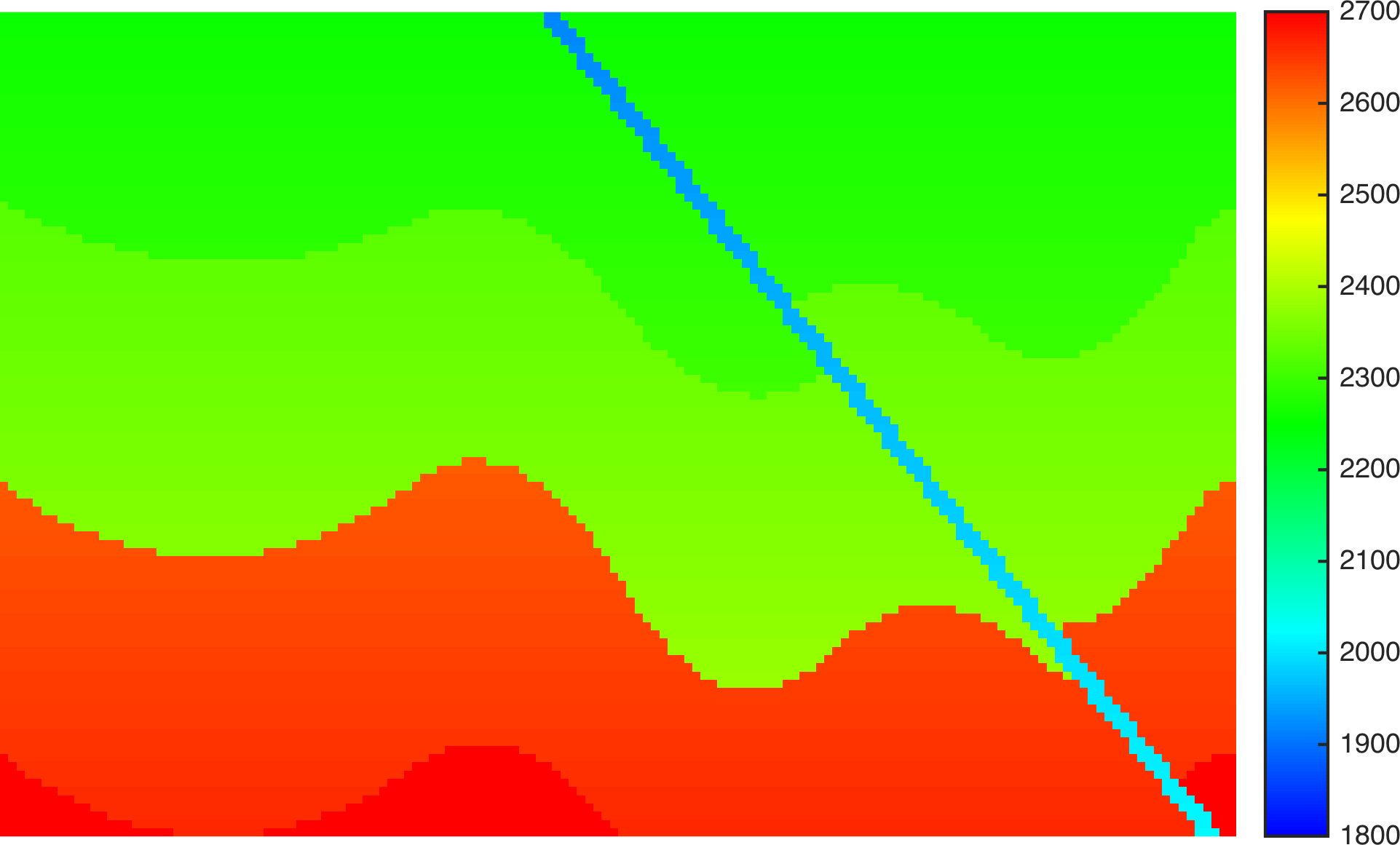}}
		\quad
		\quad
		\quad
		\subfloat[]{\includegraphics[width=.30\linewidth]{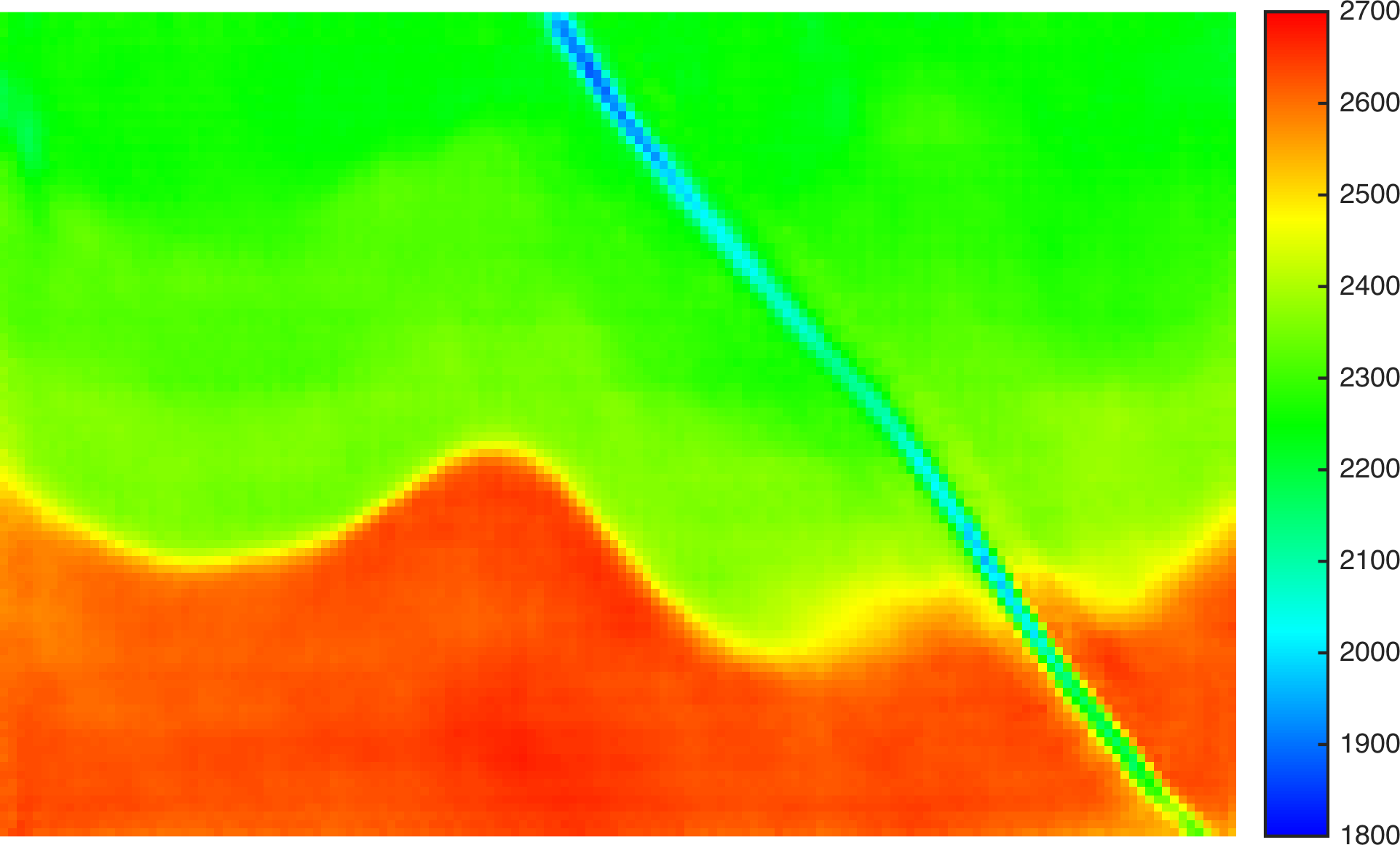}}
		\quad
		\quad
		\quad
		\subfloat[]{\includegraphics[width=.23\linewidth]{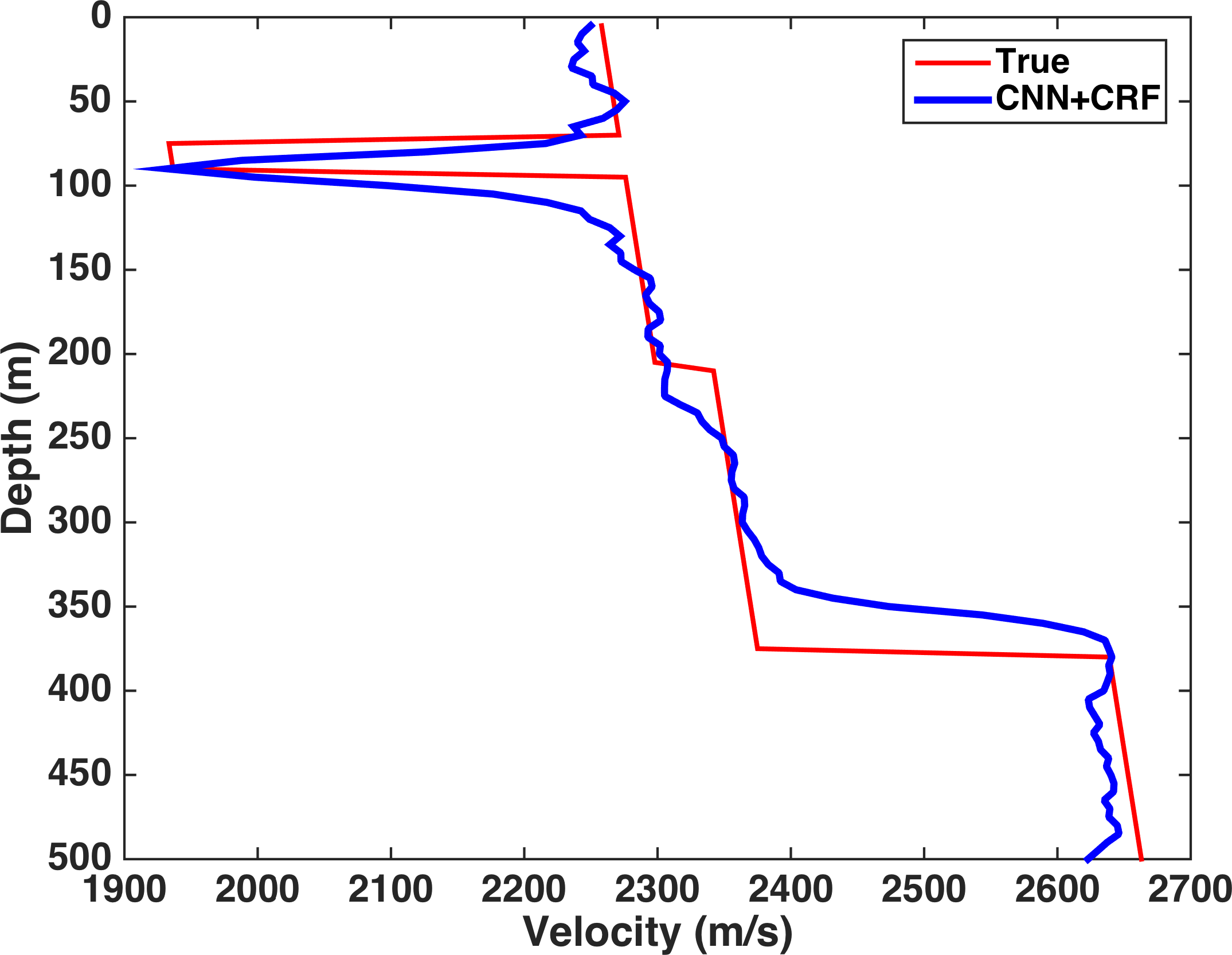}}}
	\caption{A velocity model with smoothly changed values. The ground truth is shown in (a), and the reconstruction is shown in (b). The vertical profile profile is shown in (c). Our InversionNet captures the trends of the velocity change reasonably well.}
	\label{fig:SmoothModel_Results}
\end{figure*}


\section{Experimental Results}

We compare the proposed methods with two physics-driven baselines --- full-waveform inversion with energy weighted preconditioner~(FWI-PRE), which is a wave-energy-based precondition method that aims to reduce the artifacts in the gradients caused by the geometrical spreading and defocusing effects~\cite{Zhang_seg_2012}, and full-waveform inversion with modified total-variation~(FWI-MTV), where the modified total-variation~(MTV) regularization is used in FWI optimization process~\cite{Quantifying-2015-Lin}. MTV is designed to preserve sharp interfaces in piece-wise constant structures. For all the test below, we employ a smooth model obtained by averaging the ground-truth by two wavelength as the initial guess in both these two physics-driven approaches. 

In addition to the two physics-driven baseline models we also test the performance of adding residual blocks~\cite{ResNet}, a state-of-the-art CNN building block, to the encoder. We build residual blocks as follow
\begin{equation}
    x^{(l + 1)} = x^{(l)} + \mathcal{F}^{'}(\mathcal{F}(x^{(l)})),
\end{equation}
where $\mathcal{F'}$ and $\mathcal{F}$ are two convolution blocks as defined in Eq.~\eqref{eq:conv_block}. 

\subsection{Test on FlatVel}
We show the quantitative results on FlatVel dataset in Table~\ref{table:results_FlatVel}. The two physics-driven models take more time to predict yet still have higher errors comparing with the other data-driven models. Among the data-driven models, we surprisingly find that the plain CNN outperforms its residual counterpart by a margin. For the CRF we test three different values for the window size $d$. The best performance is achieved when $d = 20$. We plot four velocity models given by each model in Fig.~\ref{fig:flat_visualize}. The two physics-driven models yield inaccurate velocity models. In particular, the PRE method performs the worst. The modified total-variation regularization helps to improve the inversion. However, the inversion results are still off compared to the ground-truth. Specifically, in the deep and boundary regions, where the data coverage becomes sparse, both PRE and MTV produce large artifacts. On the other hand, the boundaries and the fault are captured remarkably better using CNN model. The CRF ($d=20$ in the plot) further refines the velocity values within each layer by enforcing consistency. We also provide the profile (Fig.~\ref{fig:flat_profile}) at the horizontal offset 50. The profiles depict the velocity at different depths. We again observe the inaccuracy and inconsistency predictions by physics-driven models. Quantitatively, the profile of CNN matches the ground-truth, and the values near boundary and deep regions are further improved when coupled with CRF.

\subsection{Test on CurvedVel}
For CurvedVel, we also provide the quantitative results on in Table~\ref{table:results_CurvedVel}, the visualized velocity models in Fig.~\ref{fig:curved_visualize} and the profile drew on the horizontal offset~50 in Fig.~\ref{fig:curved_profile}. In this sets of subsurface velocity models, the curved geologic layers produce irregular reflection and generate significantly imbalanced data coverage, which makes the inversion much more challenging. In Table~\ref{table:results_CurvedVel}, we observe that although the accuracy decreases, the data-driven methods still outperform the physics-driven baselines. The comparison between data-driven models agrees with the test on FlatVel that adding residual blocks degrades the performance and the CNN-CRF model yields the best results. Figure~\ref{fig:curved_visualize} shows that physics-driven methods generate velocity models with large amount of artifacts. The reconstructions of the geologic faults are incomplete in the deep regions, whereas the proposed CNN-CRF model produces more accurate values within layers and the geologic fault captured is  significantly better. Furthermore, we also provide the profiles of different inversion methods in Fig.~\ref{fig:curved_profile}. Through the comparison, we observe that in general our methods still yield the most accurate velocity values compared to those physics-driven approaches. 

\subsection{Effectiveness of Conditional Random Field}

To better illustrate the effectiveness of the conditional random field, we use two velocity models from each FlatVel~(left) and CurvedVel~(right) to plot the velocity difference between the CNN and GT~(top), CNN-CRF and GT~(bottom) in Fig.~\ref{fig:cnn_crf_diff}. The red or blue regions indicate where the mean absolute error is high. We observe that the CRF further alleviate the inversion artifacts generated in the homogeneous regions, which provides better characterization of the subsurface structure.

\subsection{Tests on Robustness and Generalization}

\subsubsection{Test with Noisy Data}

\begin{table}[h]
\centering
\begin{tabular}{ c|ccccc }
& 15 dB & 20 dB & 25 dB & 30 dB & clean \\
\hline
mae & 70.30 & 70.05 & 69.01 & 68.49 & 68.70\\
\end{tabular}
\caption{The mean absolute errors of inversion results using our model under different noise levels. We demonstrate that our model is robust to some levels of additive Gaussian noise.}
\label{table:results_noise}
\end{table}

To further verify the robustness of our model to additive noise, we impose Gaussian noise to the seismic measurements using four different levels of 15~dB, 20~dB, 25~dB and 30~dB on CurvedVel. We compare the mean absolute error of the inversion results using our CNN models achieved on the testing set. The results in Table~\ref{table:results_noise} indicates that the performance of our data-driven model is robust to some levels of additive Gaussian noise.

\subsubsection{Test with Velocity Models without Fault and with two Faults}
It is important to understand the generalization ability of our InversionNet. In our training sets, we have created velocity models with only one geologic fault as shown in Fig.~\ref{fig:TrainingSeismicModel}. In this test, we provide a few velocity models either without any geologic faults~(left panel of Fig.~\ref{fig:Robustness_Tests}) or with multiple geologic faults~(right panel of Fig.~\ref{fig:Robustness_Tests}). Particularly, in the left panel of Fig.~\ref{fig:Robustness_Tests} we created two velocity models without any geologic faults. These two velocity models contain different number of layers, one with four layers and the other contain three layers. We observe that our InversionNet produces promising reconstructions even though there is not any velocity models without fault in the training set. Similarly, we provide the inversion results of velocity models with multiple faults in 
the right panel of Fig~\ref{fig:Robustness_Tests}. The reconstruction of velocity models with multiple geologic faults can be much more challenging, which can be observed in the right panel of Fig.~\ref{fig:Robustness_Tests}. Our InversionNet does provide reasonable overall reconstruction of velocity, however, the shape and velocity of the geologic faults are somehow degraded. Through these generalization tests, we conclude that our InversionNet learns the intrinsic correspondance between features and data, which is the inverse operator in our problems. 

\subsubsection{Test with Velocity Model with Smoothly Changed Values}

Subsurface model with constant velocity in each layer  is a reasonable assumption for subsurface geologic formation~\cite{Introduction-2009-Shearer}. However, in some realistic cases the velocity models might not be constant in each layer. It becomes necessary to consider the performance of our InversionNet in reconstructing velocity model with smoothly changed value. This can be another challenging task due to the fact that all the velocity models in our training set consist of constant value in each layer as shown in Fig.~\ref{fig:TrainingSeismicModel}. To test on this task, we create a velocity model with a single fault zone as shown in Fig.~\ref{fig:SmoothModel_Results}. The velocity value in each layer is gradually increased with respect to depth. We provide the inversion result using our InversionNet in Fig.~\ref{fig:SmoothModel_Results}. It can be observed that our InversionNet captures the overall geologic features including fault zone and the layers. We further provide a vertical profile located at the middle of the velocity model as shown in Fig.~\ref{fig:SmoothModel_Results}. Our InversionNet captures the trends of the velocity change reasonably well. Through this test, we show that our InversionNet yields some robustness and generalization to the smooth velocity models.




\section{Conclusions}
We develop a novel data-driven method that harnesses the power of the CNN and CRF to solve the problem of full-waveform inversion. Our CNN model consists of an encoder and a decoder. The encoder utilizes a set of convolution layers to encode seismic waves collected from multiple receivers into a high-dimensional feature vector. The decoder employs a set of deconvolution layers to decode the vector into velocity models. We further build a locally connected CRF to refine the velocity values near boundaries and faults so that the subsurface structure can be better revealed. We demonstrate through our experiments that our CNN-CRF model obtains the best results on the two synthetic velocity datasets. Through the robustness test on addictive noise, we demonstrate that our model is robust to noise. Therefore, our CNN-CRF model exhibits great potential for solving full-waveform inversion problems. 


\bibliographystyle{bst} 
\bibliography{reference}

\end{document}